%% file: main.tex
\documentclass[3p,12pt]{elsarticle}
\usepackage[fleqn]{amsmath}
\usepackage{cases}
\usepackage{booktabs}
\usepackage{multirow}
\usepackage{xcolor}
\usepackage[normalem]{ulem}
\usepackage{tabularx}
\usepackage{siunitx}
\usepackage{comment}
\usepackage[ruled, linesnumbered]{algorithm2e}
\usepackage{algpseudocode}
\usepackage{grffile}
\usepackage{rotating}
\usepackage{lineno}
\usepackage{hyperref}
\usepackage{graphicx,enumerate}
\usepackage{amssymb}
\allowdisplaybreaks[4]
\usepackage{bm}
\usepackage{subfigure}
\usepackage{caption,color}
\usepackage{subeqnarray}
\usepackage{multirow}
\usepackage{lscape}
\usepackage{rotating}
\usepackage{graphics}
\usepackage{epstopdf}
\usepackage{threeparttable,booktabs,tabularx}
\usepackage{mathtools, empheq}

\journal{Journal of Computational Physics}

\usepackage{floatrow}
\usepackage{floatrow}\newfloatcommand{capbtabbox}{table}[][\FBwidth]
\floatname{algorithm}{Algorithm}
\biboptions{numbers,sort&compress} 

\begin{document}

\begin{frontmatter}

\title{ 
General synthetic iterative scheme for multiscale radiative transfer in the finite-volume framework
}

\author{Kaiyuan Wang}
\author{Yanbing Zhang}
\author{Qi Li}
\author{Lei Wu\corref{mycorrespondingauthor}}
\cortext[mycorrespondingauthor]{Corresponding author}
\ead{wul@sustech.edu.cn}

\address{Department of Mechanics and Aerospace Engineering, 
Southern University of Science and Technology, Shenzhen 518055, China}

\newcommand{\leib}[1]{{\leavevmode\color{blue}#1}}

\begin{abstract}
Achieving efficient and accurate simulation of the radiative transfer has long been a research challenge. Here we introduce the general synthetic iterative scheme as an easy-to-implement approach to address this issue. First, a macroscopic synthetic equation, which combines the asymptotic equation at the diffusion limit and the ``high-order terms" extracted from the transport equation to account for transport effects, is introduced to accelerate the simulation of the radiative transfer equation. Second, the asymptotic preserving property is directly provided by the macroscopic process, eliminating the need for fine spatial discretization in optically thick media, as well as the need for consistency enforcement. Third, to address the issue of opacity discontinuity in the finite volume method, an adaptive least square method for gradient approximation  is proposed. 
Numerical results on several canonical tests demonstrate that, in optically thick problems, our method achieves significant speed-up over the conventional iterative schemes. Finally, with our newly developed method, we reveal the importance of resolving the Knudsen layer  in the initial stage of Tophat problem, while in steady-state the Knudsen layer can be under-resolved. 
\end{abstract}


\end{frontmatter}

\input{data/sec01}

\input{data/sec02}

\input{data/sec03}

\input{data/sec04}

\input{data/gradient}

\input{data/sec05}

\input{data/sec06}

\appendix

\input{data/appendix}

\bibliographystyle{elsarticle-num} 
\bibliography{ref/references}

\end{document}

%% file: data/sec01.tex
\section{Introduction}
\label{Introduction}

Radiation transport plays a crucial role in various fields, including astrophysics \cite{Zheleznyakov1996}, high-energy physics \cite{Kallman2023}, and inertial confinement fusion \cite{grinstein2025transition}. The interaction between radiation and media primarily manifests through emission, absorption, and scattering. Depending on the opacity of the medium, radiation transport exhibits multiscale behavior. Specifically, in optically thin media, photon propagation displays particle-like behavior, whereas in optically thick media, photons exhibits diffusive behavior due to frequent absorption and scattering. The radiative transfer equation (RTE) is used to model the interaction between radiation and matter from the diffuse to free-transport regimes.

Numerical methods for solving the transport equation have a long history of development and can be  categorized into deterministic and stochastic methods \cite{James1979}. Deterministic methods directly solve the discretized transport equation, such as the discrete ordinate method ($S_N$) \cite{osti_4376236}, spherical harmonics method ($P_N$) \cite{OU1982271}, finite-volume method \cite{Raithby1990}, and finite-element method \cite{Razzaque1983}, separated by different treatments to the solid angle. These methods transform the solution of the high-dimensional differential-integral equation into the solution of a discrete large linear system, which are solved by sophisticated numerical methods.
The stochastic Monte Carlo method \cite{FLECK1971313,howell1998monte} tracks the photon transport and scattering through statistical random sampling. Because of its high fidelity in modeling physical processes, the results obtained by the Monte Carlo method are often used as benchmark solutions. 

While both the deterministic and stochastic methods are efficient in optically thin regimes, they suffer from two major drawbacks in optically thick regimes, i.e., slow convergence and large numerical dissipation. On one hand, the strong scattering makes it difficult for local photon transport to influence the entire computational domain, causing low efficiency in algorithms based on streaming and collision at the kinetic scale. On the other hand, to reduce numerical dissipation and nonphysical solutions, the mesh grid size is constrained by the mean free path of photons \cite{larsen1987asymptotic,larsen1989asymptotic,Larsen01121992,farmer1998comparison,Erturk2018}. 
Therefore, solving the transport equation for optically thick problems usually incurs a high (and sometimes prohibitive) computational cost. 


To address the issue of slow convergence, techniques of ``synthetic acceleration" were proposed in the 1960s \cite{ADAMS20023,Kopp01091963}, including the quasi-diffusion \cite{GOLDIN1964136,Wieselquist2014}, transport synthetic acceleration \cite{Ramone01031997}, diffusion synthetic acceleration (DSA) \cite{Alcouffe01101977,Larsen01011984}, and $S_2$ synthetic acceleration \cite{Lorence01041989}. Among these methods, the DSA, which was originally developed for neutron transport and later extended to radiation transport \cite{Morel01091982,ALCOUFFE1985}, is the most well-known.  
While theoretical analysis at the continuous level indicates that the DSA is effective, its practical effectiveness and stability  depends highly on the consistency in the discretizations of the transport and diffusion equations \cite{ADAMS20023}. 
However, achieving consistency and stability is often complicated \cite{Larsen01091982, McCoy01091982,AZMY2002213, Warsa2004}, particularly in scenarios involving non-orthogonal mesh grids and discontinuous opacity. 


The high-order/low-order algorithm  represents another class of synthetic acceleration method. A moment based low-order system, i.e., the $P_1$ equation for radiation transport, is applied to accelerate the high-order RTE, and the high-order equation provides closure for the low-order system \cite{Park01052012,CHACON201721}. In the earlier work of Park \cite{Park01052012}, the $P_1$ equation is discretized in finite-volume method, and artificial consistent terms are added to enslave the solution of the synthetic equation to the RTE. Although fast convergence is achieved, the synthetic solution does not recover the correct asymptotic limit due to the ``discontinuous reconstruction for representing the emission source''. In the later work \cite{Park01112020}, the problem is fixed by redesigning the discretized $P_1$ equation. 

Since the consistency is required, the numerical error of the discrete transport equation can affect the evolution of the numerical solution at a scale much larger than the mean free path.  Larsen~\textit{et al.} analyzed several numerical schemes, thereby introducing the early concept of asymptotic preserving in optically thick regimes~\cite{larsen1987asymptotic,larsen1989asymptotic}. Subsequently, several asymptotic preserving schemes are constructed \cite{Klar1998,Jin2000,FILBET20107625}, which recover the correct asymptotic equation even on a coarse mesh that is much larger than the mean free path.

A recent significant progress in the asymptotic preserving scheme to address the issue of large numerical dissipation is the development of the unified gas kinetic scheme (UGKS) by Xu and Huang \cite{XU20107747}.
Initially applied to the BGK kinetic model in gas dynamics, UGKS was extended to linear transport equation by Mieussens \cite{MIEUSSENS2013138}, to gray and multi-frequency radiation transport equations by Sun \textit{et al.} \cite{UGKS_SUN2015,SUN2015222}, and further developed into implicit UGKS \cite{Sun_Jiang_Xu_2017}.  
The core idea of UGKS lies in the simultaneously treatment of the particle transport and collision, where the flux evaluation is based on the analytical solution of the BGK kinetic model, such that the asymptotic preserving property to the hydrodynamic level is achieved. The discrete UGKS, proposed by Guo \cite{Guo2013}, simplifies the flux calculation of UGKS, and its steady state extension has improved the computational efficiency for RTE \cite{SONG2023123799}. 
However, the efficiency of the implicit UGKS is lower than the DSA, due to its lack of ``global information exchange" in the whole computational domain.
Thus, in a recent work, the discrete UGKS is further accelerated using DSA \cite{SONG2023120349}. However, an artificial damping factor is introduced to correct the increment of radiation intensity and distribution function, which sacrifices certian acceleration rate for stability.

For various reasons, synthetic acceleration schemes that processes both the fast convergence and asymptotic preserving properties are usually designed to be complicated. 
In practical engineering applications with complicated geometry, it is crucial to achieve both properties through a single synthetic acceleration-like technique, regardless of the spatial discretization method employed. Recently, it is found that the variable Eddington Factor (VEF) method preserves both properties without enforcing the consistent discretization of the RTE and synthetic equation \cite{Olivier2017,Lou2019JCP}, but only one-dimensional problem is demonstrated.

In the field of rarefied gas dynamics, where the governing equation is the Boltzmann equation, 
 the general synthetic iterative scheme (GSIS), which possesses the fast convergence and asymptotic preserving properties, has recently been developed \cite{SU2020jcp,ZHU2021110091,Su2020siam,liu2024further, GSISPolyGas_ZENG2023, zhang2024efficient}. Initially, Valougeorgis and Naris \cite{Valougeorgis2003} proposed the synthetic iterative scheme for accelerating the simulation of special linearized rarefied gas flows (e.g., the Poiseuille flow where the flow velocity is perpendicular to the spatial coordinates, so that the resultant kinetic equation is very close to the neutron transport). Subsequently, Su \textit{et al.} and Zhu \textit{et al.} introduced the GSIS for general linear \cite{SU2020jcp} and nonlinear~\cite{ZHU2021110091} gas flows. 
For multiscale problems involving hydrodynamic flows, since GSIS can obtain the solution in dozens of iterations in the finite-difference, finite-volume and discontinuous Galerkin frameworks, with grid size much larger than the mean free path,  it has demonstrated several orders of magnitude improvement in computational time compared to the prevailing direct simulation Monte Carlo method \cite{liu2024further, GSISPolyGas_ZENG2023, zhang2024efficient}. 
In GSIS, the macroscopic synthetic equation is constructed using the Navier-Stokes equation, with corrections derived from the kinetic equation to account for rarefaction effects. This approach, where a higher-order equation provides closure to lower-order ones, is reminiscent of some moment-based acceleration methods. However, GSIS does not rely on specific discretizations \cite{Su2020siam}, and the asymptotic limit is preserved in the macroscopic synthetic equations, without adding any ``consistent terms". Additionally, GSIS is not restricted to a particular form of the collision/scattering term, making it applicable to other multiscale problems, such as dense gas flows described by the Enskog kinetic equation, as well as the multi-frequency phonon transport \cite{SHI2025113501,Zhangchuang_2021}.

In this work, we shall construct a GSIS algorithm for multiscale RTE within the finite-volume framework. We adopt the gray model in our discussion. However, the method can be straightforwardly extended to multi-frequency RTEs, as it does not rely on the specific form of the collision/scattering term \cite{SHI2025113501,Zhangchuang_2021}.
Like the VEF method, we will show that the GSIS does not necessary require the consistent discretizaiton between the RTE and synthetic equations. Also, the diffusion coefficient in the synthetic equation of GSIS is always positive, which allow stable numerical simulation, even in strong non-equilibrium photon transport. 


The rest of this paper is organized as follows. In Section \ref{RTE}, we introduce the gray RTE, and review the foundational source iteration (SI) and DSA. In Section \ref{GSIS}, we discuss the construction of GSIS, and analysis its fast convergence and the asymptotic preserving properties. Numerical schemes for both RTE and synthetic equations are presented in Section \ref{schemes}, and the special treatments of gradient are designed to handle the problem of opacity discontinuity in Section~\ref{ALSM}.  
The GSIS is tested on various conditions in Section \ref{tests}.  Section \ref{Conclusions} presents the conclusions.

%% file: data/sec02.tex
\section{Radiative Transport Equation and Iterative Methods}
\label{RTE}

In this section, the gray RTE is introduced, together with the SI and DSA \cite{LARSEN1988459,ALCOUFFE1985}. The advantages and disadvantages of the traditional DSA are discussed.

\subsection{Gray model and asymptotic behavior}

Under the frequency-independent assumption, the interaction between radiation and material is described by the gray RTE:
\begin{empheq}[left={\empheqlbrace\,}]{align}
        \frac{1}{c}\frac{\partial I}{\partial t}+\bm{\Omega}\cdot \nabla   I &=\sigma_{a}\left (\frac{1}{4\pi} \phi - I\right ) 
    + \sigma_{s}\left (\frac{1}{4\pi} \int_{\mathcal{S}^{2}} I d \bm{\Omega}- I\right ) , \label{RTE_g} \\
     C_{v}\frac{\partial T}{\partial t} &= \sigma_{a} 
        \left (\int_{\mathcal{S}^{2}} I d \bm{\Omega} - \phi \right), \label{Tm_g}
\end{empheq}
where the radiation is characterized by the intensity distribution function $I(\bm{x},\bm{\Omega},t)$, with $\bm{x}$ being the spatial position. $\bm{\Omega}$ is unit directional vector in the solid angle space $\mathcal{S}^{2}$. The speed of light is $c$. The left-hand side of Eq.~\eqref{RTE_g} represents the advection of radiation, while the right-hand side represents the interaction of radiation with the medium. The parameters $\sigma_a$ and $\sigma_s$ are cross sections accounting for emission-absorption and scattering, respectively. In this work, the emission, absorption and scattering are isotropic, and the optical parameters are formally depended on position $\bm{x}$ and material temperature $T$. The energy density of black body emission is 
\begin{equation}
    \phi=a c T^4,
\end{equation}
where $a$ is the radiation constant. The effect on material temperature, caused by energy exchange with radiation, is described by Eq.~\eqref{Tm_g}, where $C_v$ is the volume specific heat.

Let $\theta \in[0,\pi]$ and $\varphi\in[0,2\pi]$ denote the polar angle and the azimuthal angle, respectively, and $d\mathbf{\Omega} = \sin \theta d\theta d\varphi$. The radiation macroscopic quantities are obtained by integrating the distribution function in solid angle space, e.g., the radiation energy density $\rho$ is the zeroth order moment: 
\begin{equation}
    \label{def_rho}
    \rho = \int_{\mathcal{S}^{2}} I d \bm{\Omega},
\end{equation}
while the first order moment $\bm{F}$ is defined as:
\begin{equation}
    \label{def_F}
    \bm{F} = \int_{\mathcal{S}^{2}} \bm{\Omega} I d \bm{\Omega} .
\end{equation}
$\bm{F}$ is a vector, and if multiplied by $c$, it represents the radiation heat flux or the photon momentum density. For the sake of convenience, we do not distinguish between these two interpretations here. 

The total cross section is defined as $\sigma_t = \sigma_s + \sigma_a$. In optically thick regimes, the mean free path of photon is much smaller than the characteristic length, i.e., ${1}/{\sigma_t} \ll x_{\infty}$, and the evolution of physical quantities has a much larger time scale than photon transport, i.e., $t_{\infty} \gg {x_{\infty}}/{c}$. Selecting $\sigma_{\infty}$ as the reference cross section, a dimensionless parameter 
\begin{equation}
\epsilon = \frac{1}{\sqrt{c\sigma_{\infty}t_{\infty}}}
\end{equation}
is introduced, which is analogous to the Knudsen number in rarefied gas flows. By applying replacements as follows \cite{larsen1987asymptotic,MOREL1996445}:
\begin{equation}
    \label{scale_rep}
    \frac{\partial }{\partial t} \to \epsilon\frac{\partial }{\partial t}, \quad \sigma_t \to \frac{\sigma_t}{\epsilon}, \quad \sigma_a \to \frac{\sigma_t}{\epsilon} - \sigma_s,
\end{equation}
Eqs.~\eqref{RTE_g} and \eqref{Tm_g} are normalized:
\begin{empheq}[left={\empheqlbrace\,}]{align}
    \frac{\epsilon}{c}\frac{\partial I}{\partial t}+\bm{\Omega}\cdot \nabla I &= -\frac{\sigma_{t}}{\epsilon} I + \frac{1}{4\pi} \sigma_s \rho +  \frac{1}{4\pi} \left( \frac{\sigma_t}{\epsilon} - \sigma_s \right) \phi,
    \label{RTE_g_dimless} \\
     \epsilon C_{v}\frac{\partial T}{\partial t} &= \left ( \frac{\sigma_t}{\epsilon} - \sigma_s \right ) 
       \left ( \rho -\phi \right ) . \label{Tm_g_dimless}
\end{empheq}

Multiplying Eq.~\eqref{RTE_g_dimless} with 1 and $\bm{\Omega}$, and integrating the resultant equations with respect to the solid angle, we obtain the zeroth- and first-order moment equations:
\begin{equation}\label{RTE_M0}
    \begin{aligned}
    \frac{\epsilon}{c} \frac{\partial \rho}{\partial t} + \nabla \cdot \bm{F} &= \left ( \frac{\sigma_t}{\epsilon} - \sigma_s \right )  \left( \phi - \rho \right), \\
    \frac{\epsilon}{c} \frac{\partial \bm{F}}{\partial t} + \nabla \cdot \bm{P} &= -\frac{\sigma_t}{\epsilon} \bm{F},
    \end{aligned}
\end{equation}
which are not closed since the tensor $\bm{P}=\int_{\mathcal{S}^{2}} \bm{\Omega}\bm{\Omega} I d \bm{\Omega}$ is not expressed in terms of lower-order moments $\rho$ and $\bm{F}$. A series of macroscopic moment equations are obtained if one continues to take higher-order moments to Eq.~\eqref{RTE_g_dimless}. However, they are not solvable before making a closure. 

To analyze the asymptotic behavior of the RTE in the diffusion limit $\epsilon\rightarrow0$ \cite{LARSEN1983285,larsen1987asymptotic}, the distribution function is expanded into power series of $\epsilon$ as $
    I=\sum_{n=0}^{\infty } \epsilon^{n}I^{(n)}$.
Substituting this expansion into Eq.~\eqref{RTE_g_dimless}, and collecting terms with the same order of $\epsilon$, the radiant intensity is approximated by
$I = \frac{1}{4\pi}\rho - \frac{\epsilon}{4\pi\sigma_t}\bm{\Omega}\cdot\nabla \rho+ O(\epsilon^2)$,
and the Fourier's law of heat conduction is obtained: 
\begin{equation}
    \label{k-1,o0,F1}
    \bm{F} = -\frac{\epsilon}{3\sigma_t}\nabla \rho,
\end{equation}
which enables us to close the zeroth moment equation~\eqref{RTE_M0}. Setting $\epsilon=1$, the asymptotic diffusion equation is given by:
\begin{equation}
    \label{noneq_dae}
    \begin{cases}
        \begin{aligned}
        \frac{1}{c} \frac{\partial \rho}{\partial t}-\nabla \cdot \left(\frac{1}{3\sigma_{t}}\nabla\rho\right)
        &=\sigma_{a}\left ( \phi- \rho\right ) ,\\
        C_{v}\frac{\partial T}{\partial t}&= \sigma_{a}  
        \left ( \rho -\phi \right ) .
        \end{aligned}
    \end{cases}
\end{equation}

In Section~\ref{GSIS}, we preserve the form of Eq.~\eqref{noneq_dae} to construct the GSIS method for RTE, in the diffusion limit. In other regimes, the heat flux $\bm{F}$ will be designed to account for the non-equilibrium effects, based on the numerical solution of the RTE, rather than analytical truncations that work only when $\epsilon$ is not large.

\subsection{Source iteration}\label{SI_problem}

The presence of the integral term  makes iteration necessary when solving the RTE~\eqref{RTE_g}. SI is the fundamental iterative approach. Starting from an initial guess $I^0$, each iteration updates the radiation intensity distribution by streaming and collision process, until convergence is achieved. Using $m$ as the inner iteration index, the updated index following a SI is temporarily denoted as $m+\frac{1}{2}$. 
To derive the iterative update scheme, we first discrete Eqs.~\eqref{RTE_g} and \eqref{Tm_g} implicitly in time: 
\begin{equation}
  \label{RTE_g_semi}
  \frac{1}{c}\frac{I^{m+\frac{1}{2}} - I^{n}}{\varDelta t}+\bm{\Omega}\cdot \nabla   I^{m+\frac{1}{2}} =\sigma_{a}^{m}\left (\frac{1}{4\pi} \phi^m - I^{m+\frac{1}{2}}\right ) 
  + \sigma_{s}^{m}\left (\frac{1}{4\pi} \rho^{m}- I^{m+\frac{1}{2}}\right ) , 
\end{equation}
\begin{equation}
  \label{Tm_g_semi}
  C_{v}\frac{T^{m+\frac{1}{2}}-T^{n}}{\varDelta t} = \sigma_{a}^{m} 
  \left ( \rho^{m+\frac{1}{2}} - \phi^{m+\frac{1}{2}} \right ) ,
\end{equation}
where $n$ denotes the previous time step (outer iteration). 

By replacing $T^{m+\frac{1}{2}}$ in Eq.~\eqref{Tm_g_semi} by the Fleck-Cummings linearization~\cite{FLECK1971313}: $\phi^{m+\frac{1}{2}} = 4ac(T^n)^3 T^{m+\frac{1}{2}}-3\phi^n$, a linear relation between $\rho^{m+\frac{1}{2}}$ and $\phi^{m+\frac{1}{2}}$ is obtained:
\begin{equation}
  \left ( \frac{C_v}{4a c \left ( T^n\right )^3 \varDelta t} + \sigma_{a}^{m} \right ) \phi^{m+\frac{1}{2}} = \sigma_{a}^{m} \rho^{m+\frac{1}{2}} + \frac{C_v T^n}{4\varDelta t}
  .\label{Tm_semi_linear}
\end{equation}
Combining Eqs.~\eqref{RTE_g_semi} to~\eqref{Tm_semi_linear}, the SI typically takes the following form:
\begin{equation} \label{RTE_SI}
  \left ( \hat{\sigma}_t^m + \bm{\Omega}\cdot \nabla \right ) I^{m+\frac{1}{2}} = \frac{1}{4\pi}\hat{\sigma}_s^m \rho^m + Q,
\end{equation}
where  the artificial cross sections with a hat mark are 
\begin{equation}\label{arti_sigmaT}
  \hat{\sigma}_s = \sigma_s + \mu\sigma_a, \quad
  \hat{\sigma}_t = \frac{1}{c\varDelta t } + \sigma_t,
\end{equation}
with $\mu = {a c T^3 \varDelta t \sigma_a}/{(C_v + a c T^3 \varDelta t \sigma_a)}$. Correspondingly, the artificial absorption cross section is defined as $\hat{\sigma}_a = \hat{\sigma}_t - \hat{\sigma}_s$.
On the right-hand side of Eq.~\eqref{RTE_SI}, the source term $Q$ contains information from only the previous time step, whose value is considered constant during the inner iteration.

In SI, the update is straightforward,
\begin{equation}
  \label{SI_update}
  \rho^{m+1}_{SI} = \rho^{m+\frac{1}{2}}, \quad I^{m+1}_{SI} = I^{m+\frac{1}{2}}.
\end{equation}
Assuming $\sigma_a$ and $\sigma_s$ are constant, the convergence of the SI is evaluated by the Fourier analysis \cite{ADAMS20023}. We take the one-dimensional case as example, using the Fourier transform:
\begin{equation}
  \label{fourier_transform}
  I^m \left (x, \theta \right ) \to \tilde{I}^m \left (\lambda, \theta \right ) \exp \left (\mathrm{i} \hat{\sigma}_t \lambda x \right ), \quad 
  \rho^m \left (x\right ) \to \tilde{\rho}^m \left (\lambda\right ) \exp \left (\mathrm{i} \hat{\sigma}_t \lambda x \right ), \quad
  \frac{\partial}{\partial x} \to i\hat{\sigma}_t \lambda,
\end{equation}
where $i$ is the imaginary unit, and $\bm{\Omega}\cdot \nabla=\cos\theta$ is determined by the polar angle  $\theta$ only. At each wave number $\lambda$, we have $
  \left ( \hat{\sigma}_t + \mathrm{i} \hat{\sigma}_t \lambda \cos\theta  \right ) \tilde{I}^{m+1} = \hat{\sigma}_s \tilde{\rho}^m$.
Define $\gamma_{\hat{s}} = {\hat{\sigma}_s}/{\hat{\sigma}_t}$ as the effective scattering ratio, the error decay rate in SI is
\begin{equation}
  \label{SI_1d_rho}
  \omega_{SI} (\lambda) = \left | \frac{\tilde{\rho}^{m+1}} {\tilde{\rho}^m} \right | 
  =\frac{1}{2} \int_{0}^{\pi} \frac{ \gamma_{\hat{s}} \sin \theta } { 1 + \mathrm{i} \lambda \cos\theta  } d\theta
  = \gamma_{\hat{s}} \frac{\arctan \lambda}{\lambda} .
\end{equation}
The rate of convergence of SI is determined by the maximum of $\omega_{SI}$, that is, the spectral radius. Clearly, the spectral radius is bounded by $\gamma_{\hat{s}}$ at $\lambda\to0$, which implies the iteration is stable and convergent. However, as $\gamma_{\hat{s}} \to 1$, i.e., when $\varDelta t\to \infty$, the spectral radius of SI also tends to 1. If at the same time the solution is dominated by waves with small $\lambda$, which is in a much larger scale relative to $1/\hat{\sigma}_t$, the convergence could be rather slow. 

\subsection{Two versions of DSA}


The enhance the convergence, diffusion synthetic equations are used to facilitate ``global information exchange" in the whole computational domain. 
The first version of DSA solves the diffusion equation of the iteration error after SI \cite{Larsen01011984}: 
\begin{equation}
  \label{DSA_dif}
  -\nabla \cdot \left (\frac{1}{3\hat{\sigma}_t^m} \nabla \Phi ^{m+1}\right ) + \hat{\sigma}_a^m \Phi ^{m+1} = \hat{\sigma}_s^m \left ( \rho^{m+\frac{1}{2}} -  \rho^{m}\right ),
\end{equation}
whose solution is added to correct the next step of iteration:
\begin{equation}
  \label{DSA_update}
  \rho^{m+1}_{DSA} = \rho^{m+\frac{1}{2}} + \Phi ^{m+1} , \quad I^{m+1}_{DSA} = I^{m+\frac{1}{2}} + \frac{1}{4 \pi} \Phi^{m+1}.
\end{equation}

\begin{figure}[t]
  \centering
\includegraphics[width=0.6\textwidth,trim={30 20 60 30},clip]{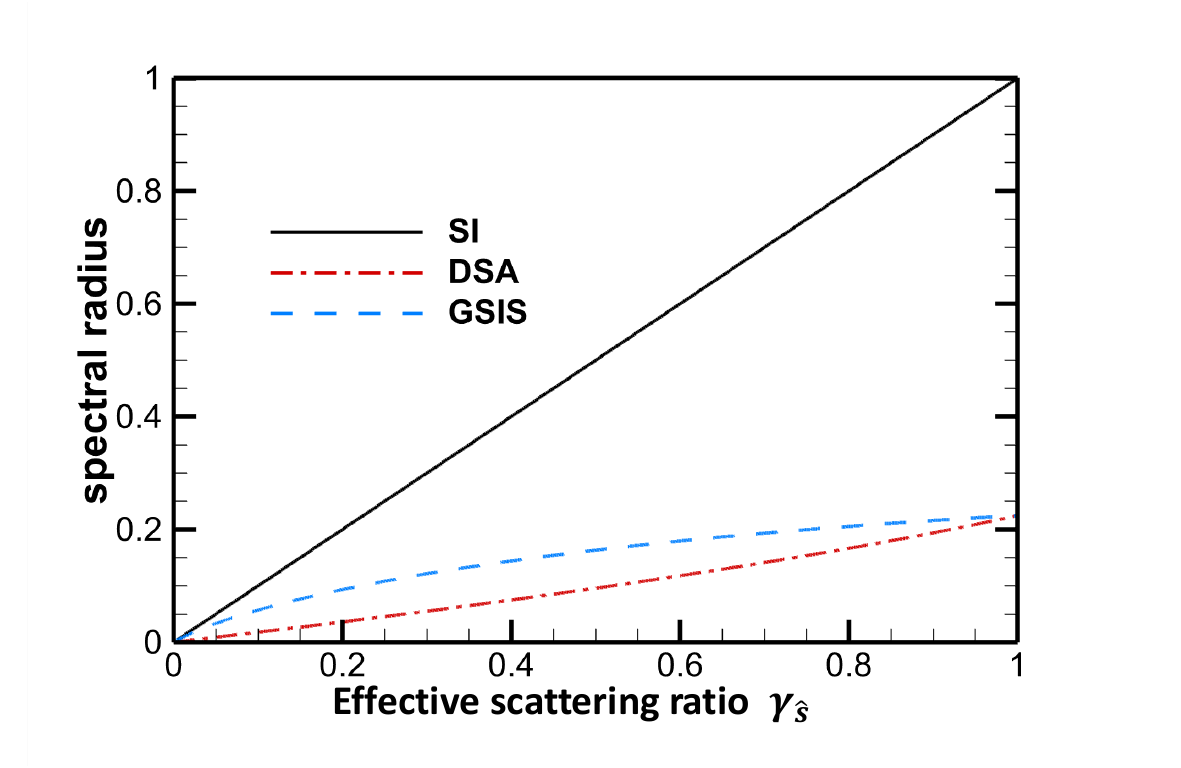}
  \caption{The spectral radius of SI, DSA, and GSIS with respect to effective scattering ratio $\gamma_{\hat{s}}$, in a continuous problem. 
  }
  \label{fig:spr_with_gammas}
\end{figure}

For infinite homogeneous media, the Fourier analysis at the continuous level shows that the error decay rate of DSA is \cite{ADAMS20023}:
\begin{equation}
  \label{DSA_1d_rho}
  \omega_{DSA} (\lambda) = \left | \frac{\tilde{\rho}^{m+1}}{\tilde{\rho}^{m}}\right | = \gamma_{\hat{s}} \left [ \frac{\arctan \lambda}{\lambda} + \frac{\gamma_{\hat{s}}\arctan \lambda - \lambda}{\lambda \left (1 + \frac{1}{3} \lambda^2 - \gamma_{\hat{s}} \right )} \right ].
\end{equation}
Figure~\ref{fig:spr_with_gammas} shows that the maximum spectral radius is about 0.22, 
which implies that DSA is theoretically efficient, regardless of the transport regime and time step. 

However, at the discrete level, achieving  accurate and efficient acceleration requires a consistently discretized diffusion equation, which is supposed to be derived from the discrete RTE using the $P_1$ approximation \cite{ADAMS20023}.  From Eq.~\eqref{DSA_update} we know that, as the SI converges, the diffusion equation provides no more correction. Hence, the DSA itself is a pure acceleration technique, and the converged solution is controlled solely by the RTE. This suggests that constructing a multiscale scheme overcoming the slow convergence and large numerical dissipation requires the discretization of RTE to be asymptotic preserving. It is not difficult to derive a consistent discretization using diamond-difference. 
While for other schemes, it might be rather complicated; this is reflected in the half-century-long development of synthetic schemes.

While the diamond-difference method for the RTE can recover the diffusion limit, it has its own drawbacks. As shown in~\ref{Appdx:vrf}, in slab geometry, it must use a suitable damping factor (like that adopted in Ref.~\cite{SONG2023120349}) to ensure numerical stability, which, however, significantly reduces the acceleration rate. Furthermore, designing consistent discretizations for non-rectangular mesh grids is challenging. The complexity of discretization is an ineligible factor for the application of DSA.

Alcouffe introduced another version of the DSA in neutron transport, where the diffuse equation for $\rho^{m+1}$ rather than the increment $\Phi= \rho^{m+1} -\rho^{m+\frac{1}{2}}$ is used.  For the transport equation~\eqref{RTE_SI}, the diffusion equation can be heuristically written as \cite{Alcouffe01101977}:
\begin{equation}
  \label{DSA_Alcouffe}
  -\nabla \cdot \left (\frac{1}{3\hat{\sigma}_t^m} \nabla \rho^{m+1}\right ) + \hat{\sigma}_a^m \rho ^{m+1} = {4\pi Q}  -\nabla \cdot \left (
  \bm{F}^{m+\frac{1}{2}}+\frac{1}{3\hat{\sigma}_t^m} \nabla \rho^{m+\frac{1}{2}}\right ).
\end{equation}
Although Alcouffe introduced the diamond-difference method to maintain consistency in the discretization of the transport and synthetic equations, our numerical tests in~\ref{Appdx:vrf} show that, unlike the first version of DSA in Eq.~\eqref{DSA_dif}, when the synthetic equation is written in its full form (like Eq.~\eqref{DSA_Alcouffe} and that in the VEF method~\cite{Lou2019JCP}) without the consistent term~\cite{Park01052012}, one can use any method to discretize the synthetic and transport equations. In this case, the transport equation is enslaved to the correct diffusion equation in the diffusion limit, regardless of how the transport equation is discretized\footnote{However, it is rather surprising that Alcouffe's method has not been extended to the RTE.}. 
On the contrary, in Park's method~\cite{Park01052012}, the synthetic equation is enslaved to the transport equation by adding the consistent term; when the latter is not solved with an asymptotic preserving scheme, it exhibits large numerical dissipation~\cite{Park01112020}. Designing consistent terms in the synthetic equation and the asymptotic preserving scheme for the kinetic equation is rather complicated, which we believe limits the application of the high-order/low-order methods~\cite{Park01052012,CHACON201721}.

%% file: data/sec03.tex
\section{General Synthetic Iterative Scheme for RTE}
\label{GSIS}

In this section, following the recent success of GSIS in solving the rarefied gas flows~\cite{SU2020jcp,ZHU2021110091,liu2024further, GSISPolyGas_ZENG2023, zhang2024efficient}, we first construct a synthetic acceleration method for the RTE. 
Second, through an asymptotic analysis, we explain why it is possible for GSIS to recover the asymptotic equation, even if the RTE is discretized in a simple upwind scheme which naturally aligns with the transport of photons.
Finally, we analyze the convergence rate of GSIS on the continuous scale. The convergence property of the discretized RTE will be shown in the numerical simulation in Section~\ref{tests}.

\subsection{A simple construction}

Inspired by {Alcouffe's DSA for neutron transport} and the GSIS in rarefied gas dynamics~\cite{Alcouffe01101977,ZHU2021110091}, 
we separate the constitutive relation for heat flux into the diffusion part and higher-order terms ($\text{HoT}_{\bm{F}}$) that describes transport (rarefaction) effects:
\begin{equation}  \label{GSIS_HoT}
   \bm{F}^{m+1}=-\frac{1}{3\sigma_t}\nabla \rho^{m+1}+\underbrace{ \bm{F}^{m+\frac{1}{2}} + \frac{1}{3{\sigma}_t^{m}}\nabla \rho^{m+\frac{1}{2}} }_{\text{HoT}^{m+1}_{\bm{F}}},
\end{equation}
and the zeroth moment equation \eqref{RTE_M0} of the RTE is recast into a synthetic form:
\begin{equation}
  \label{GSIS_M0}
  \frac{1}{c} \frac{\partial \rho}{\partial t} + \nabla \cdot \left ( -\frac{1}{3\sigma_{t}}\nabla\rho + \text{HoT}_{\bm{F}} \right ) = \sigma_{a}\left ( \phi - \rho \right ).
\end{equation}

During inner iteration, like other synthetic acceleration methods, GSIS alternately solves the transport equation~\eqref{RTE_g_semi} and the macroscopic equations:
\begin{equation}
  \label{GSIS_dse}
  \begin{cases}
      \begin{aligned}
      \frac{1}{c} \frac{\rho^{m+1}-\rho^{n}}{\varDelta t} + \nabla \cdot \left ( -\frac{1}{3\sigma_{t}^{m+1}}\nabla\rho^{m+1} + \text{HoT}^{m+1}_{\bm{F}} \right )
      &= \sigma_{a}^{m+1}\left ( \phi^{m+1} - \rho^{m+1} \right ), \\
      C_{v}\frac{T^{m+1} - T^n}{\varDelta t} &= \sigma_{a}^{m+1}  \left ( \rho^{m+1} - \phi^{m+1} \right ) .
      \end{aligned}
  \end{cases}
\end{equation}
The next iteration is then corrected as:
\begin{equation}
  \label{GSIS_update}
  \rho^{m+1}_{GSIS} = \rho^{m+1}, \quad I^{m+1}_{GSIS} = I^{m+\frac{1}{2}} + \frac{1}{4 \pi} \left ( \rho^{m+1} - \rho^{m+\frac{1}{2}} \right ).
\end{equation}

We believe that the constitutive relation~\eqref{GSIS_HoT} is better than that used in the VEF method, since the diffusion equation~\eqref{GSIS_M0} always has positive diffusion coefficient so that it permits the stable numerical simulation. However, in the VEF method, the effective diffusion coefficient is proportional to $\int_{\mathcal{S}^{2}} \bm{\Omega}\bm{\Omega} I d \bm{\Omega}$, which would be negative (i.e., anti-Fourier's law of heat conduction) in strong non-equilibrium transport (e.g., see one of the evidences in rarefied gas flows~\cite{Emerson2010}). This will cause numerical instability. Moreover, the synthetic equation in GSIS is easier to be solved than that of the VEF method.
Other advantages of the present GSIS method are that it does not require the Fleck-Cummings linearization~\cite{FLECK1971313}, and can be straightforwardly extended to multi-frequency problems~\cite{ShiYi2023JCP}.

\subsection{Analysis of convergence}

Since the discretization is not determined, we present the analysis in the continuous formula, in a one-dimensional space; that of the high-dimensional discretized system will be presented only in numerical simulations. 

Equation~\eqref{GSIS_dse} is linearized to:
\begin{equation}
  \label{dse_iter}
  \left ( \hat{\sigma}_t - \hat{\sigma}_s - \nabla \cdot \frac{1}{3{\sigma}_t}  \nabla \right )  \rho^{m+1} =  4\pi Q -  \nabla \cdot \left ( \bm{F}^{m+\frac{1}{2}} + \frac{1}{3{\sigma}_t}  \nabla \rho^{m+\frac{1}{2}}\right ) .
\end{equation}
The Fourier transform yields: 
\begin{equation}
  \label{SI_ft_F}
  \tilde{F}^{m+\frac{1}{2}} = 2\pi \int_0^{\pi} \tilde{I}^{m+\frac{1}{2}}\cos \theta \sin \theta d\theta =  \mathrm{i} \gamma_{\hat{s}} \frac{-\lambda + \arctan \lambda}{\lambda^2}\tilde{\rho}^m.
\end{equation}
At each wave number $\lambda$, we have:
\begin{equation}
  \label{dse_ft}
  \left ( \hat{\sigma}_t + \frac{\hat{\sigma}_t^2}{3{\sigma}_t}\lambda^2 -\hat{\sigma}_s \right )  \tilde{\rho}^{m+1} = -\mathrm{i}\lambda \hat{\sigma}_t\left ( \mathrm{i} \gamma_{\hat{s}} \frac{-\lambda + \arctan \lambda}{\lambda^2} + \mathrm{i}\frac{\hat{\sigma}_t}{3{\sigma}_t} {\gamma_{\hat{s}}}\arctan \lambda \right ) \tilde{\rho}^m .
\end{equation}
We define the ratio $\gamma_{t} = \sigma_t/\hat{\sigma}_t$, which is relevant to the time step $\varDelta t$ and within the range $ \gamma_{\hat{s}} \le \gamma_{t} \le 1$. The rate of convergence is:
\begin{equation}
  \label{dse_ft_rho}
  \begin{aligned}
    \omega_{GSIS}  
    = \gamma_{\hat{s}} \left [\frac{\arctan \lambda}{\lambda} + \frac{\gamma_{\hat{s}} \arctan \lambda - \lambda}{\lambda \left ( 1 + \frac{1}{3\gamma_t} \lambda^2 - \gamma_{\hat{s}}\right )}\right ].
  \end{aligned}
\end{equation}



The spectral radius is shown in Fig.~\ref{fig:spr_with_gammas}. Although the spectral radius is slightly higher than that of the DSA after using the Fleck-Cummings linearization, 
the global upper bound of $\omega_{GSIS}$, i.e., for arbitrary values of $\lambda$, $\gamma_{\hat{s}}$, and $\gamma_{t}$, is the same as DSA. 
Numerical simulations in \ref{Appdx:vrf} and Section~\ref{tests} show examples that the fast convergence are maintained in discretized system.

\subsection{Analysis of asymptotic preserving property}

To investigate the impact of spatial discretization at the diffusion limit, we consider the steady-state discretized RTE with the scaling parameter $\epsilon$:
\begin{equation}
  \label{RTE_steady_d}
  \bm{\Omega} \cdot \nabla_{\bm{\delta}} \bar{I} = \frac{\sigma_t}{\epsilon}\left ( \frac{1}{4\pi}\bar{\rho} - \bar{I} \right ).
\end{equation}
Assuming that the transport iteration is sufficiently converged, yields the numerical solution $[\bar{I}, \bar{\rho}]$, the numerical error is produced by inaccurate gradient calculation, which is denoted by the operator $\nabla_{\bm{\delta}}$. 

We are concerned with whether the diffusion equation can be derived from Eq.~\eqref{RTE_steady_d}, or equivalently, whether $\text{HoT}_{\bm{F}}$ tend to \(O(\epsilon^2)\) as \(\epsilon \to 0\), so that Eq.~\eqref{GSIS_M0} turns to the first equation in Eq.~\eqref{noneq_dae}; we are not concerned with how the diffusion equation is solved. This is because, even in optically thick media, when the diffusion equation is given, the numerical error for solving it, albeit small, always exists. If the diffusion equation can be derived from Eq.~\eqref{RTE_steady_d}, we say the numerical scheme is asymptotic preserving.

For volume-based schemes, 
the gradient operator can be expressed as the sum of an exact or accurate enough gradient operator, and an error operator $\varDelta x^{k} \bm{\delta}$, where $k$ is the order of accuracy:
\begin{equation}
  \label{grad_operator}
  \nabla_{\bm{\delta}}\left ( \bm{x}, \bm{\Omega}\right ) = \nabla + \varDelta x^{k} \bm{\delta} \left ( \bm{x}, \bm{\Omega}\right ).
\end{equation}
To show that a specially designed discretization is not necessary in GSIS, we here presuppose a conventional and simple upwind scheme, which naturally satisfies: (i)
the local numerical gradient $\nabla_{\bm{\delta}} \bar{I}\left ( \bm{x}, \bm{\Omega}\right )$ is represented as a certain linear combination of discrete values in each direction $\bar{I}(\bm{\Omega})$; and (ii) the direction dependency of $\nabla_{\bm{\delta}}$ arises solely from the upwind property.
Based on this assumptions, an angle independent gradient operator is obtained by averaging $\nabla_{\bm{\delta}}$ from an arbitrary direction and its opposite direction, and correspondingly an angle-independent error operator is produced:
\begin{equation}
  \label{grad_operator_mean}
  \bar{\nabla}_{\bm{\delta}}\left ( \bm{x}\right ) = \frac{1}{2} \left [\nabla_{\bm{\delta}}\left ( \bm{x}, \bm{\Omega}\right ) + \nabla_{\bm{\delta}}\left ( \bm{x}, -\bm{\Omega}\right ) \right ] = \nabla + \varDelta x^{k} \bar{\bm{\delta}} \left ( \bm{x}\right ).
\end{equation}
The symmetry is apparent in regular meshes. For more general cases, we give a brief proof in \ref{Appdx:proof}, according to the cell-centered scheme used in this paper.

At the diffusion limit $\epsilon \to 0$, $\bar{\rho}$ and $\bar{I}$ are expanded inter power series of $\epsilon$ as $\bar{\rho}=\bar{\rho}^{(0)}+\epsilon \bar{\rho}^{(1)}+\cdots$ and $\bar{I}=\bar{I}^{(0)}+\epsilon \bar{I}^{(1)}+\cdots$; then, substituting both expansions into Eq.~\eqref{RTE_steady_d}  and collecting terms of orders $\epsilon^{-1}$ and $\epsilon^{0}$ gives:
\begin{eqnarray}
   \label{RTE_steady_d_om1}
  \epsilon^{-1}: \quad  \frac{1}{4\pi}\bar{\rho}^{(0)} = \bar{I}^{(0)},\\
   \label{RTE_steady_d_o0}
  \epsilon^{0}: \quad \bm{\Omega} \cdot \nabla_{\bm{\delta}} \bar{I}^{(0)} = \sigma_t \left ( \frac{1}{4\pi}\bar{\rho}^{(1)} - \bar{I}^{(1)} \right ),
\end{eqnarray}
where $\bar{I}^{(0)}$ is isotropic, and the first-order moment of $\bar{I}^{(1)}$ is $\bar{\bm{F}}^{(1)}$. If the solid angle is discretized sufficiently, the angle-dependent part of $\nabla_{\bm{\delta}}$ vanishes in even order moments of $\nabla_{\bm{\delta}} \bar{I}^{(0)}$, since that arbitrary direction $\bm{\Omega}$ corresponds to its opposite $-\bm{\Omega}$ in the integral. Hence we have,
\begin{equation}
  \label{grad_operator_int}
  \int \bm{\Omega}\left ( \bm{\Omega} \cdot {\nabla}_{\bm{\delta}}\bar{I}^{(0)} \right ) d\Omega = \int \bm{\Omega}\left ( \bm{\Omega} \cdot \bar{\nabla}_{\bm{\delta}}\bar{I}^{(0)} \right ) d \bm{\Omega} = \bar{\nabla}_{\bm{\delta}} \frac{1}{3}\bar{\rho}^{(0)}.
\end{equation}

Taking the first-order moment of $\bar{I}$ directly from the expansions yields:
\begin{equation}
  \label{RTE_steady_d_o0_m1}
  \begin{aligned}
    \bar{\bm{F}} = \int \bm{\Omega} \left (\bar{I}^{(0)} + \epsilon \bar{I}^{(1)} \right ) d \bm{\Omega} + O(\epsilon^2)
    &= - \frac{\epsilon}{3\sigma_t}\bar{\nabla}_{\bm{\delta}} \bar{\rho}^{(0)} + O(\epsilon^2)\\
    &= - \frac{\epsilon}{3\sigma_t}\left ( \nabla + \varDelta x^{k}\bar{\bm{\delta}} \right ) \bar{\rho} + O(\epsilon^2).
  \end{aligned}
\end{equation}
To recover the asymptotic limit, the error term $- \frac{\epsilon}{3\sigma_t}\varDelta x^{k}\bar{\bm{\delta}}\bar{\rho}$ implies that a restriction on grid size $O(\epsilon \varDelta x^{k}) + O(\epsilon^2) \sim O(\epsilon^2)$ is imposed for conventional numerical schemes, that is, 
$\varDelta x \sim O(\epsilon^{{1}/{k}})$. 
Thus, the conventional transport discretizations have their asymptotic behavior affected by errors at the diffusion limit, depending on the precision of the scheme. No matter how accurate the scheme is, there always exists a sufficiently small $\epsilon$ that makes the numerical solution fail to describe the physical process.

In GSIS, the higher order part of heat flux is extracted as follows\footnote{In GSIS, the synthetic equation and kinetic equation can be solved by different numerical methods with different order of accuracy. Here, we assume that the accuracy of solving the synthetic equation is the same as that of solving the kinetic equation.}:
\begin{equation}
  \label{RTE_steady_d_HoT}
  \begin{aligned}[b]
    \text{HoT}_{\bm{F}} = \bar{\bm{F}} + \frac{\epsilon}{3\sigma_t} \bar{\nabla}_{\bm{\delta}}\bar{\rho} = \bar{\bm{F}} +   \frac{\epsilon}{3\sigma_t}\bar{\nabla}_{\bm{\delta}} \bar{\rho}^{(0)} + O(\epsilon^2),
  \end{aligned}
\end{equation}
which is at the order of $\epsilon^2$ according to the second line of Eq.~\eqref{RTE_steady_d_o0_m1}. Thus, 
the asymptotic equation for $\rho$ is recovered correctly at the diffusion limit:
\begin{equation}
  \label{dse_steady_d}
   \nabla \cdot \left (- \frac{\epsilon}{3\sigma_t} \nabla \rho +O(\epsilon^2) \right ) = 0 
   \quad \Rightarrow \quad
   \nabla \cdot \left (- \frac{\epsilon}{3\sigma_t} \nabla \rho  \right ) = 0.
\end{equation}
This demonstrates that in GSIS, even when $\Delta{x}\sim O(1)$, even though the flux $\bm{\bar{F}}$ extracted from the RTE is affected by discretization error to the order of $\epsilon$, the asymptotic diffusion equation is still recovered. That is, in the diffusion limit, solving the RTE via GSIS is as accurate as solving the diffusion equation~\eqref{dse_steady_d}. The numerical simulation in \ref{Appdx:vrf} presents one such example.

Finally, it should be emphasized that, the above analysis is applied to the bulk region where the variation of physical quantities is small. In regions with steep variations \cite{Su2020siam}, such as the shock wave structure in rarefied gas flows, and the Knudsen layer near the solid wall, the spatial resolution should be at scale of the mean free path, to capture the real physics; the latter will be illustrated in the numerical example in Section~\ref{Tophat_num}. 

%% file: data/sec04.tex
\section{Numerical Schemes}
\label{schemes}

\begin{figure}[t]
  \centering
\includegraphics[width=0.9\textwidth]{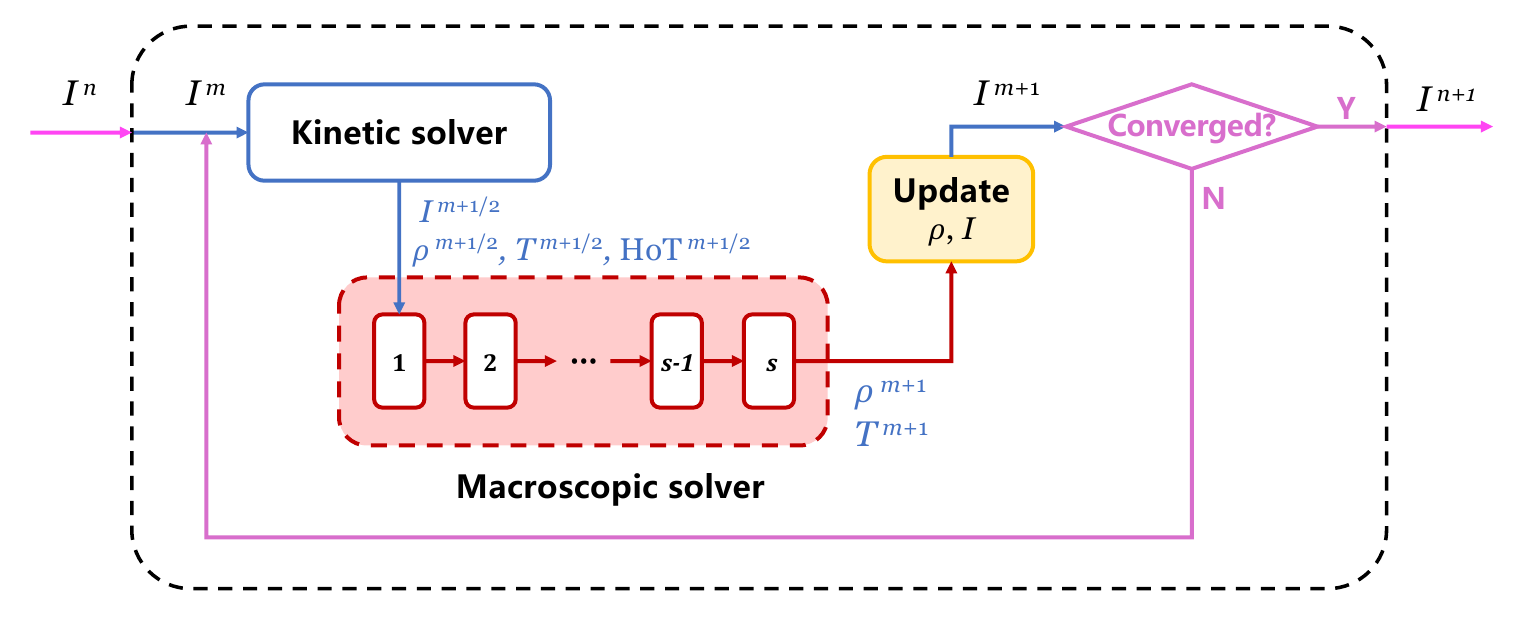}
  \caption{Flowchart of the GSIS for time-dependent RTE, where $n$ is the time step, $m$ is the inner iteration index to solve the transport equation, and $s$ is the sub-inner iteration index to solve the synthetic equation.
  }
  \label{fig:flowrate}
\end{figure}

Numerical schemes are constructed for the mesoscopic RTE and the macroscopic synthetic equations, respectively. They are both implemented via conventional discretizations and hence are not complex. Critical steps in GSIS are highlighted in Fig.~\ref{fig:flowrate} and explained in the following sections.

\subsection{Discretization of the RTE}


The unit sphere $\bm{\Omega}$ is divided into $K$ solid angular segments represented by the unit vectors $\bm{\Omega}_1, \bm{\Omega}_2, \dots,\bm{\Omega}_{K}$, 
associated with a set of integration weights $w_1, w_2, \dots, w_{K} $.
These weights are derived from numerical integration methods such as the Gauss-Legendre quadrature and the $S_n$ quadrature.
The numerical integration over the solid angle is performed by summing the products of these weights and corresponding function values.
For spatial discretization, a cell-centered finite-volume scheme is adopted.
The average radiation intensity within the control volume $V_i$ in the direction of $\bm{\Omega}_k$ is denoted by $I_{i,k}$.
The integral of the radiative intensity $I(\bm{x},\bm{\Omega})$ over the control volume $V_i$ 
and the solid angle is approximated by the following summation:
\begin{equation}
\int_{V_i}\int_{\mathcal{S}^{2} } p(\bm{\Omega}) I(\bm{x},\bm{\Omega}) d\bm{\Omega} dV
    \approx V_i \sum_{k=1}^{K} p_k w_k I_{i,k},
\end{equation}
where $p(\bm{\Omega})$ is a polynomial function defined on the unit sphere $\mathcal{S}^{2}$. For the zeroth 
and the first order moments, we set $p=1$ and $p=\bm{\Omega}$, respectively. 

The optical parameters $\sigma_a$ and $\sigma_s$ are assumed to be uniform within each control volume. 
For the sake of simplicity, we use the backward Euler method to discretize Eq.~\eqref{RTE_g} implicitly in time. 
Given a time increment $\varDelta t = t^{n+1} - t^{n}$, the discretized  RTE is expressed as:
\begin{equation}
    \label{RTE_d}
    \frac{1}{c} \frac{I_{i,k}^{n+1}-I_{i,k}^{n}}{\varDelta t} +\frac{1}{V_{i}}\sum_{j\in N(i)}
    I_{ij,k}^{n+1} \bm{\Omega}_k\cdot \bm{A}_{ij}=
    \sigma^{n+1}_{a,i} \left ( \frac{\phi_{i}^{n+1}}{4\pi}-  I_{i,k}^{n+1}\right )  
    +\sigma^{n+1}_{s,i} \left ( \frac{\rho_{i}^{n+1}}{4\pi}-  I_{i,k}^{n+1}\right ),  
\end{equation}
where the Gauss theorem has been applied to convert the volume integral into a surface integral of fluxes, and $\bm{A}_{ij} = \bm{n}_{ij}A_{ij}$ represents the product of the area and the unit normal vector of the face $ij$, pointing from the cell $i$ to cell $j$.

The distribution of radiation intensity within each control volume $I_{i,k}(\bm{x})$ is assumed to be linear:
\begin{equation}\label{Ii,k}
       I_{i,k}(\bm{x})= I_{i,k}+ \psi \nabla I_{i,k}\cdot(\bm{x}-\bm{x}_i),
\end{equation}
where the spatial gradient $\nabla I_{i,k}$ is determined using the least squared method. The Barth-Jespersen limiter \cite{Brath1989}, denoted by $\psi$, is applied to avoid oscillation. 
The fluxes at the interfaces, approximated by the face center values $I_{ij,k}$ are constructed by a second-order upwind scheme:
\begin{equation}
    \label{Iij}
       I_{ij,k}=
       \begin{cases}
        I_{i,k}(\bm{x}_{ij}), \quad \bm{\Omega}_k \cdot \bm{n}_{ij}\ge 0,
        \\
        I_{j,k}(\bm{x}_{ij}), \quad \bm{\Omega}_k \cdot \bm{n}_{ij}<0.
        \end{cases}
\end{equation}

To facilitate an iterative and matrix-free approach for implicit methods, we introduce the following delta-form of the discretized transport
equation:
\begin{equation}
    \label{RTE_d_increment}
    \left (\frac{1}{c\varDelta t}+\sigma^m_a + \sigma^m_s\right )
    \varDelta I_{i,k}^{m} +\frac{1}{V_{i}}\sum_{j\in N(i)}\varDelta I_{ij,k}^{m} \bm{\Omega}_k\cdot \bm{A}_{ij}=r_{i,k}^{m} ,
\end{equation}
where $\varDelta I_{i,k}^{m}= I_{i,k}^{m+\frac{1}{2}}-I_{i,k}^{m}$
is the iteration's
incremental change. The increments in flux $\varDelta I_{ij,k}$ on the left-hand side are approximately expressed in a first-order upwind scheme (when the iteration converges, it does not affect the overall accuracy),
\begin{equation}
    \label{delta_Iij}
    \varDelta I_{ij,k}=\begin{cases}
        \varDelta I_{i,k}, \quad \bm{\Omega}_k \cdot \bm{n}_{ij}\ge 0,
        \\
        \varDelta I_{j,k}, \quad \bm{\Omega}_k \cdot \bm{n}_{ij}<0.
        \end{cases}
\end{equation}
On the right hand side, the mesoscopic residual $r_{i,k}^m$ is defined as follows: 
\begin{equation}
    \label{RTE_d_res}
    r_{i,k}^{m} =-\frac{1}{c} \frac{I_{i,k}^{m}-I_{i,k}^{n}}{\varDelta t}
    -\frac{1}{V_{i}}\sum_{j\in N(i)}I_{ij,k}^{m}\bm{\Omega}_k\cdot \bm{A}_{ij}
    +\sigma^m_{a,i} \left ( \frac{\phi_{i}^{m}}{4\pi}-  I_{i,k}^{m}\right )
    +\sigma^m_{s,i} \left ( \frac{\rho_{i}^{m}}{4\pi}-  I_{i,k}^{m}\right ).
\end{equation}

Equation~\eqref{RTE_d_increment} is then converted to the following linear system: 
\begin{equation}
    \label{RTE_d_mtx}
    d_{ii}\varDelta I_{i}^{m} +\sum_{\substack{j \in N(i) \\ \bm{\Omega}_k \cdot \bm{n}_{ij} < 0}} d_{ij}\varDelta I_{j}^{m}=r_{i}^{m},
\end{equation}
where the diagonal and non-diagonal elements are, respectively:
\begin{equation}
    \label{elem_mtx}
    \begin{aligned}
    d_{ii} &= \frac{1}{c\varDelta t} + \sigma^m_{t,i} + \frac{1}{V_i} \sum_{\substack{j \in N(i) \bm{\Omega}_k \cdot \bm{n}_{ij} > 0}} \bm{\Omega}_k \cdot \bm{A}_{ij},\\ 
    \quad d_{ij} &= \frac{1}{V_i} \bm{\Omega}_k \cdot \bm{A}_{ij}.
    \end{aligned}
\end{equation}

Equation~\eqref{RTE_d_mtx} is solved by the lower-upper symmetric Gauss–Seidel technique. Convergence is achieved when $\varDelta I_{i}^m $ approaches zero,  which corresponds to the mesoscopic residuals $r_{i}^{m}$ also tending towards zero. 


\subsection{Discretization of macroscopic equations}

In the same finite-volume framework, the macroscopic equations \eqref{GSIS_dse} are discretized as follows:
\begin{equation}\label{MSE_d}
    \frac{1}{c} \frac{\rho_i^{n+1}-\rho_i^{n}}{\varDelta t}+\frac{1}{V_{i}}\sum_{j\in N(i)}\bm{F}_{ij}^{n+1}\cdot \bm{A}_{ij}
    = \sigma_{a,i}^{n+1}\left ( \phi_i^{n+1} - \rho_i^{n+1}\right ),
\end{equation}
\begin{equation}\label{Tm_d}
    C_v \frac{T_i^{n+1}-T_i^{n}}{\varDelta t}
    = \sigma_{a,i}^{n+1}\left ( \rho_i^{n+1} - \phi_i^{n+1}\right ).
\end{equation}

To capture the asymptotic behavior as the Knudsen number approaches zero, the construction of macroscopic interface flux $\bm{F}_{ij}$ is crucial. 
We adopt the diffusive flux scheme from conventional computational fluid dynamics method: 
\begin{equation}
    \label{Fij_d}
    \begin{split}
        \bm{F}_{ij} = \frac{1}{2}\left(\bm{F}_{i}+\bm{F}_{j}\right) = -\frac{1}{2}\left( \frac{\nabla\rho_i}{3\sigma_{t,i}}+\frac{\nabla\rho_j}{3\sigma_{t,j}}\right) +
        \frac{1}{2}\left(\text{HoT}_{\bm{F},i}^{m+\frac{1}{2}}+\text{HoT}_{\bm{F},j}^{m+\frac{1}{2}}\right),
    \end{split}
\end{equation}
where the macroscopic radiation heat flux at interface is approached by the average of the values at neighboring cell centers. The flux is composed of two components: the low-order diffusive flux, which varies with $\rho$ during the iteration, and the flux contributed by the high-order term. The latter is calculated at the cell center before the sub-inner iteration of the macroscopic solver (see Fig.~\ref{fig:flowrate}) and remains constant.
Additionally, an adaptive least squares method is employed to compute the spatial gradient of $\rho$, taking into account the spatial variation of $\sigma_t$, see details in Section \ref{ALSM}.

Like the numerical solving of mesoscopic equation, we also rewrite the macroscopic equations in delta-form:
\begin{align}
 \left(\frac{1}{c\varDelta t }+\sigma_a^s\right)\varDelta \rho_i^s + \frac{1}{V_i}\sum_{j\in N(i)}\varDelta \bm{F} _{ij}^s\cdot\bm{A}_{ij}  &= R^s_{1,i} ,\label{MSE_d1}\\ 
    \left(\frac{C_v}{\varDelta t}+4ac(T^{s}_{i})^{3}\sigma_a^s \right )\varDelta T_i^s &= R^s_{2,i},
    \label{MSE_d2}
\end{align}
where $\varDelta \rho_i=\rho_i^{s+1}-\rho_i^s$, $\varDelta T_i=T_i^{s+1}-T_i^s$, and the sub-inner iteration count is $s$. The source term on the right-hand side of Eq.~\eqref{Tm_d}, which represents the rate of energy exchange, is linearized with respect to $\varDelta T$.
The macroscopic residuals $R^s_{1,i}$ and $R^s_{2,i}$ are defined as:
\begin{align}
    R^s_{2,i} &= -\frac{1}{c} \frac{\rho_i^{s}-\rho_i^{n}}{\varDelta t}-\frac{1}{V_{i}}\sum_{j\in N(i)}\bm{F}_{ij}^{s}\cdot \bm{A}_{ij} 
    +\sigma_{a,i}^{s}\left ( \phi_i^{s} - \rho_i^{s}\right ), \label{R1_def}\\ 
    R^s_{i,2} &= -C_v \frac{T_i^{s}-T_i^{n}}{\varDelta t} +\sigma_{a,i}^{s}\left ( \rho_i^{s} - \phi_i^{s}\right ) .\label{R2_def}
\end{align}

The change in surface flux $\varDelta \bm{F}$ is induced by $\varDelta \rho$. 
The incremental normal flux of face $ij$ is approximated in terms of $\varDelta \rho_i$ and $\varDelta \rho_j$:
\begin{equation}
    \label{Gamma_ij_def}
\varDelta\bm{F}_{ij}\cdot\bm{n}_{ij} = \Gamma_{ij}\left( \varDelta \rho_i - \varDelta \rho_j\right).
\end{equation}
$\Gamma_{ij}$ is an estimation to face flux change, see details around Eq.~\eqref{Gamma_ij_com2} in Section \ref{ALSM}. 

Finally, the discretized equations are converted into a coupling linear system with respect to $\varDelta \rho_i$ and $\varDelta T_i$:
\begin{align}
    D_{1,ii}\varDelta \rho_i^s + \sum_{j\in N(i)}D_{1,ij}\varDelta \rho_j^s &= R^s_{1,i} ,\label{MSE_d_mtx1}\\ 
    D_{2,ii}\varDelta T_i^s  &= R^s_{2,i}, \label{MSE_d_mtx2}
\end{align}
where  
\begin{equation}
    \label{elem_mtx_M}
    \begin{aligned}
    D_{1,ii} &=\frac{1}{c\varDelta t }+\sigma_a^s + \frac{1}{V_i} \sum_{j \in N(i) } \Gamma_{ij}A_{ij},
    \quad D_{1,ij} = -\frac{1}{V_i} \Gamma_{ij}A_{ij}, \\
    \quad
    D_{2,ii}&=\frac{C_v}{\varDelta t}+4ac(T^{s}_{i})^{3}\sigma_a^s.  
    \end{aligned}
\end{equation}

The system is solved using the lower-upper symmetric Gauss–Seidel technique. When it converges or reaches the maximum iteration steps, the value of $\rho^{s+1}$ is assigned to $\rho^{m+1}$, and the density and distribution function are updated as per Eq.~\eqref{GSIS_update}.

\subsection{Local pseudo-time method}\label{pseudo_time}

The theoretical analysis of stability is based on assumptions of linearity and infinite uniform media. In practice, however, many factors (e.g., nonlinearities, boundary conditions, and discontinuities in opacity) may lead to instability. In particular, opacity discontinuity has a significant negative impact on synthetic acceleration methods such as DSA. Generally speaking, a simple way to stabilize the iteration is to limit the increment.


In traditional computational fluid dynamics, the pseudo-time method is employed to enhance the convergence of calculations.  The pseudo-time step is locally set in each grid cell and is typically determined by the characteristic length and flow speed, controlled by a global parameter \(\text{CFL}_p\). Accordingly, in this work, an additional term \({1}/{c \varDelta t_{p}}\) is introduced to the main diagonal \(d_{ii}\), where \(\varDelta t_{p}\) is given as:
\begin{equation}
  \label{cfl_pse}
  \varDelta t_{p,i} = {\text{CFL}}_p \frac{\varDelta x_{ref,i}}{c }.
\end{equation}
As shown in Eq.~\eqref{RTE_d_mtx}, if the solution is sufficiently converged, it is fully determined by the right-hand side term. Therefore, this does not affect the final solution. 


As long as $\text{CFL}_p$ is kept small, the increment is effectively limited. By setting this parameter appropriately, the method is applicable to problems involving nonlinearity.
However, $\text{CFL}_p$ is typically required to be as small as the grid Knudsen number at the optically thick side, which results in an excessive sacrifice of efficiency. From our observation, at the beginning, the instability is generated at the discontinuous interface, while there is no significant influence in the interiors of optically thin and optically thick zones. Therefore, it is better to use the local pseudo time. 

Suppose neighboring cells $i$ and $j$ have disparate optical thicknesses, where $\sigma_{t,i}  \varDelta x_{ref,i} \ll  \sigma_{t,j} \varDelta x_{ref,j}$, and their corresponding diagonal elements $d_{ii} \ll  d_{jj}$. Since the optically thin cell $i$ is sensitive to upstream flux change,  a local time $\tau_p$ is defined as:
\begin{equation}
  \label{tau_pse}
  \tau_{p,i} = \min_{\substack{j \in N(i)}} \frac{\beta \varDelta x_{ref,i}}{ \left |\sigma_{t,i} \bm{r}_{ij}^L - \sigma_{t,j} \bm{r}_{ij}^R \right |  c},
\end{equation}
where $\beta$ is the control parameter, $\bm{r}_{ij}^L$ and $ \bm{r}_{ij}^R$ represent for vectors pointing form the center of face $ij$ to centers of cell $i$ and $j$, respectively, see Fig.~\ref{fig:face_reconstruction_1d_example}(b). Instead of a global limiting, the main purpose of adopting this form is to limit the increments in the first layer of optically thin cells at discontinuous interfaces. 
This constitutes a capture for neighboring cells with a large divergence in optical thicknesses, such that if adding ${1}/{c\varDelta \tau_{p}}$ to the diagonal, $d_{ii}$ and $d_{jj}$ will be in the same order.

One may use both pseudo-time settings simultaneously to achieve a general control for instabilities and a specialized control for discontinuities in opacity. The diagonal element in Eq.~\eqref{elem_mtx} is then modified to 
\begin{equation}
  \label{elem_main_shifted}
  d_{ii} = \frac{1}{c\varDelta t_{p,i}} + \frac{1}{c\varDelta \tau_{p,i}} +  \frac{1}{c\varDelta t}  + \sigma^m_{t,i} + \frac{1}{V_i} \sum_{\substack{j \in N(i) \\ \bm{\Omega}_k \cdot \bm{n}_{ij} > 0}} \bm{\Omega}_k \cdot \bm{A}_{ij}.
\end{equation}
However, so far as we test, GSIS is stable in most cases, including some challenging ones. For more discussions, we give an example in Section \ref{tests}.

    

\subsection{Boundary condition}
The boundary condition is a key factor describing how radiation enters or leaves the computational domain. The incoming radiation is specified by an external source term $Q(\bm{x},\bm{\Omega})$. 
For diffuse reflection, the intensity distribution function at boundary face $iw$ is constructed as follows:
\begin{equation}
    \label{micro_bc}
    I_{iw,k} =\begin{cases}
        I_{i,k}(\bm{x}_{iw}), &\bm{\Omega}_k \cdot \bm{n}_{iw}\ge 0,
        \\
        Q_{iw,k} + \frac{\alpha_w}{2\pi}\int_{\bm{\Omega} \cdot \bm{n}_{iw}\ge 0}I_{i}(\bm{x}_{iw},\bm{\Omega}) d\bm{\Omega}, &\bm{\Omega}_k \cdot \bm{n}_{iw}<0,
        \end{cases}
\end{equation}
where the wall face normal vector, pointing outward from the domain, is denoted by $\bm{n}_{iw}$. 

The boundary flux $\bm{F}_{iw}$ is integrated and serves as the initial boundary condition for the synthetic equations:
$\bm{F}_{iw}^{init} = \int \bm{\Omega} I_{iw} d \bm{\Omega}$. 
To enhance the efficiency of alternating iterations in the transport and synthetic equations, updating the boundary flux $\bm{F}_{iw}$ is necessary during the macroscopic solver~\cite{liu2024further}. It is implemented according to changes in macroscopic quantities at the boundary cell, thus is not as straightforward as using the intensity function. Unlike the original diffusion equation, the boundary condition of the synthetic equation is supposed to contain the multiscale feature. We give the expression as:
\begin{equation}
    \label{macro_bc_delta_flux}
    \varDelta \bm{F}_{iw} = \left [ \frac{\bm{x}_{iw}-\bm{x}_i}{3\sigma_{t,i} \left | \bm{x}_{iw}-\bm{x}_i  \right |^2  }
    + \frac{1}{4}\bm{n}_{iw}\right ]  \varDelta \rho_{i}.
\end{equation}
The first term accounts for the impact of the gradient change in $\rho$ on the diffusion flux at the boundary face, while the second term represents the momentum of isotropic radiation with energy $\varDelta \rho_i$ that escapes the computational domain:
\begin{equation}
    \label{macro_bc_delta_flux2}
    \frac{1}{4\pi} \varDelta \rho_{i} \int_{n_{iw}\cdot \bm{\Omega}>0} \bm{\Omega} d \bm{\Omega} = \frac{1}{4}\varDelta \rho_{i}.
\end{equation}
The sum of these two terms provides a larger estimate of the actual change in the boundary flux, and this formulation is primarily adopted for stability considerations. In current work, no reflection is assumed on the inner side of boundary.

%% file: data/gradient.tex
\section{An adaptive least square method for discontinuous opacity}
\label{ALSM}

The numerical approximation of the gradient operator is crucial for both the transport and synthetic equations, especially when the opacities of adjacent cells vary abruptly. In this section, we first review the traditional methods for gradient calculation within the finite-volume framework, then propose an adaptive least square method to handle problems involving disparate opacities in adjacent cells. 

\subsection{Traditional methods}

Traditionally, once the linear approximation of an arbitrary quantity $U$ is assumed within cell $i$, the least square method is applied to calculate the spatial gradient at the cell-center. From each neighboring cell $j$, the change in $U$ is expressed in a first-order Taylor approximation:
\begin{equation}
\label{Uj-Ui}
  \varDelta U_{j} \approx \nabla_i U \cdot \varDelta \bm{x}_j,
\end{equation}
where $\varDelta U_{j} = U_j - U_i$ and $\varDelta \bm{x}_j = [\varDelta x_j, \varDelta y_j] = \bm{x}_j - \bm{x}_i$. In the finite-volume framework, this will lead to an overdetermined system if the number of neighboring cells is larger than the number of dimensions. The least squares function is thus defined as a quadratic sum of residuals:
\begin{equation}
\label{LSF}
  Y  = \sum_{j\in N(i)} \left ( \nabla_i U \cdot \varDelta \bm{x}_j - \varDelta U_{j} \right )^2 .
\end{equation}
Considering the two-dimensional case for simplicity, let $[G_x, G_y]$ denote the components of $\nabla_i U$, $Y$ is minimized when
\begin{equation}
  \label{LSM_mtx}
  \begin{bmatrix}
    \sum_{j\in N(i)} \varDelta x_j^2 & \sum_{j\in N(i) }\varDelta x_j \varDelta y_j\\
    \sum_{j\in N(i)} \varDelta x_j \varDelta y_j & \sum_{j\in N(i)} \varDelta y_j^2
  \end{bmatrix}
  \begin{bmatrix}
   G_x\\
  G_y
  \end{bmatrix}
  =\begin{bmatrix}
   \sum_{j\in N(i) }\varDelta x_j \varDelta U_j\\
   \sum_{j\in N(i) }\varDelta y_j \varDelta U_j
  \end{bmatrix}.
\end{equation}


Unlike the diamond difference method where the flux is continuous at cell edges, or the linear discontinuous method \cite{larsen1989asymptotic} that uses an additional equation to determine the variation within each cell, the finite-volume method typically results in a gradient operation that considers spatial relations only. This raises an issue on accuracy when the opacity has discontinuities in the computational domain. In addition, there are stability concerns associated with the current synthetic method. Specifically, the calculation of the gradient impacts the upwind flux of the transport equation, the diffusion flux, and the higher-order terms of the synthetic equation. However, these issues can be resolved by incorporating an opacity parameter into the gradient calculations.

For example, in a one-dimensional geometry, when the two adjacent cells $i$ and $i+1$ have disparate opacities,  in analogy with the harmonic mean method \cite{Chang01041992} of heat conductivity,  the diffusion flux from both side should be equal:
\begin{equation}
\label{face_coe_harmonic_diffusion_flux_eq}
    F_{i+\frac{1}{2}} ^D = -\frac{\rho_{i+1}-\rho_{i+\frac{1}{2}}}{3\sigma_{t,i+1} \left(x_{i+1}-x_{i+\frac{1}{2}} \right)}
    = -\frac{\rho_{i+\frac{1}{2}}-\rho_{i}}{3\sigma_{t,i}  \left(x_{i+\frac{1}{2}}-x_{i} \right)},
\end{equation}
where $\varDelta x_i$ and $\varDelta x_{i+1}$ are the cell lengths, so that
the face value $\rho_{i+\frac{1}{2}}$ is uniquely determined:
\begin{equation}
    \label{face_rho_harmonic}
    {\rho_{i+\frac{1}{2}}} = \rho_i + \frac{1}{2}\varDelta x_{i}\frac{\sigma_{t,i+1}\left (\rho_{i+1} - \rho_{i} \right)}{\sigma_{t,i}\varDelta x_i + \sigma_{t,i+1}\varDelta x_{i+1}}
    = \rho_{i+1} + \frac{1}{2}\varDelta x_{i+1}\frac{\sigma_{t,i}\left (\rho_{i} - \rho_{i+1} \right)}{\sigma_{t,i}\varDelta x_i + \sigma_{t,i+1}\varDelta x_{i+1}}.
\end{equation}
As a result, the diffusion flux at the face is expressed as:
\begin{equation}
    \label{face_coe_harmonic_diffusion_flux}
    F_{i+\frac{1}{2}} ^D = -\frac{\rho_{i+1}-\rho_i}{3\sigma_{t,i+\frac{1}{2}} (x_{i+1}-x_i)},
\end{equation}
where the equivalent opacity at the face ${i+\frac{1}{2}}$ is given by: 
\begin{equation}
    \label{face_coe_harmonic}
    {\sigma_{t,i+\frac{1}{2}}} = \frac{\sigma_i \varDelta x_i + \sigma_{i+1} \varDelta x_{i+1}}{\varDelta x_i + \varDelta x_{i+1}}.
\end{equation}


\subsection{New methods}

In GSIS, simply applying the harmonic mean method to the diffusion equation is not sufficient. The reconstruction of the upwind flux of the RTE cannot be modified using the same approach. Alternative modifications to the discrete gradient operator $\nabla_i$, which will simultaneously affect the upwind flux of the RTE, the computation of higher-order terms, and the diffusion flux, are needed.

It should be noted that,  however, an appropriate gradient algorithm is not unique. Generally, there are some principles that should be satisfied. First, at discontinuous interface, the optically thick side dominates the effective optical thickness. Second, the resulting numerical gradient operator $\nabla_i$ is linear. Third, it degenerates to a certain gradient calculation algorithm, such as the original least square method, when the opacity is uniform or vary slowly in space.





\begin{figure}[t]
    \centering 
    \subfigure[]{\includegraphics[width=0.53\textwidth,trim={50 15 50 35},clip]{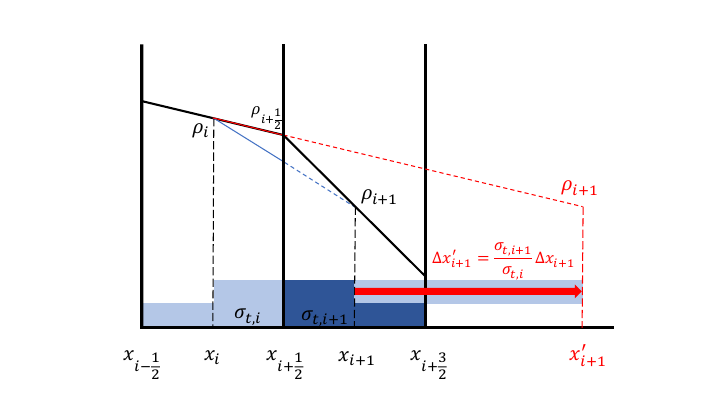}}
   \subfigure[]{ \includegraphics[width=0.44\textwidth,trim={20 -10 20 0},clip]{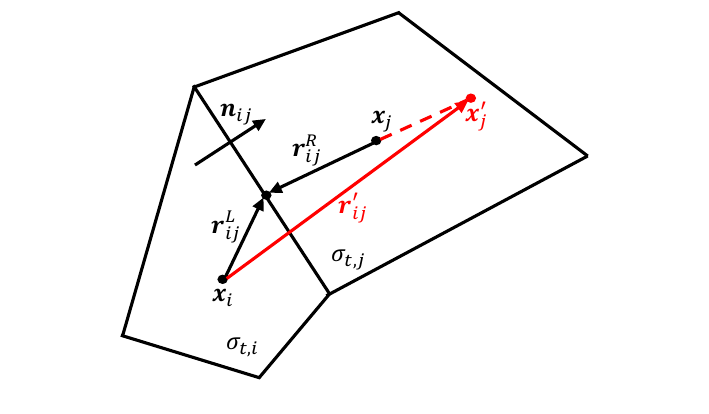}}
    \caption{
    Schematic of the flux reconstruction in (a) one-dimensional and (b) two-dimensional  finite-volume method when adjacent cells have disparate opacities.
    }
\label{fig:face_reconstruction_1d_example}
\end{figure}


The harmonic mean method offers us some inspiration and guidance. We interpret it in Fig.~\ref{fig:face_reconstruction_1d_example}(a) in another way by introducing the effective position, rather than the equivalent opacity. 
The radiation energy density, distribution function, and opacity at the cell $i$ is $\rho_i$, $I_i$, and $\sigma_{t,i}$, perspectively. The straightforward linear approximation for gradient, shown in blue dashed lines, does not provide consistent results in the two adjacent cells. 
The reconstruction suggested by the harmonic mean method, when $\rho_{i+\frac{1}{2}}$ is given by Eq.~\eqref{face_rho_harmonic}, is considered to be optimal.
If one insists to use the linear approximation (shown in red lines) in the right hand side of the interface, then
an auxiliary point $x'_{i+1}$ is created by shifting $x_{i+1}$ according the ratio of opacity in both cells:
\begin{equation}
    \label{shift_disp}
     \frac{x'_{i+1} - x_{i+\frac{1}{2}}}{x_{i+1} - x_{i+\frac{1}{2}}} = \frac{\sigma_{t,i+1}}{\sigma_{t,i}},
\end{equation}
such that the face value $\rho_{i+\frac{1}{2}}$ is exactly the same as that of the harmonic mean method. 

This idea can be easily extended to the finite-volume framework in high-dimensional space. For example, for general two-dimensional cases, as shown in Fig.~\ref{fig:face_reconstruction_1d_example}(b), a feasible modification imitating Eq.~\eqref{shift_disp} is to shift the adjacent cell center point $\bm{x}_j$ to $\bm{x}'_j$, in the direction of the vector $\bm{r}_{ij}^{R}$ pointing to the face midpoint $\bm{x}_{ij}$, which satisfies:
\begin{equation}
    \label{shifted_j}
    \bm{x}'_j - \bm{x}_{ij}  = \frac{\sigma_{t,j}}{\sigma_{t,i}} \bm{r}_{ij}^{R}.
\end{equation}
Meanwhile, the original approximation in  Eq.~\eqref{Uj-Ui} is replaced by: 
\begin{equation}
\label{new_vector}
    U_j - U_i \approx  \nabla_i U \cdot \left( \bm{x}'_j - \bm{x}_i\right )  .
\end{equation}
Correspondingly, the least squares function becomes: 
\begin{equation}
    \label{LSF_mdf}
    Y \left( \nabla_i U \right) = \sum_{j\in N(i)} \left [ \nabla_i U \cdot \left ( \bm{x}'_{j} - \bm{x}_i\right ) - \left ( U_j - U_i \right ) \right]^2,
\end{equation}
and the gradient can be calculated in analogy to Eq.~\eqref{LSF}.  Accordingly, the estimation of the diffusion flux changes in Eq.~\eqref{Gamma_ij_def} is given as:
\begin{equation}
  \label{Gamma_ij_com2}
  \Gamma_{ij} = -\frac{\left ( \sigma_{t,i} \bm{r}_{ij}^L - \sigma_{t,j}\bm{r}_{ij}^R  \right )  \cdot \bm{n}_{ij}}{3 \left |\sigma_{t,i} \bm{r}_{ij}^L - \sigma_{t,j}\bm{r}_{ij}^R \right |^2},
\end{equation}
where $\left |\sigma_{t,i} \bm{r}_{ij}^L - \sigma_{t,j} \bm{r}_{ij}^R \right |$ is regarded as the optical thickness between adjacent cell grids.

The effectiveness of the resulting gradient algorithm is shown in Section~\ref{tests}.

%% file: data/sec05.tex
\section{Numerical Examples}
\label{tests}

In this section, several numerical tests are presented, including one- and two-dimensional problems, steady-state and transient problems, as well as in homogeneous and heterogeneous media. These cases are designed to highlight the characteristics of multi-scale problems, featuring a wide range of optical coefficient variations. Comparisons of the efficiency and accuracy between conventional iteration methods and the synthetic iteration method are also provided. 
The convergence criterion is defined as the volume-weighted relative error of a macroscopic quantity $W$ between successive iterations:
\begin{equation}
    \label{Err_W}
    \text{Err}(W^m) = \sqrt{\frac{\sum_i \left ( W_i^m - W_i^{m-1} \right )^2 V_i}{\sum_i \left (W_i^{m} \right ) ^2 V_i}}.
\end{equation}
Convergence is deemed to be achieved when $\max \left [\text{Err}(\rho^m), \text{Err}(T^m) \right ] < 10^{-6}$, unless otherwise specified.

The test cases include both dimensioned and dimensionless problems. The normalized form \eqref{RTE_g_dimless} with the scaling parameter $\epsilon$, is used. For dimensionless cases, unless otherwise specified, the radiation constant $a$ and the speed of light $c$ are both set to $1$. For dimensional cases, $\epsilon$ is set to $1$, while other parameters are physical: the radiation constant is $a = 0.01372 \, \text{\text{GJ}/cm}^3/\text{\text{keV}}^4$, and the speed of light is $c = 29.98 \, \text{cm/ns}$. The CFL number is defined as CFL$= {\epsilon \varDelta x_{ref}}/{(c \varDelta t)}$, where $\varDelta x_{ref}$ is the characteristic length of spatial grid. It is noted that except in case in Section~\ref{test_pseudo}, the pseudo-time method in Section~\ref{pseudo_time} is actually not used, by setting the control parameters to be sufficiently large.


\subsection{One-dimensional steady-state transport}

For one-dimensional steady-state cases, the time derivative is dropped since $\varDelta t \to \infty$. The transport equation is given as:
\begin{equation}
\cos\theta\frac{\partial I}{\partial x} = \frac{\sigma_t}{\epsilon}\left(  \frac{\rho}{4\pi}     -I\right),
\end{equation}
where the computation domain is $x \in \left [ 0,1 \right ]$. 
At the left (right) boundary, there is a hot (cold) wall with an isotropic radiation energy source $\rho = 1$ ($\rho = 1\times 10^{-6}$). The walls are assumed to be non-reflective. The solid angle is discretized by the Gauss-Legendre quadrature with $N_{\theta}\times N_{\varphi}=16\times24$ nodes. In this sequence of tests, all parameters are dimensionless.

\begin{figure}[t]
    \centering
   {\includegraphics[width=0.48\textwidth,trim={10 10 10 10},clip]{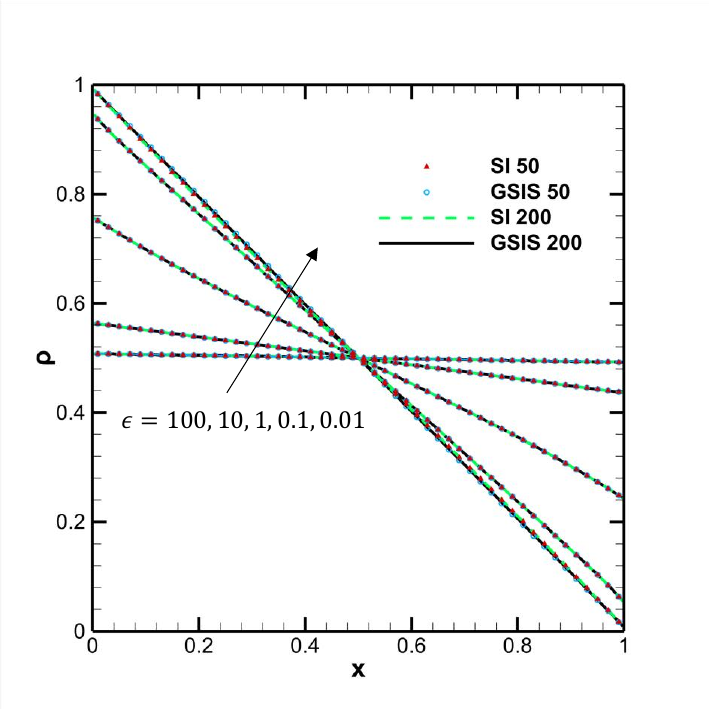}}
   {\includegraphics[width=0.48\textwidth,trim={10 10 10 10},clip]{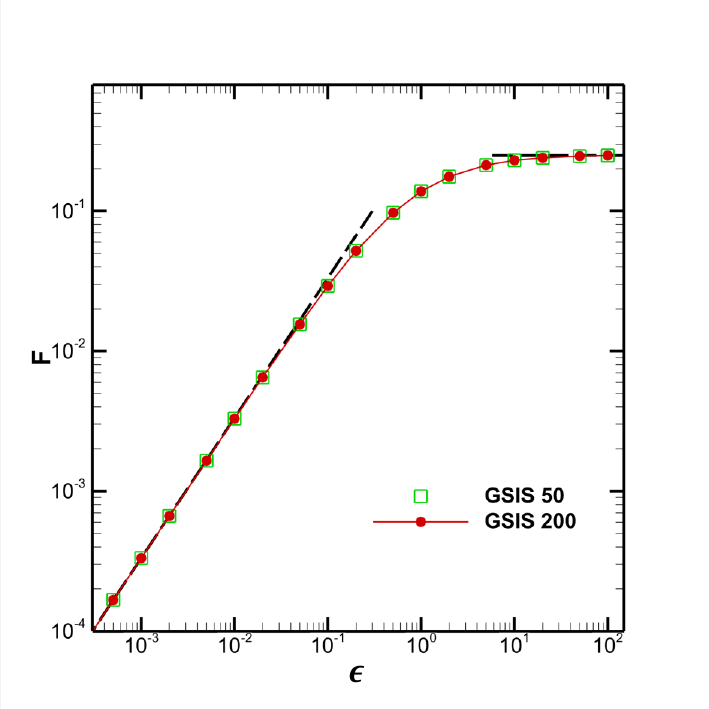}}
    \caption{Profiles of radiation energy and heat flux in the one-dimensional case. The heat flux is obtained from GSIS only.  
    }
    \label{fig:1d_const}
\end{figure}

\begin{table}[t!]
	\centering
    \caption{A comparison of computational efficiency of SI and GSIS in the one-dimensional case: iteration steps (index m) and computation time.  ``-- --" means that the iteration takes too long.
    }
\label{table:1d_const}
\begin{tabular}{|c|c|c|c|c|c|c|c|}
    \hline
    \multicolumn{2}{|c|}{$\epsilon$}            & 100 & 10 & 1 & 0.1 & 0.01 & 0.001 \\
    \hline
    \multirow{2}{*}{SI 50}  & iteration steps &  3 & 5 & 9 & 91 & 3048 & -- --\\
    \cline{2-8}              & CPU time (s)& 0.0130 & 0.0204 & 0.0332 & 0.260 & 8.43 & -- --\\
    \hline
    \multirow{2}{*}{GSIS 50}  & iteration steps &  3 & 5 & 9 & 23 & 16 & 34 \\
    \cline{2-8}              & CPU time (s)& 0.0136 & 0.0317 & 0.0474 & 0.0979 & 0.0774 & 0.158\\
    \hline
    \multirow{2}{*}{SI 200}  & iteration steps &  3 & 4 & 10 & 91 & 3070 & -- --\\
    \cline{2-8}              & CPU time (s) & 0.0471 & 0.0553 & 0.120 & 0.984 & 32.5 & -- --\\
    \hline
    \multirow{2}{*}{GSIS 200}  & iteration steps &  3 & 4 & 9 & 50 & 61 & 19 \\
    \cline{2-8}              & CPU time (s)& 0.0480 & 0.0580 & 0.141 & 0.728 & 0.877 & 0.331\\
    \hline
\end{tabular}
\end{table}

\subsubsection{Constant opacity}

We keep $\sigma_t = 1$, but vary the value of $\epsilon$. For comparison between SI and GSIS, calculations are performed on uniform mesh with $N=50$ and $N=200$ cells, respectively. 
 
Profiles of radiation energy for $\epsilon = 100, 10, 1, 0.1, 0.01$ are plotted in Fig.~\ref{fig:1d_const}, where both SI and GSIS provide accurate solutions when $N=50$. 
The computational efficiency, shown in Table~\ref{table:1d_const}, is assessed by iteration steps and computation time to reach convergence. 
When $\epsilon = 100, 10, 1$, the SI is efficient, and the additional solving of the synthetic equation in GSIS causes a slight increase in computation time when compared to that of the SI. 
When $\epsilon$ is small, SI exhibits slow convergence, while GSIS demonstrates its superb  convergence acceleration, e.g., about one or two orders of magnitude speed-up for $\epsilon \sim 0.01$. 
If $\epsilon$ is further decreased, SI would reach its limitation in grid size and become dissipative, together with the slow convergence rate at the analytical level.  
However, the heat flux in Fig.~\ref{fig:1d_const} shows that the diffusion limit (shown as asymptotic lines in black dashes) is recovered in GSIS successfully. 
A more striking manifestation of the asymptotic preserving property of the GSIS can be observed in \ref{Appdx:vrf}, where the correct solution is achieved even when the cell size exceeds the mean free path by approximately four orders of magnitude.

\subsubsection{Spatial-varying opacity}


$\epsilon$ is set to $1$, while the opacity $\sigma_t$ is given by a piecewise function:
\begin{equation}
    \sigma_t (x) =\begin{cases}
        1, &x \le  0.2, \\
        0.1, & 0.2 < x \le 0.5, \\
        \exp{(10x+1)}, & 0.5 < x \le 0.8, \\
        1000, & x > 0.8.
        \end{cases}
\end{equation}
The discontinuities in opacity at $x=0.2$, 0.5 and 0.8 result in transport-transport, transport-diffusion, and diffusion-diffusion interfaces, respectively.
Between $x=0.5$ and 0.8, there is a continuous but steep variation of opacity. Both factors introduce difficulties in the numerical simulation.
The test is presented on uniform mesh with $N=50$ and $N=200$. 
We compare the results of either the original least square method  (denoted by ``N") or the improved least square method  (denoted by ``Y")  used in GSIS.

\begin{figure}[t]
    \centering
{\includegraphics[width=0.48\textwidth,trim={10 10 10 10},clip]{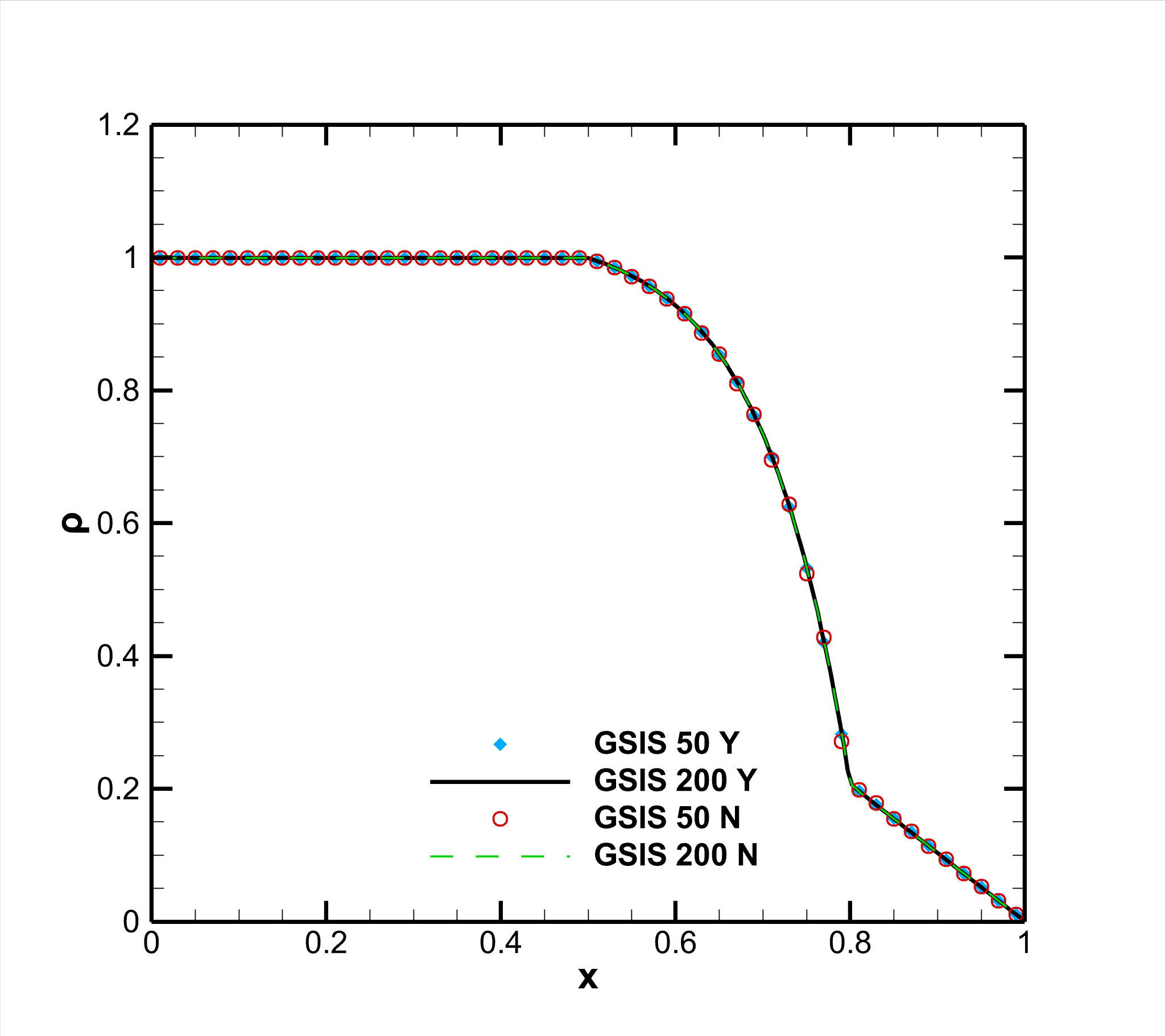}}
{\includegraphics[width=0.48\textwidth,trim={10 10 10 10},clip]{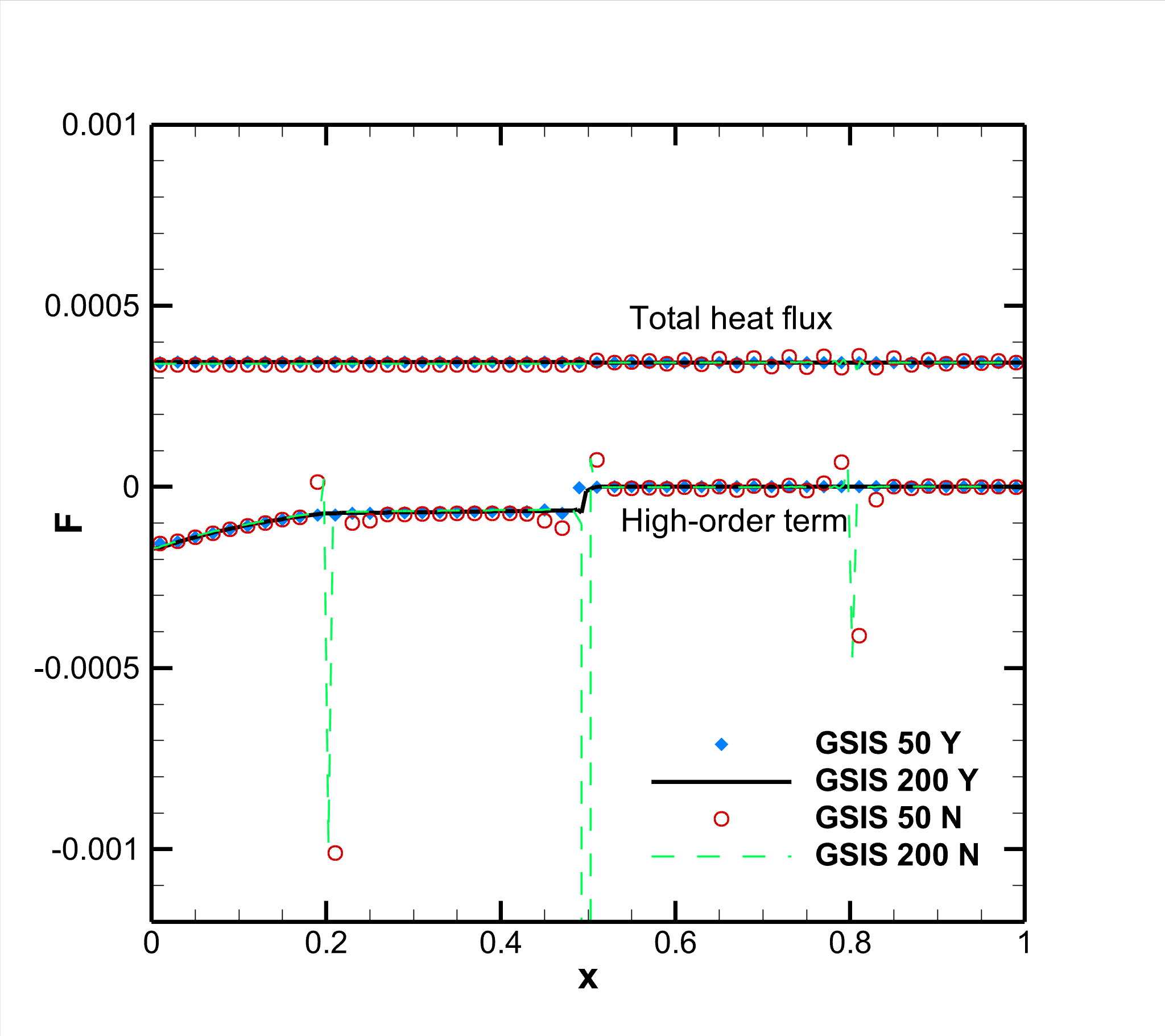}}
    \caption{Radiation energy and heat flux (total as well as the high-order component) in the one-dimensional heat conduction with spatial-varying opacity.}
    \label{fig:1d_vrb}
\end{figure}

Figure~\ref{fig:1d_vrb} shows the distributions of radiation energy,  the total heat flux distribution, and the contributions of high-order terms.
The heat flux distributions highlights the advantages of the improved method, especially in handling problem with coefficient variations. 
 The total heat flux is supposed to be constant, and the results using the improved method, despite fewer grids (GSIS 50 Y), still accurately portray it, even with significant discontinuities and spatial variations in opacity. 
In contrast, results without the improved method show oscillations near the interfaces where one side is in the diffusion regime. 
This is due to the wrong capture of higher-order terms at discontinuous optical coefficient interfaces, which would introduce extra perturbations. 
We can infer that these perturbations propagate off the interface in solving the macroscopic synthetic equation, under the effect of the diffusion operator, diminishing with respect to the number of grids.
Hence it impacts calculation with fewer grids more significantly. For transport-transport interfaces, the solution of the transport equation dominates in the synthetic iteration, and gradient calculations only affect the interface reconstruction accuracy, which differs from transport-diffusion and diffusion-diffusion interfaces. 

In Fig.~\ref{fig:1d_vrb}, the higher-order term $\text{HoT}_{\bm{F}}$ calculated from Eq.~\eqref{RTE_steady_d_HoT} confirms the GSIS as an effective multiscale method: as $\epsilon$ approaches 0, the higher-order terms describing transport effect, should also approach $0$, which is particularly important in radiative transport problems facing discontinuities of opacity. 
Additionally, near the transport-diffusion interface at $x=0.5$, the variation in high-order terms is steep. Although using 50 grids is sufficient for the calculation, employing more grids is a better choice to accurately portray the variation of high-order terms.

\subsection{Two-dimensional steady-state transport}

The rectangular computation domain is bounded by $x=0$, $y=1$, $x=1$, and $y=0$. The bottom  wall is hot with an isotropic source $\rho_h = 1$, and the rest are cold walls with $\rho_c=1\times10^{-6}$. 
All the boundaries are non-reflective. 
The governing equation is 
\begin{equation}
\cos \theta \frac{\partial I}{\partial x}+ \sin \theta \cos \varphi \frac{\partial I}{\partial y}= {\sigma_t}\left(  \frac{\rho}{4\pi}     -I\right), 
\end{equation}
and the solid angle is discretized by the Gauss-Legendre quadrature with $N_{\theta}\times N_{\varphi} = 36 \times 48$ nodes. 

\begin{figure}[!t]
    \centering
{\includegraphics[width=0.32\textwidth,trim={20 80 20 270},clip]{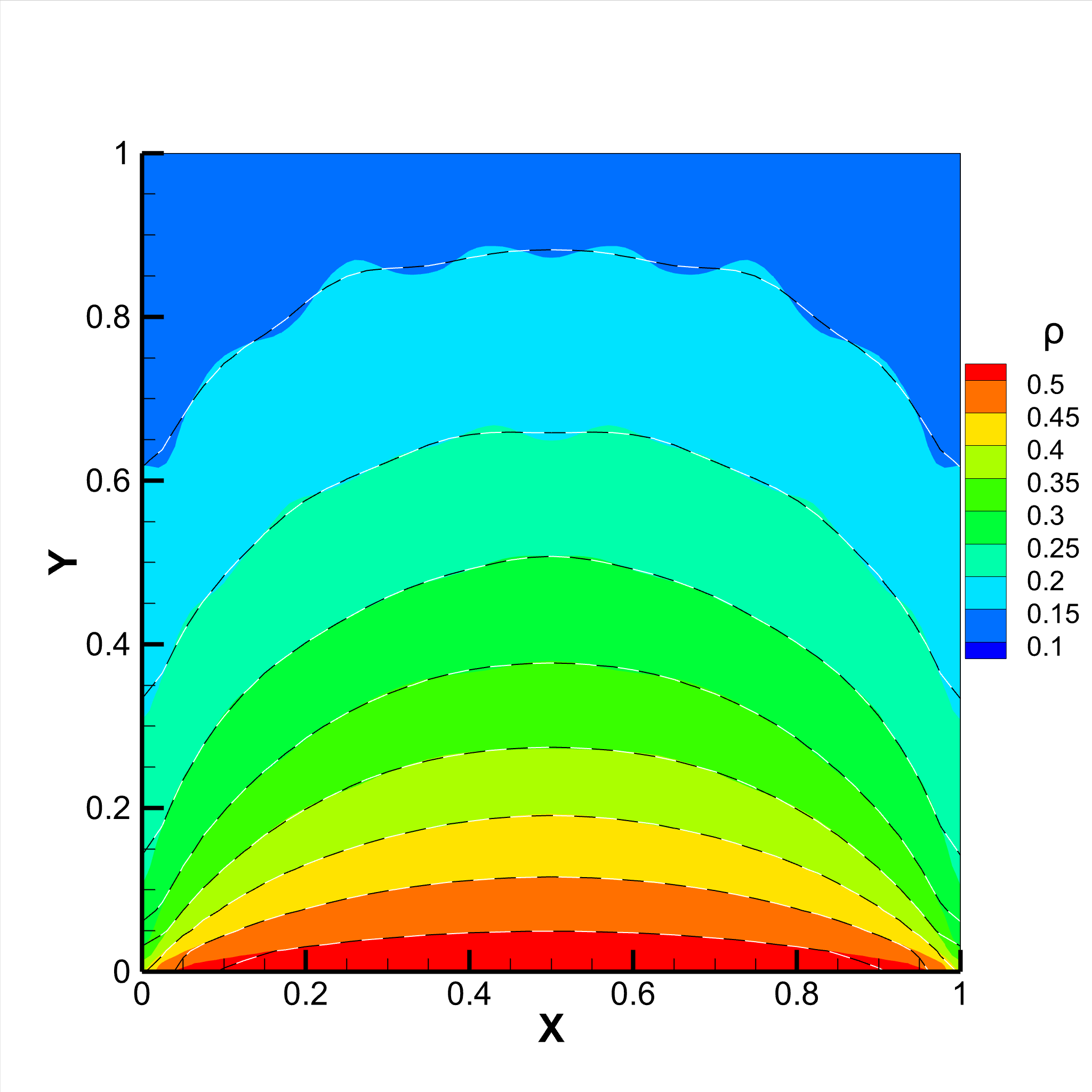}} 
{\includegraphics[width=0.32\textwidth,trim={20 80 20 270},clip]{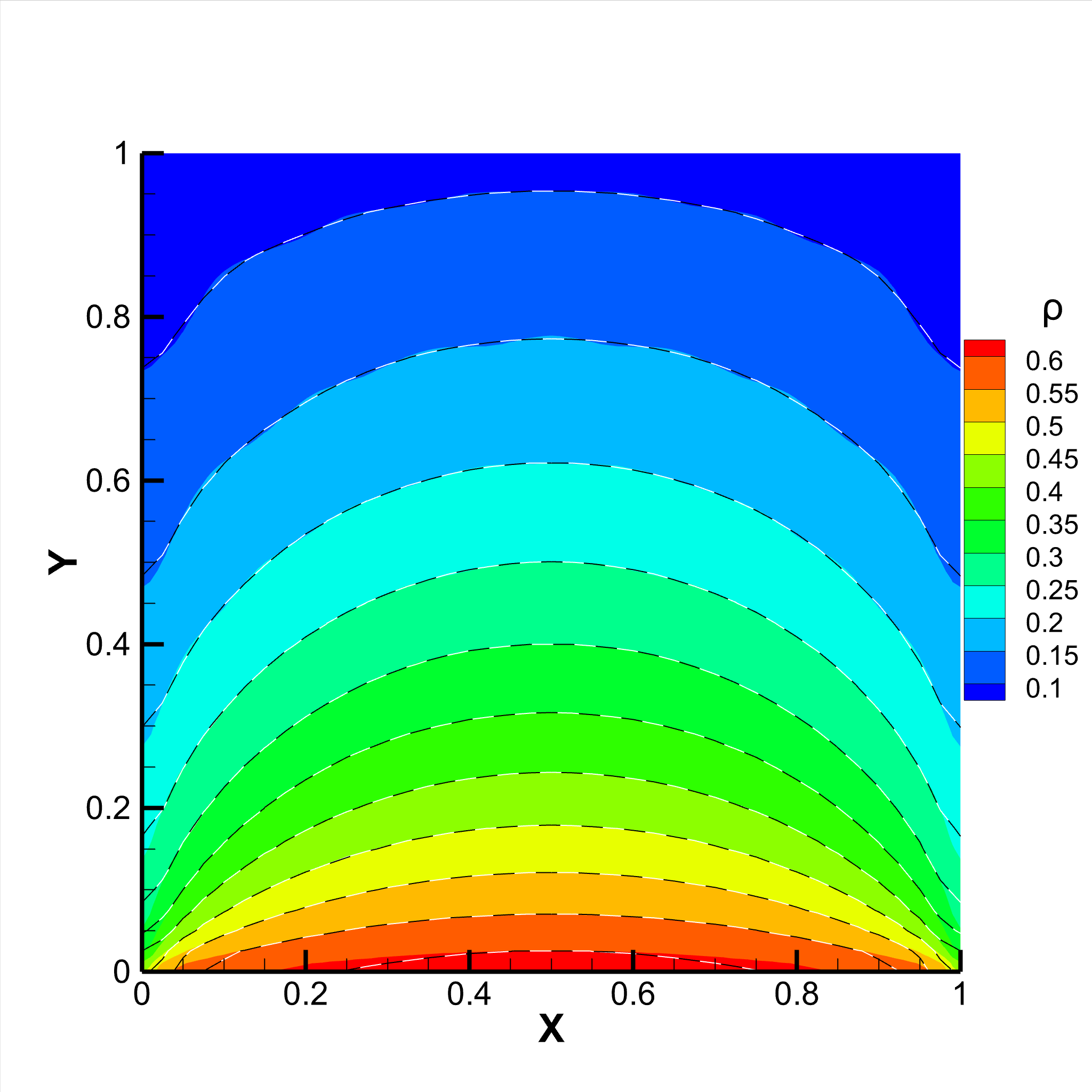}}
{\includegraphics[width=0.32\textwidth,trim={20 80 20 270},clip]{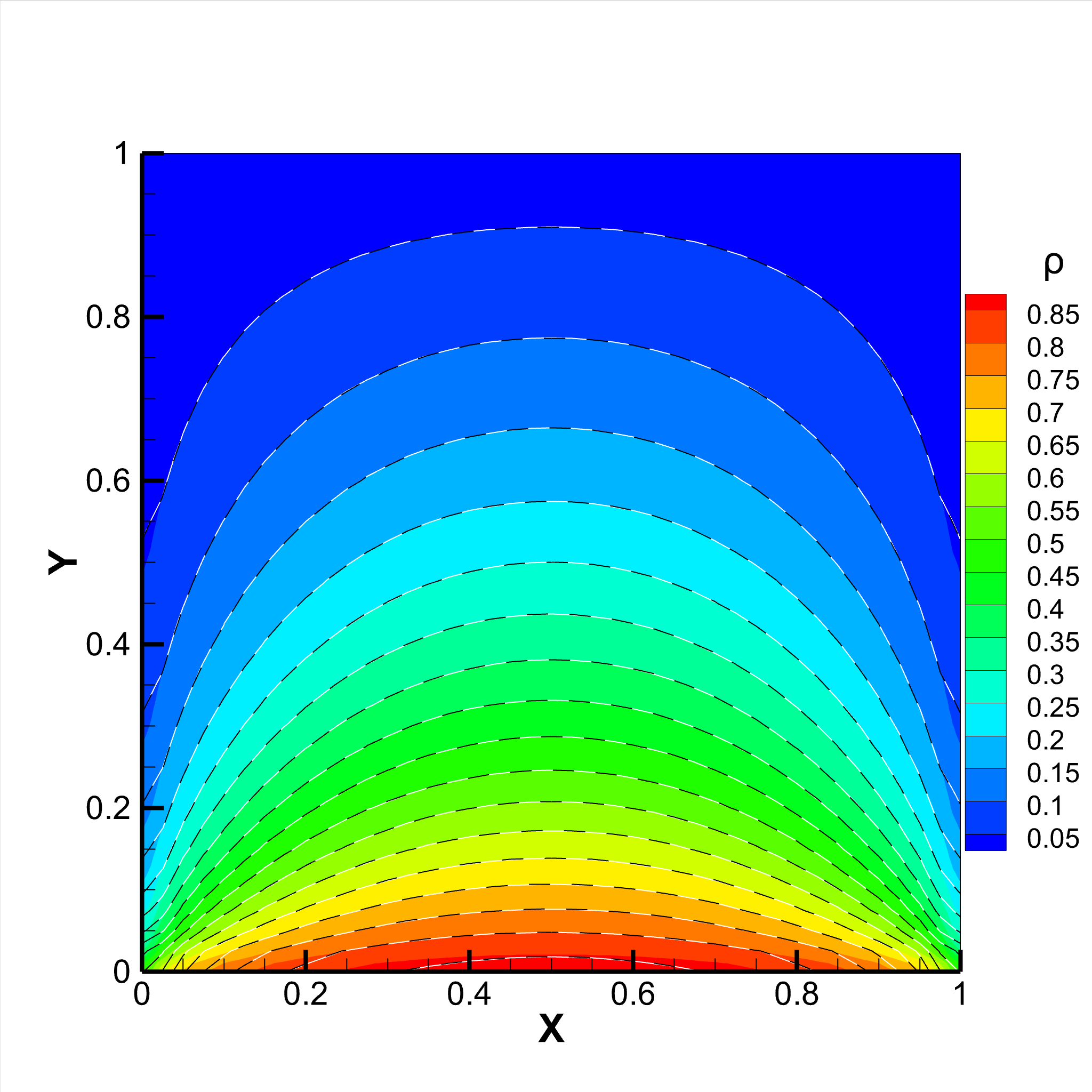}}\\
\vspace{0.3cm}
{\includegraphics[width=0.32\textwidth,trim={20 80 20 260},clip]{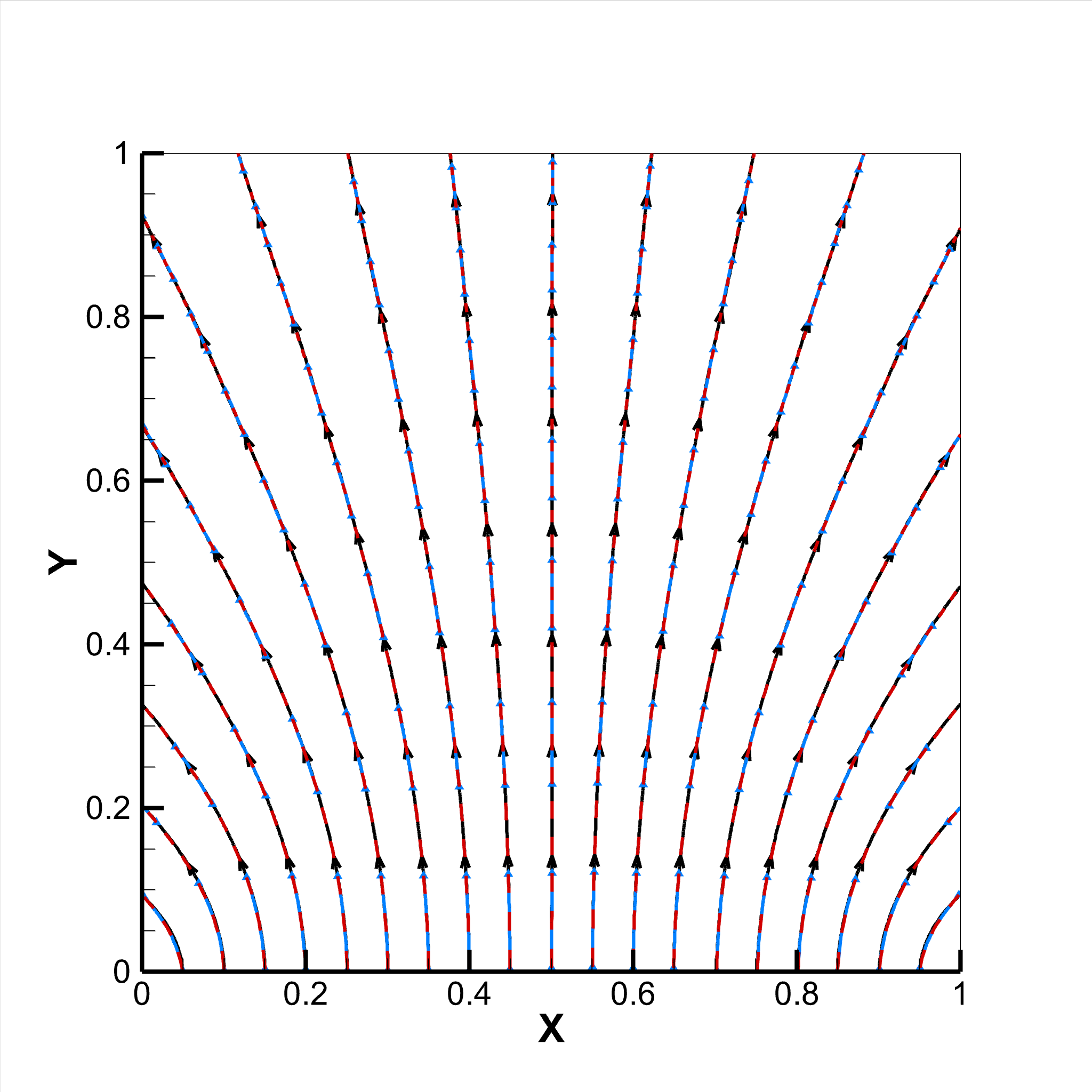}}
{\includegraphics[width=0.32\textwidth,trim={20 80 20 260},clip]{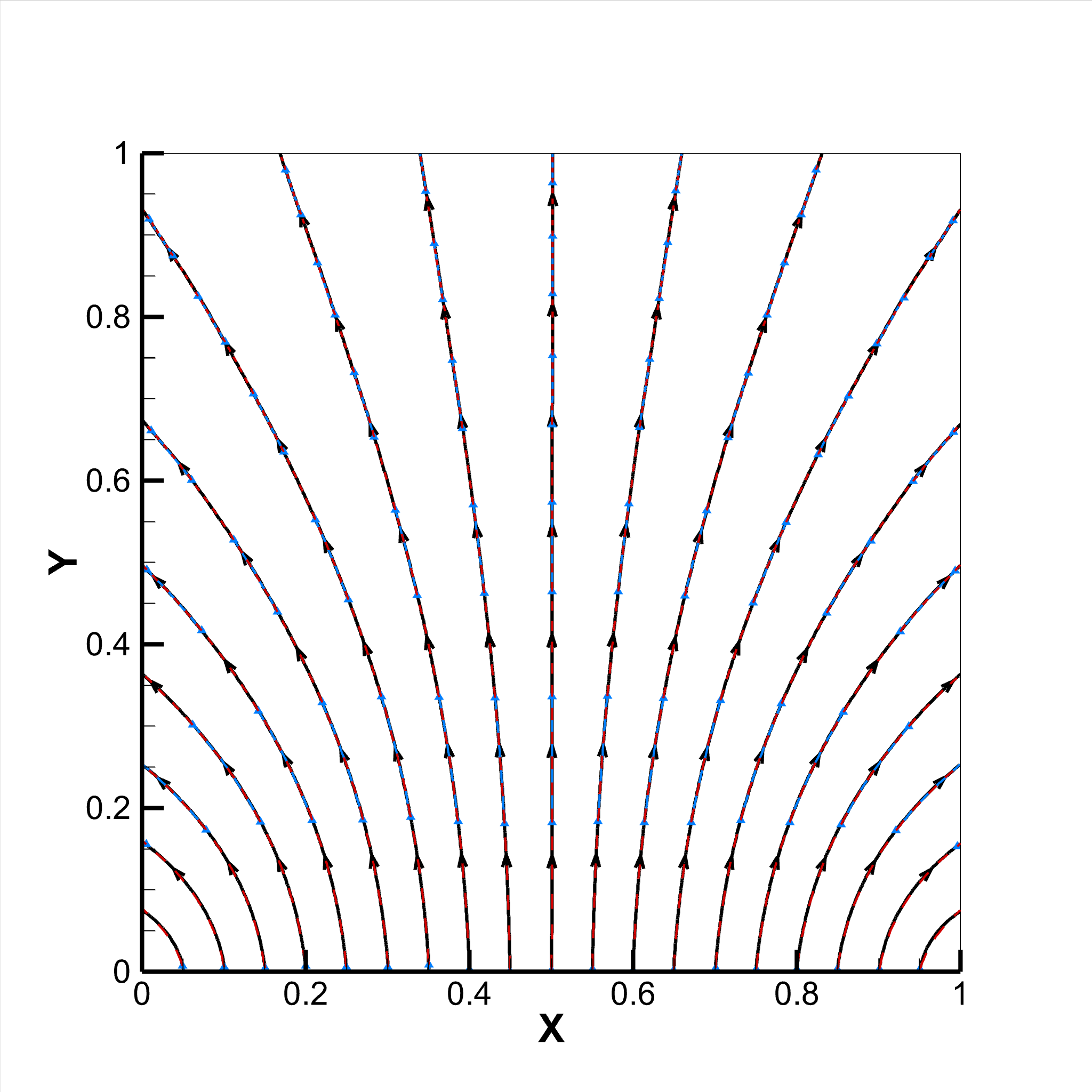}}
{\includegraphics[width=0.32\textwidth,trim={20 80 20 260},clip]{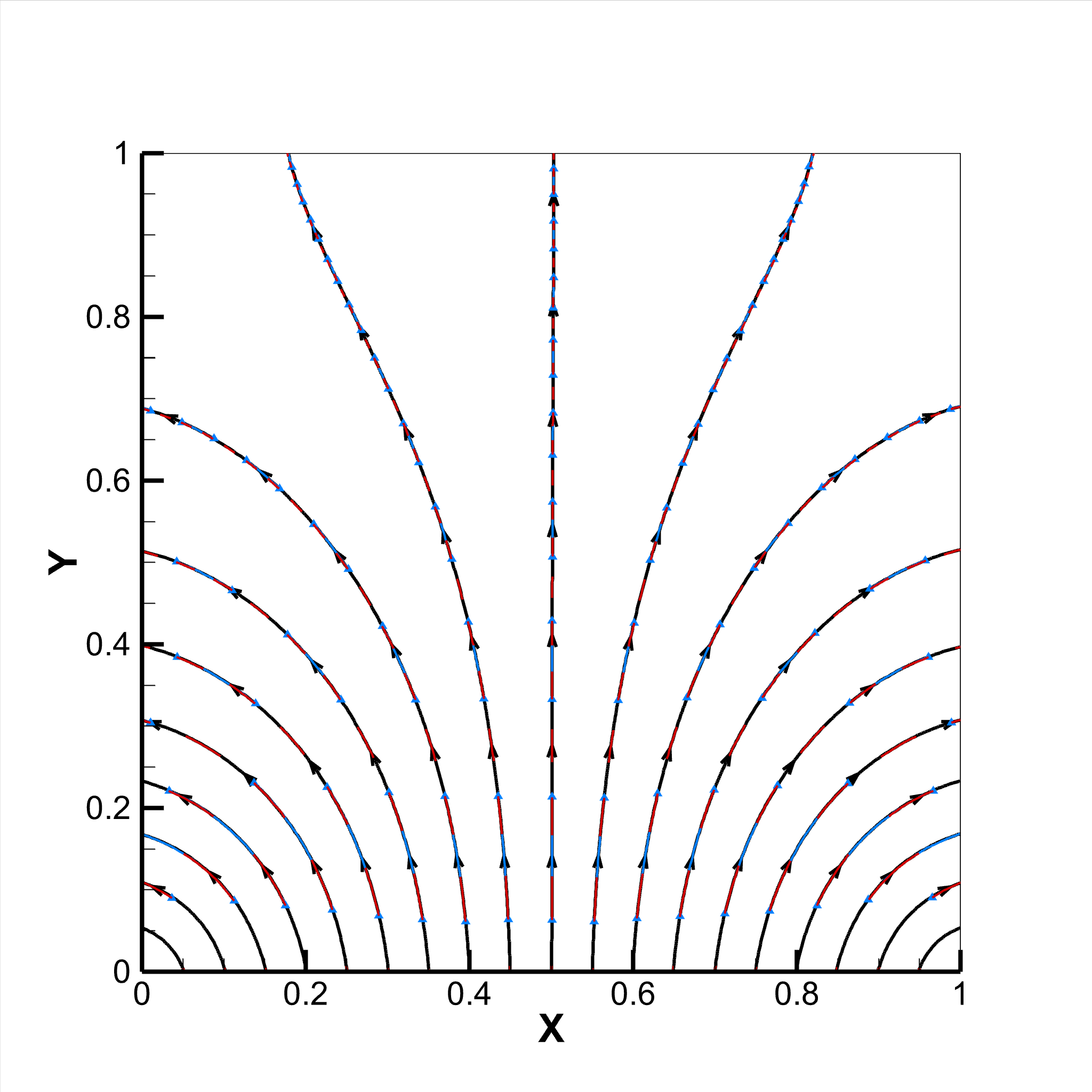}}
    \caption{(First row) Contours of radiation energy at different constant opacity $\sigma_t$. The colored background, dashed lines and white solid lines are results obtained from the SI $100$, SI $40$ and GSIS $40$, respectively. 
    (Bottom row) Heat flux streamlines for different constant opacity $\sigma_t$. The black solid line with arrow, red dashed line and blue dashed line with triangle are SI $100$, SI $40$ and GSIS $40$ results, respectively.
    From the left to right column, $\sigma_t = 0.2$, 1, and 10, respectively.
    }
    \label{fig:2d_const_contour}
\end{figure}


\subsubsection{Radiation in uniform black square}

The computational domain is filled of  media with uniform opacity $\sigma_t$. Uniform meshes with $N_x\times N_y=40\times 40$ and $100 \times 100$ are used. 
The distributions of radiation energy are shown in Fig.~\ref{fig:2d_const_contour}.  For optically thin media, the interaction between photons and the medium is weak. The radiation density becomes sparse along the path that photons transport away from the source. 
Since few photons are reflected back, the energy at the bottom tends to $0.5$ on the inner side of the domain. In contrast, photons are strongly scattered during transport in optically thick media. The radiation energy near the hot and cold walls approaches $1$ and 0, respectively. At the two bottom corners, the radiation energy exhibits significant spatial variation in optically thick media, as indicated by the dense contour lines. Using a $100 \times 100$ grids provides a better representation of large variations and boundary layers than using $40 \times 40$ grids. However, increasing the number of grids cells does not always lead to better results. In optically thin media, for the same angular discretization, the ray effect is more pronounced in Fig.~\ref{fig:2d_const_contour}(a) with the $100 \times 100$ grids, while the coarse spatial grid has more numerical dissipation, which smooths out the ray effect~\cite{RayEffect_ZHU2020}. 
Apart from these two factors due to discretization, results from GSIS 40, SI 40 and SI 100 are in a good agreement.

\begin{table}[t!]
	\centering
    \caption{ A comparison of computational efficiency of SI and GSIS in the two-dimensional case: iteration steps  and computation time. }
\label{table:2d_const}
\begin{tabular}{|c|c|c|c|c|c|c|c|}
    \hline
    \multicolumn{2}{|c|}{$\sigma_t$}            & 0.2 & 1 & 10 & 20 & 50 & 100 \\
    \hline
    \multirow{2}{*}{SI 40}  & iteration steps &  21 & 25 & 181 & 491 & 2005 & 5588\\
    \cline{2-8}              & CPU time (s)& 9.27 & 11.0 & 78.3 & 212 & 864 & 2403\\
    \hline
    \multirow{2}{*}{GSIS 40}  & iteration steps &  21 & 20 & 29 & 29 & 22 & 27 \\
    \cline{2-8}              & CPU time (s)& 10.2 & 10.1 & 14.5 & 14.6 & 11.1 & 13.3\\
    \hline
    \multirow{2}{*}{SI 100}  & iteration steps &  43 & 48 & 214 & 531 & 2045 & 5629\\
    \cline{2-8}              & CPU time (s) & 116 & 128 & 573 & 1436 & 5552 & 15088\\
    \hline
    \multirow{2}{*}{GSIS 100}  & iteration steps &  43 & 36 & 75 & 91 & 97 & 94 \\
    \cline{2-8}              & CPU time (s)& 133 & 116 & 223 & 269 & 284 & 275\\
    \hline
\end{tabular}
\end{table}

The heat flux streamlines are shown in the bottom row of Fig.~\ref{fig:2d_const_contour}. As the optical thickness increases, the heat flux vectors pointing out of the boundaries tend to be more perpendicular, the streamlines directed towards the top (side) boundary become sparse (denser). The results show good agreement with both SI and GSIS.

\begin{figure}[t!]
    \centering
   {\includegraphics[width=0.45\textwidth,trim={20 80 20 260},clip]{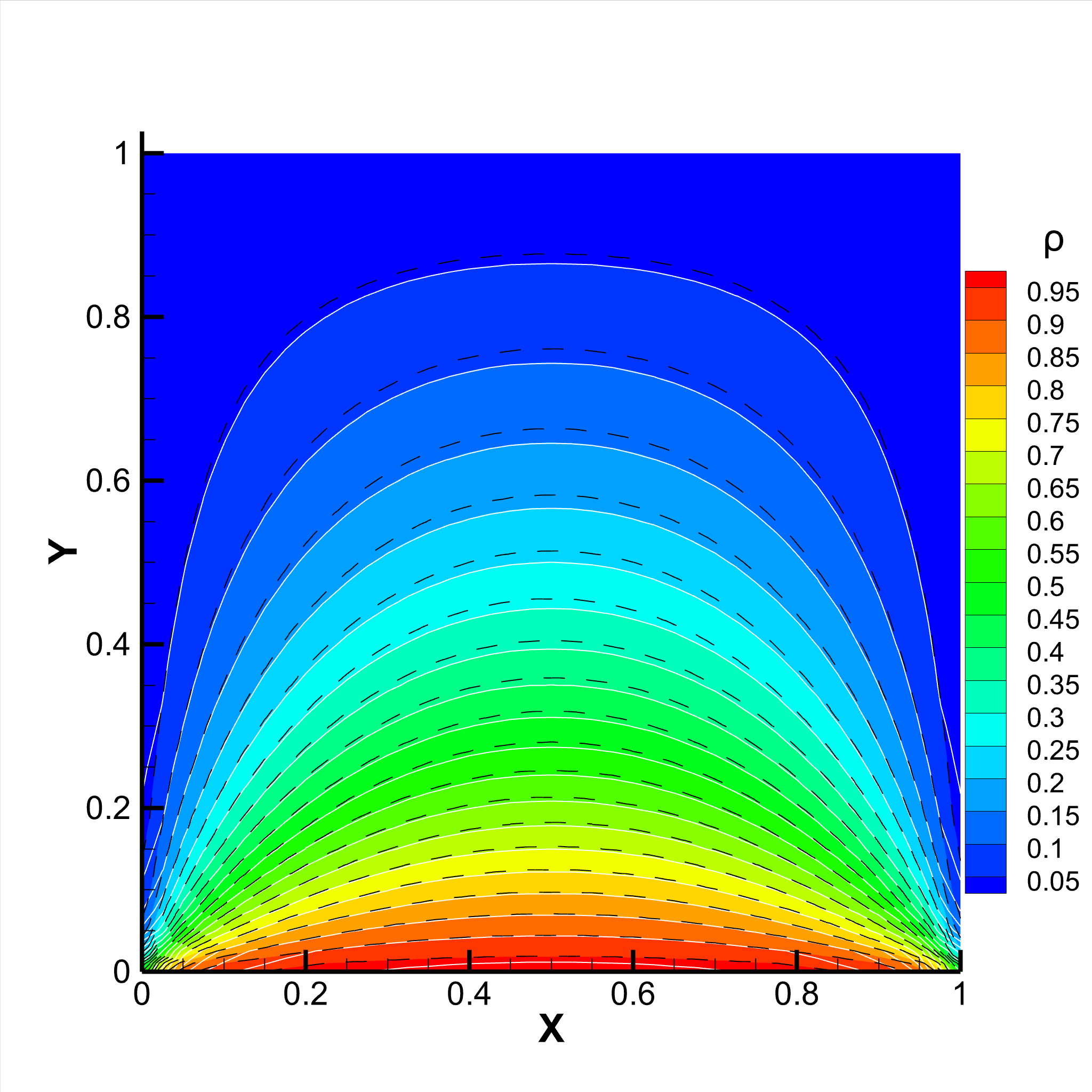}}
   {\includegraphics[width=0.45\textwidth,trim={20 80 20 260},clip]{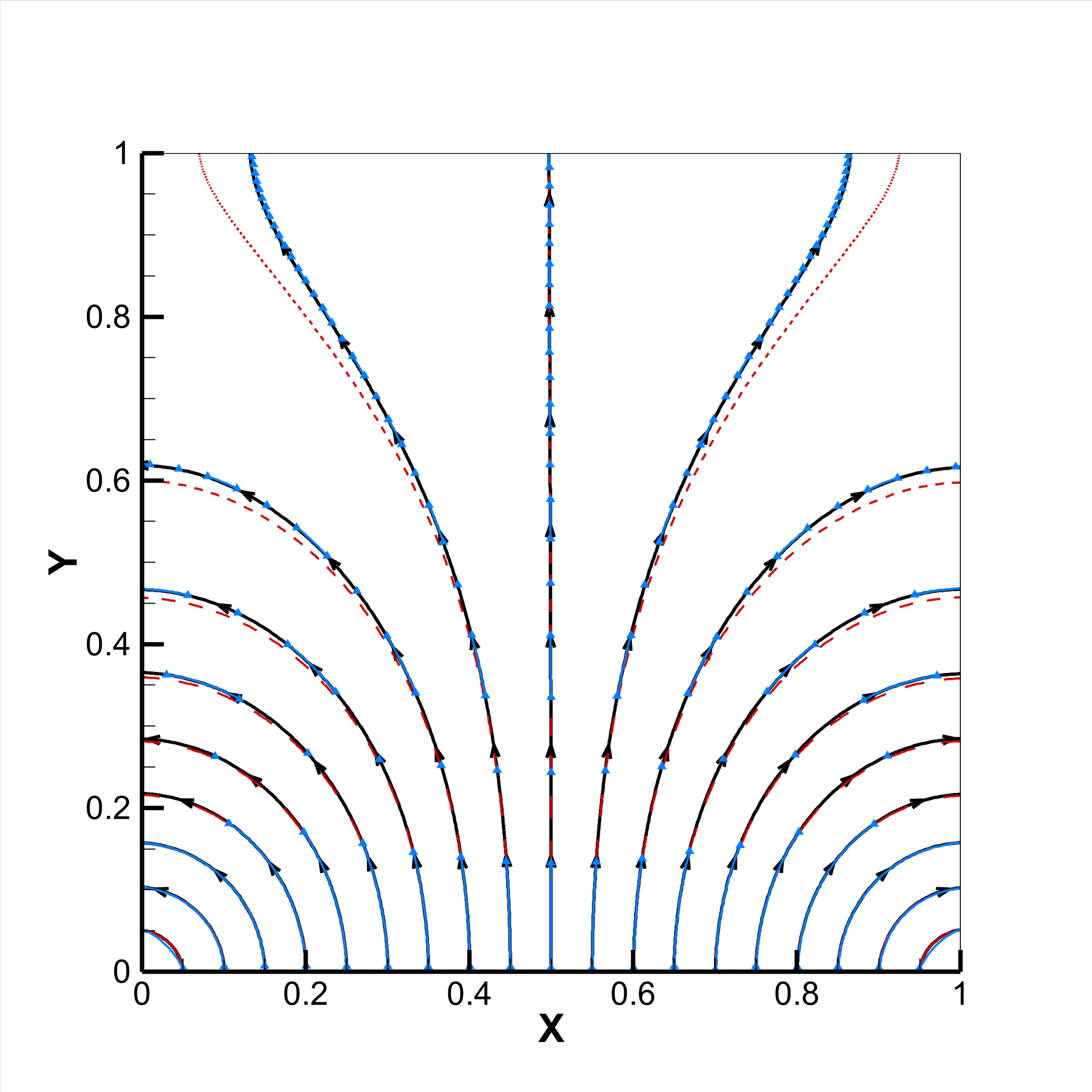}}
    \caption{(Left) Contour of radiation energy (colored background: GSIS $100$; black dashed line: SI $100$; white solid line: GSIS $40$) and (Right) streamlines of heat flux  (black solid line with arrow: GSIS $100$; red dashed line: SI $100$; blue dashed line with triangle: GSIS $40$) for $\sigma_t=100$.}
    \label{fig:2d_const_sigma100}
\end{figure}

The computational efficiency is compared in Table~\ref{table:2d_const}. For large optical thickness, the SI converges slowly, whereas the GSIS converges within 100 iterations for all cases considered. When \(\sigma_t = 100\), Fig.~\ref{fig:2d_const_sigma100} shows that the SI result exhibits a noticeable discrepancy from that of GSIS. This suggests that SI has not fully converged due to the false convergence~\cite{ADAMS20023}, despite meeting the convergence criterion in Eq.~\eqref{Err_W}. Even in this case, the GSIS has demonstrated a computational time savings of approximately two orders of magnitude. 

\subsubsection{A test for pseudo-time method}\label{test_pseudo}

\begin{figure}[t!]
    \centering
\subfigure[]{\includegraphics[width=0.45\textwidth,trim={20 80 20 100},clip]{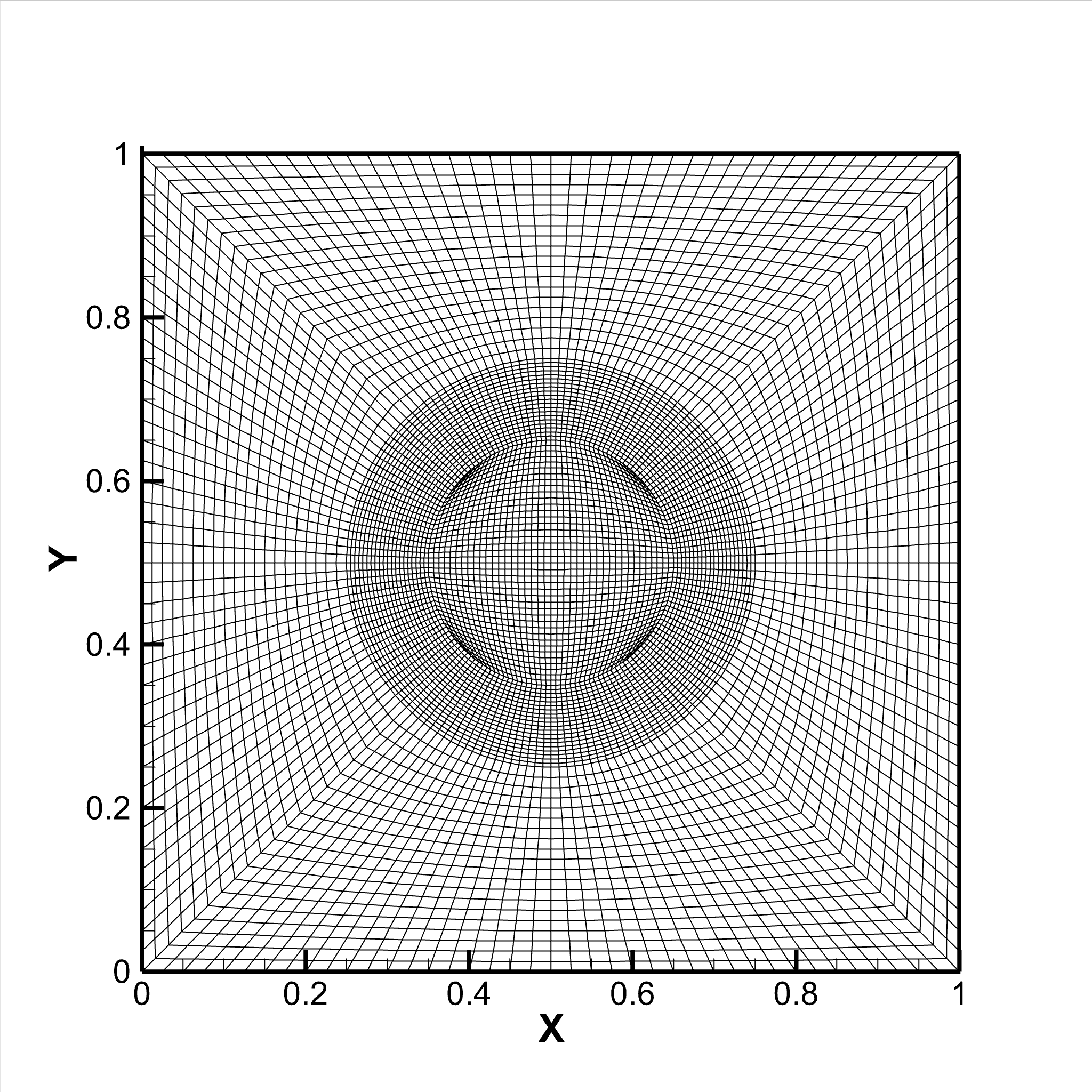}}
\subfigure[]{\includegraphics[width=0.45\textwidth,trim={20 80 20 100},clip]{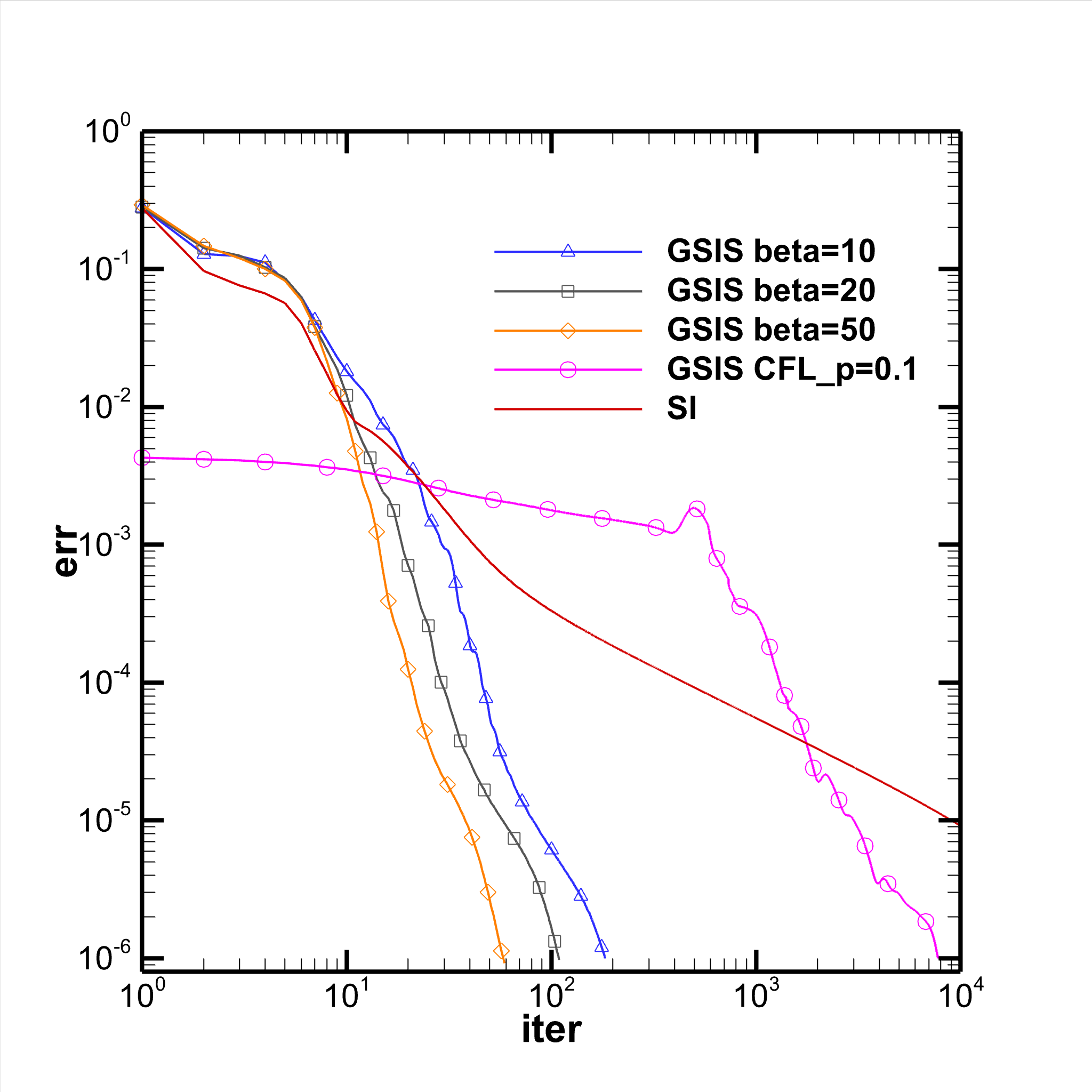}}
\\
\subfigure[]{\includegraphics[width=0.45\textwidth,trim={20 80 20 100},clip]{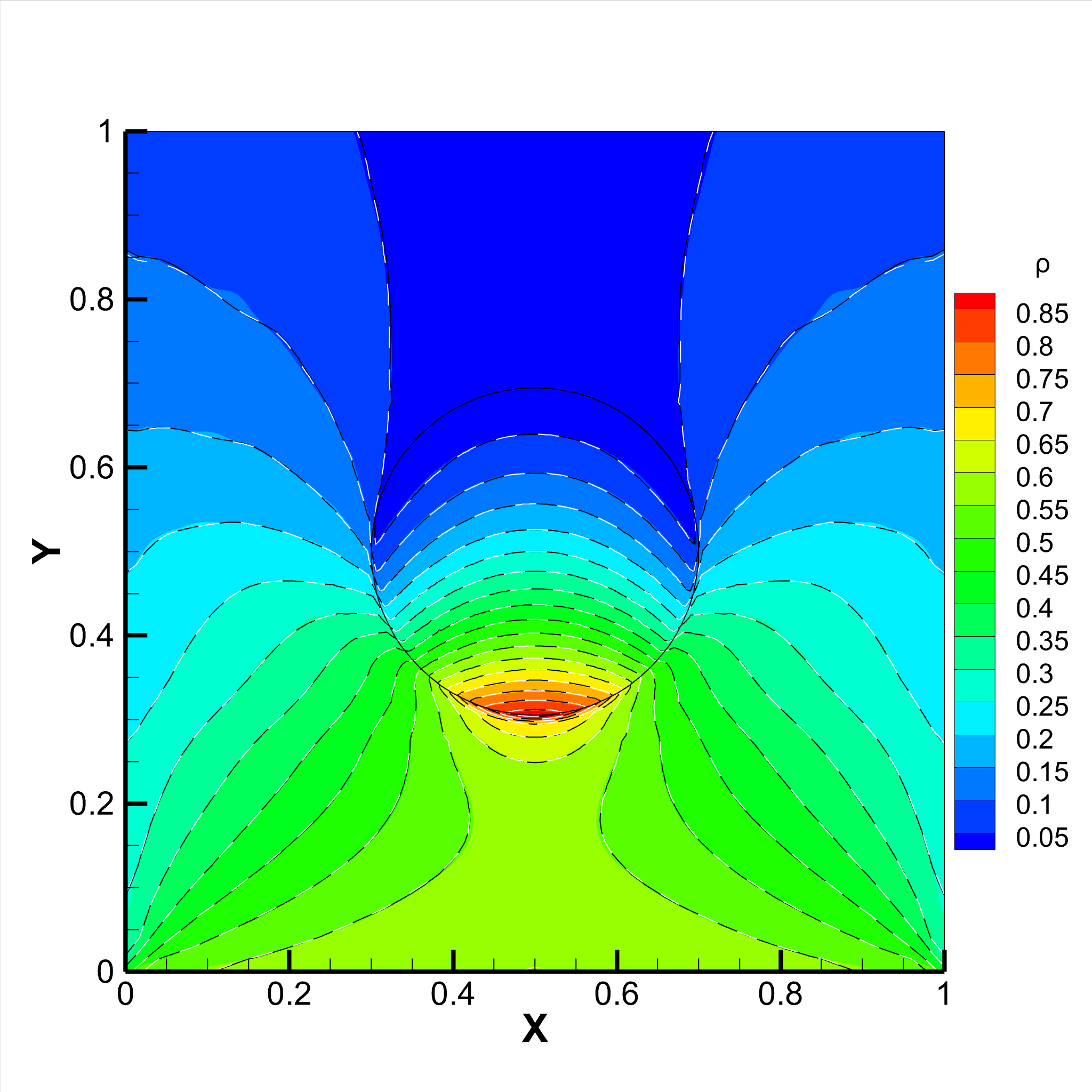}} 
\subfigure[]{\includegraphics[width=0.45\textwidth,trim={20 80 20 100},clip]{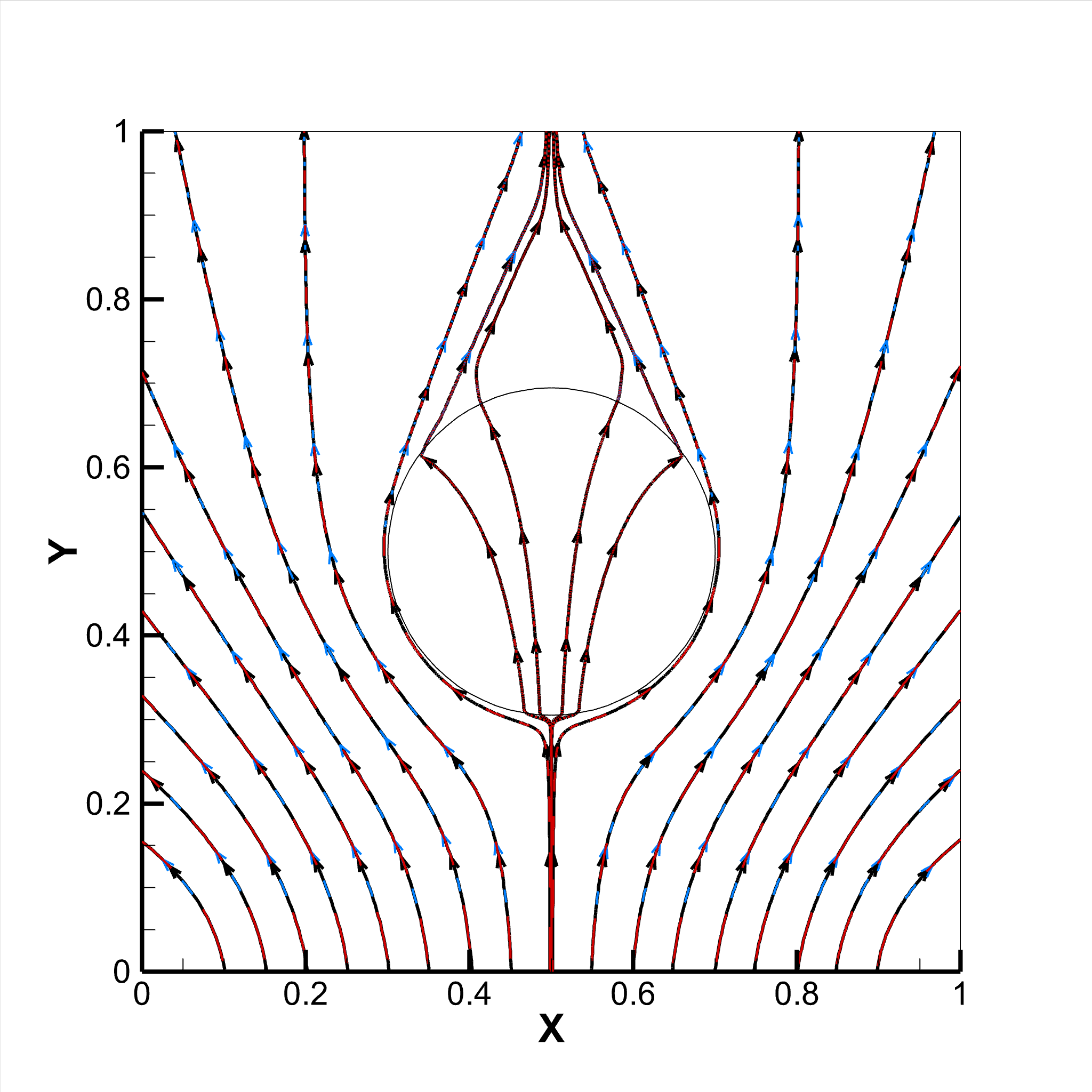}}
    \caption{(a) Geometry and  mesh. (b) Error convergence curves for GSIS of different pseudo-time stepping strategies.
    (c) Contour of radiation energy. Calculation on a double refined mesh is shown as reference solution. Colored background, black solid lines, and white dashed lines are numerical results from the GSIS refined, GSIS $\beta=50$, and GSIS CFL$_p$=0.1, respectively. (d) Heat flux streamlines. Black solid lines with arrows, blue dashed lines with arrows, and red dashed lines correspond to GSIS refined, GSIS $\beta=50$, and GSIS CFL$_p$=0.1, respectively.
    }
    \label{fig:2regions_circle}
\end{figure}

The central optical thick medium region is a circle of radius 0.2, with $\sigma_{t, \text{inner}} = 1000$, while the rest domain is filled with optical thin medium of $\sigma_{t, \text{outer}} = 0.01$. The geometry and non-orthogonal mesh are shown in Fig.~\ref{fig:2regions_circle}(a). We have refined the mesh at the interface between the optically thick and thin media, and the optical thickness of a single grid in the central circular region is approximately 5. Under these conditions, conventional iteration methods fail to produce accurate results and converge slowly. Moreover, directly solving the steady-state equation with synthetic iteration is unstable, necessitating a pseudo-time stepping method to limit the iteration increment at each step.

Two strategies of pseudo-time stepping are considered: the first is based on the local transport characteristic time defined by in Eq.~\eqref{cfl_pse}, controlled by the global parameter CFL$_p$, while the second is  based on the optical thickness between grids defined in Eq.~\eqref{tau_pse}, controlled by the global parameter $\beta$. Due to the significant difference in optical coefficients between the two regions, the first strategy requires limiting the increment in all grids in each iteration step. Tests show that setting CFL$_p=0.1$ ensures stable convergence. For the second strategy, the maximum value of $\beta$ is about 50.  

Figure~\ref{fig:2regions_circle}(b) shows that, the SI is stable but fails to meet the convergence criteria even after 10,000 iterations, while for GSIS, the iteration steps are 59 for GSIS $\beta=10$, 109 for GSIS $\beta=20$, 183 for GSIS $\beta=50$, and 7749 for GSIS $\text{CFL}_p$=0.1. Digging into details, we find that, 
within the initial 10 iterations, the transport equations of the optically thin region dominate the iteration increments, {as the photon is transporting from the boundary to interior}. The second strategy indirectly separates the optically thick and thin regions, reducing cross-region perturbations while maintaining the transport states of each region. Thus, the error reduction trend largely aligns with SI.  After approximately 10 iterations, the influence of $\beta$ value becomes evident. Under stable conditions, simulations with lesser restrictions (GSIS $\beta=50$) evolve toward the convergent solution more swiftly. In contrast, applying a small global pseudo-time step {in the first strategy} restricts the initial iteration increments, leading to a slow error decay.

In Fig.~\ref{fig:2regions_circle}(c, d), we present the contours of radiation energy density distribution and the streamlines of radiation heat flux. The two strategies converge to the same result. The radiation energy reaches a global maximum at the bottom of the optically thick circular region, with dense contour lines nearby, indicating significant spatial gradients. The point of maximum temperature (or energy density) is not necessarily closest to the heat source but rather located at the shallow surface of the optically thick region where radiation energy is absorbed, a characteristic of radiation heat transfer. Few heat flux streamlines penetrate the circular region, with most of the entering streamlines tending to leave normally. The final convergent results from different pseudo-time stepping methods show good consistency.


\subsection{Time-dependent GSIS at the free transport and diffusion limits}

Two cases with extremely small or large opacity are discussed~\cite{FEM-UGKS_XU2020}, corresponding to the free transport limit and the diffusion limit, showing how the synthetic solution behaves at both ends. We keep $\sigma_a = 0$ and set $\sigma_s$ to be a constant in Eq.~\eqref{RTE_g_dimless}, such that the RTE is reduced to a linear transfer model, independent of the material temperature. 
The computational domain spans $x \in [0,1]$. The left boundary is an isotropic radiation source with $\rho_h=1$, and the right boundary is cold with $\rho_c=1\times 10^{-6}$, both boundaries have no reflection. The solid angle is discretized by the Gauss-Legendre quadrature with $N_{\theta}\times N_{\varphi}=16\times24$ nodes.

\subsubsection{Free transport limit}

Initially, the domain is in equilibrium at $\rho = \rho_c$. All parameters are dimensionless, with $\epsilon=1$ and a small scattering coefficient $\sigma_s = 10^{-4}$. One may derive an approximate solution with the collisionless assumption, at a given time $\tau$:
\begin{equation}
    \label{I_ft}
    I(t, x, \bm{\Omega}) = I( t - \tau, x - c \tau, \bm{\Omega}) + O(\sigma_s c \tau).
\end{equation}
Radiation with separate angular positions $\theta$ has different axial velocity. We set the reference length as $L=c \tau$, then at some position $x_1(x_1 < L)$, the distribution of radiation is $\rho_h/4\pi$ when $\cos \theta  > {x_1}/{L}$, and  $\rho_c/4\pi$ when $\cos  \theta  < {x_1}/{L}$.
Integrating over the solid angle space yields:
\begin{equation}
    \label{rho_ft}
    \begin{aligned}[b]
         \rho(x_1) &= 2 \pi \int_{0}^{\pi} I(x_1, \theta) \sin \theta d \theta\\
         &=  \frac{\rho_h}{2} \int_{0}^{\arccos \left ( \frac{x_1}{L} \right )}
         \sin \theta d \theta + \frac{\rho_c}{2} \int_{\arccos \left ( \frac{x_1}{L} \right )}^{ \pi } \sin \theta  d \theta \\
         &= \frac{1}{2} \left ( \rho_h - \rho_c \right ) \left ( 1 - \frac{x_1}{L}\right ) + \rho_c .
    \end{aligned} 
\end{equation}
Therefore, the radiation energy distribution is formally resemble a polyline, with its vertex advancing along with the propagation distance of the $\theta=0$ direction radiation. 

\begin{figure}[t]
    \centering
    \subfigure[]{\includegraphics[width=0.48\textwidth,trim={40 60 40 120},clip]{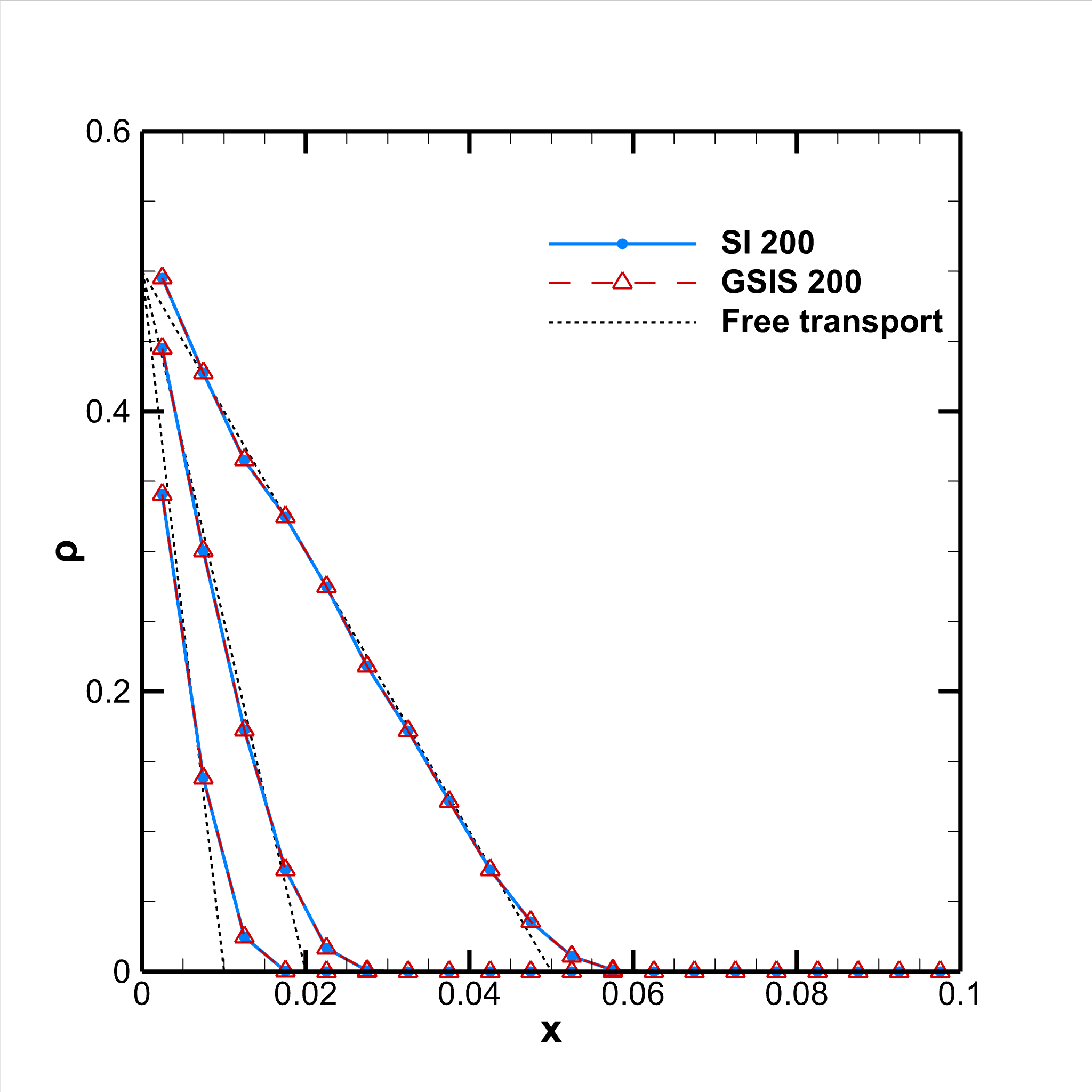}}
    \subfigure[]{\includegraphics[width=0.48\textwidth,trim={40 60 40 120},clip]{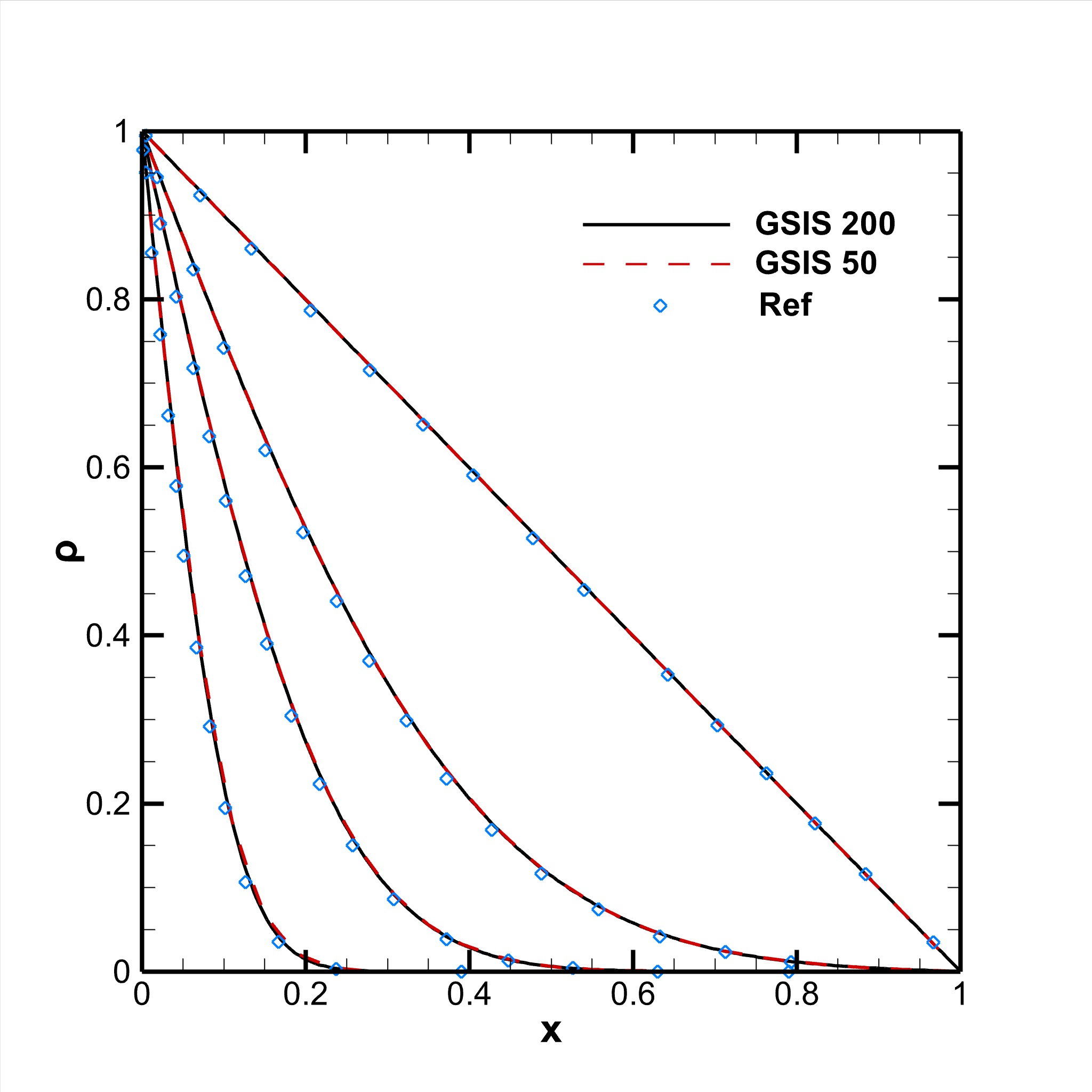}}
    \caption{(a) Radiation energy at $t=0.01$, $0.02$, and $0.05$ at the free-transport limit. The analytical solution \eqref{rho_ft} is indicated by black dashed lines.
    (b) Radiation energy distribution at $t=0.01$, $0.05$, $0.15$, and $2.0$ under the diffusion limit. The reference solution is indicated by diamond-shaped markers.
    }
    \label{fig:freetransport}
\end{figure}

The computational domain is discretized into a uniform grids with $N_x = 200$, and the time step $\varDelta t$ is set to $1\times 10^{-6}$. The radiation energy at $t=0.01$, $0.02$, and $0.05$ are depicted in Fig.~\ref{fig:freetransport}(a). The data points from SI and GSIS almost overlap, which validates that synthetic iteration can recover the numerical solution of the conventional discrete ordinates method under the free-transport limit. 

As indicated in Eq.~\eqref{I_ft}, the radiation intensity distribution has discontinuities at the propagation front. Due to the limited number of grids and the dissipative nature of upwind fluxes at discontinuities, some discrepancies may arise between the numerical and analytical solutions. However, the numerical solution still captures the overall evolutionary trend and the conservative property of the finite volume method. 

It should be emphasized that the selected time for this problem is relatively short, with the radiation energy only propagating across a few grid distances. As the number of grids traversed increases, if the number of discrete units in the solid angle space is not increased, the ray effect will become pronounced. Actually, the ray effect begins to emerge at $t=0.05$, manifesting as non-smoothness in the data point connections.

\subsubsection{Diffusion limit}

We choose $\epsilon=10^{-8}$ and $\sigma_s=10^8$, and the optical thickness of a single grid is much larger than 1, indicating a diffusive radiation transport regime. The evolution in the diffusion state is extremely slow. Due to the scaling parameter $\epsilon$, one unit of time corresponds to the time it takes for radiation to travel $10^8$ units of length. With $\varDelta t=10^{-6}$, the equivalent CFL number is on the order of $10^{4}$. The conventional iteration methods are no longer effective for this problem, and the macroscopic diffusion equation within the synthetic iteration method will govern the evolution of the numerical solution.

Calculations are performed using uniform grids with $N_x=50$ and 200. The radiation energy distributions at $t=0.01$, $0.05$, $0.15$, and $2.0$ are shown in Fig.~\ref{fig:freetransport}(b). Both results agree well with the reference solution from UGKS \cite{UGKS_SUN2015}, indicating that the synthetic iteration method can accurately recover the correct time evolution relationship under the diffusion limit. Moreover, except for the initial short period, the synthetic iteration  converges within the minimum number of iterations, i.e., 3 sweeps of the kinetic equation. It can be inferred that during the initial stage of the computation, more iterations are required, which is related to the left boundary layer grids. Once their values approach 1, the overall evolution of the solution is almost entirely governed by the diffusion equation. The computational times for GSIS with 50 and 200 grids are 85.5 minutes and 373 minutes, respectively. 

\subsection{Marshak Waves}

The Marshak wave problem is a classic test case in radiation transport simulations and has been extensively studied~\cite{LARSEN201382,UGKS_SUN2015,FEM-UGKS_XU2020,IUGKWP_photon,AP-IMEX}. In this problem, a one-dimensional radiation wave is driven by a constant and isotropy source at the left boundary. The strong radiation-matter coupling, opacity variations, non-linear and time-dependent behavior induce a certain degree of challenges for accurate and efficient numerical simulation. In the following two cases, the solid angle is discretized by $N_{\theta} \times N_{\varphi} = 16 \times 24$.

\begin{figure}[t!]
    \centering
    \subfigure[]{\includegraphics[width=0.45\textwidth,trim={20 80 20 200},clip]{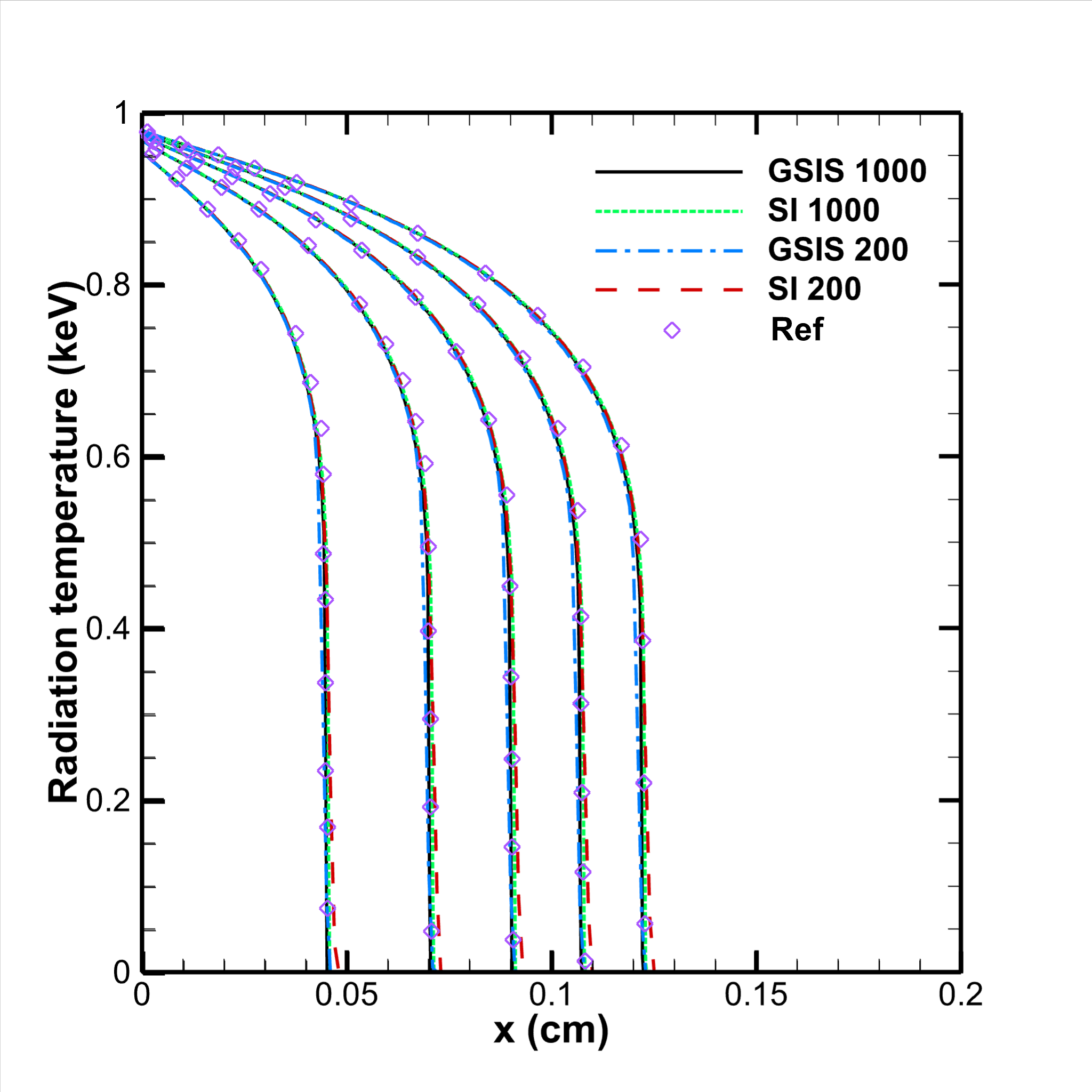}}
    \subfigure[]{\includegraphics[width=0.45\textwidth,trim={20 80 20 200},clip]{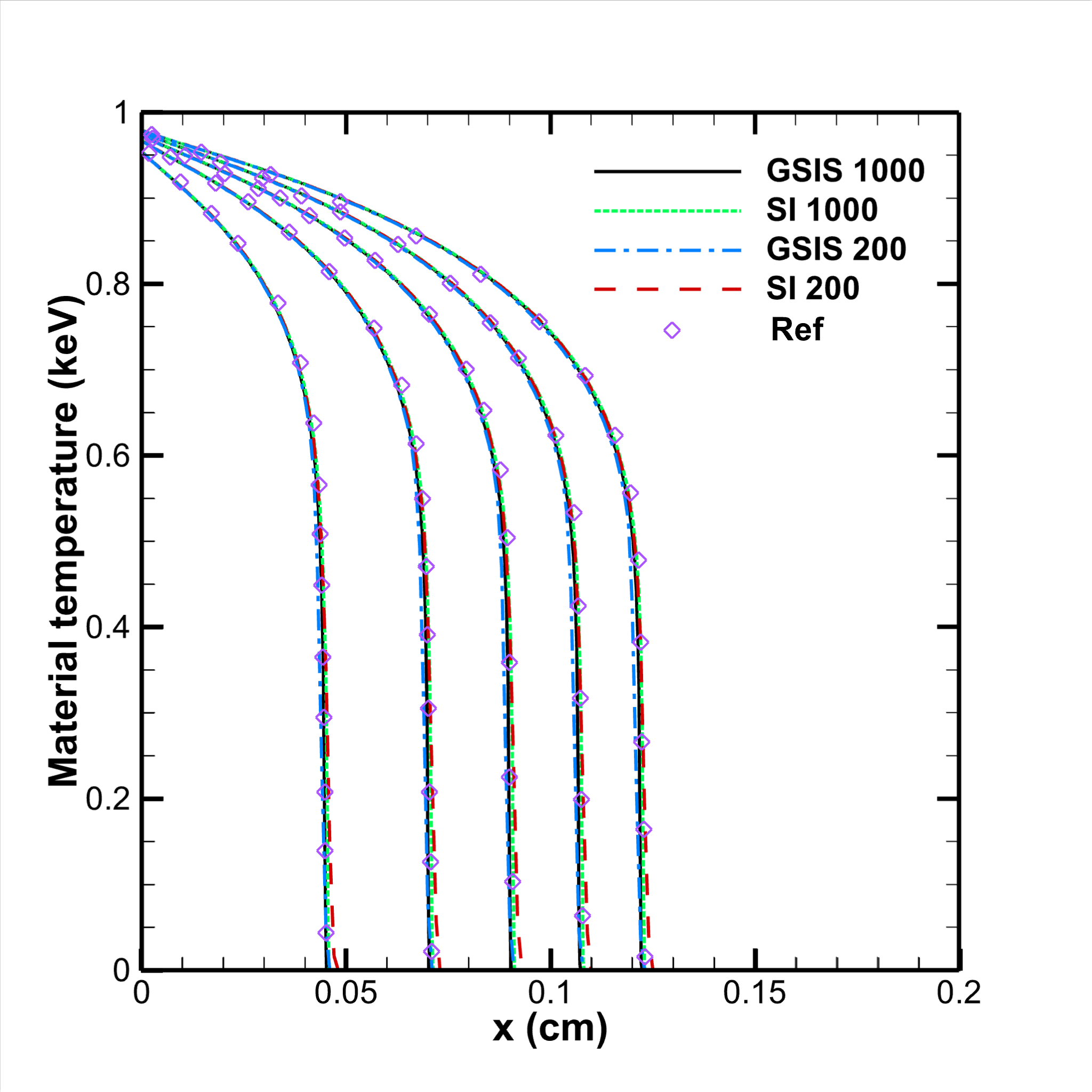}}\\
    \subfigure[]{\includegraphics[width=0.45\textwidth,trim={20 80 20 200},clip]{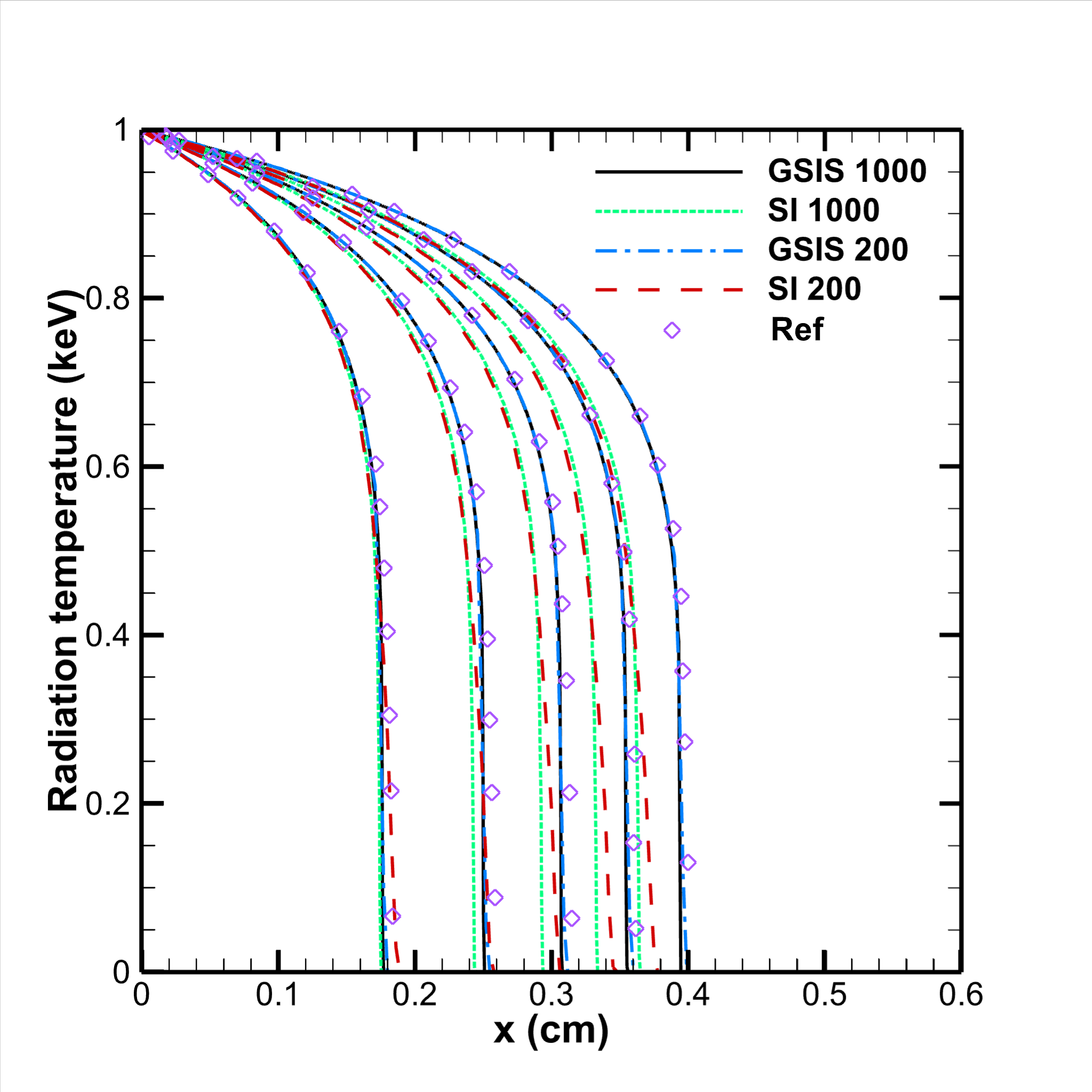}}
    \subfigure[]{\includegraphics[width=0.45\textwidth,trim={20 80 20 200},clip]{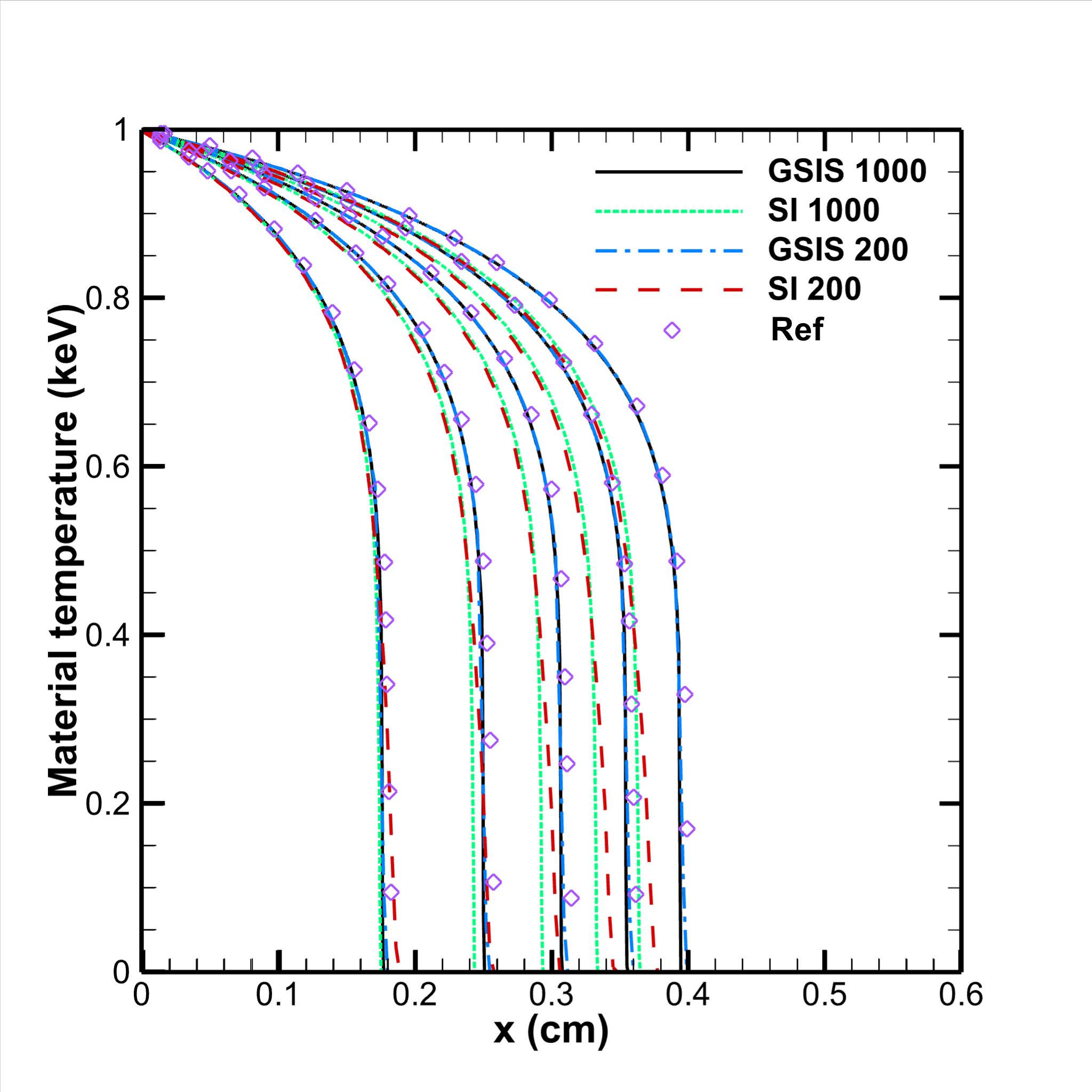}}
    \caption{(a, b) Marshak wave-2A problem. Lines from left to right stand for $t = 0.2, 0.4, 0.6, 0.8, 1$~ns. 
    (c, d) Marshak wave-2B problem. Lines from left to right correspond to $t=15,30,45,60,74$~ns. Reference solution obtained from the IUGKWP method \cite{IUGKWP_photon} is used for validation.
    }
    \label{fig:Mrashak2A}
\end{figure}

\subsubsection{Marshak Wave-2A}

The Marshak wave-2A problem represents an optically thin case. The absorption coefficient is taken to be $\sigma_a = 30/T^3 \text{cm}^{-1}$, where $T$ is the material temperature measured in $\text{keV}$. The temperature of radiation source at the left boundary is $1~\text{keV}$. Initially, both the radiation and material (the specific heat is $0.3 ~\text{GJ}/(\text{keV}\cdot\text{cm}^3$) are in equilibrium at temperature of $10^{-6}~\text{keV}$. 
The calculations are preformed in the domain $x \in [0,0.2]~\text{cm}$, which is uniformly discretized by $N_x=200$ and $1000$ grids. The time step is fixed at $\varDelta t = 3 \times 10 ^{-4}$~ns. 

The radiation and material temperatures are depicted in Fig.~\ref{fig:Mrashak2A}(a, b). 
Both SI and GSIS achieve a good approximation of the traveling wave. Compared to SI, a slight difference emerges at the wave front, where radiation-matter interaction is intense. This discrepancy arises due to the modifications to the fluxes in the GSIS procedure. However, it does not impact the overall behavior, as validated by the reference solution. The computational times from from $t = 0$ to $t = 1$ ns for SI 200, GSIS 200, SI 1000, and GSIS 1000 are $5.84$, $5.35$, $28.47$, and $25.23$ minutes, respectively. GSIS demonstrates a slight advantage, primarily attributed to the synchronous update of material temperature during macroscopic iterations.

\subsubsection{Marshak Wave-2B}

In this problem, the absorption coefficient is increased to $\sigma_a = 300/T^3 \text{cm}^{-1}$. The computation domain is $0.6~\text{cm}$ in length, which is discretized by $N_x=200$ and $1000$ grids, respectively. The time step is $\varDelta t = 1 \times 10 ^{-3}$~ns. 

The distributions of radiation and material temperatures at $t=15 ,30, 45, 60, 74$~ns are plotted in Fig.~\ref{fig:Mrashak2A}(c, d). 
A significant difference is observed between the GSIS and SI solutions. The SI solutions exhibit a lag relative to the GSIS solutions during propagation. This discrepancy arises because SI converges slowly in the optically thick regime, leading to potential false convergence at each time step. As a result, the accumulated time-marching increments produce incorrect long-time behaviors. Worse still, at the wave front, SI 200 is notably more dissipative than SI 1000. In contrast, by applying the synthetic equation, the correct propagation speed is restored in GSIS, and the dissipation at the wave front is effectively controlled; the GSIS solution nearly overlap the multiscale solution from the IUGKWP method \cite{IUGKWP_photon} The computational times for SI 200, GSIS 200, SI 1000, and GSIS 1000 are 160, 96.3, 780, and 515 minutes, respectively.

\subsection{Tophat test}\label{Tophat_num}

\begin{figure}[t]
    \centering
    \subfigure[Geometry]{\includegraphics[width=0.4\textwidth,trim={40 -200 40 140},clip]{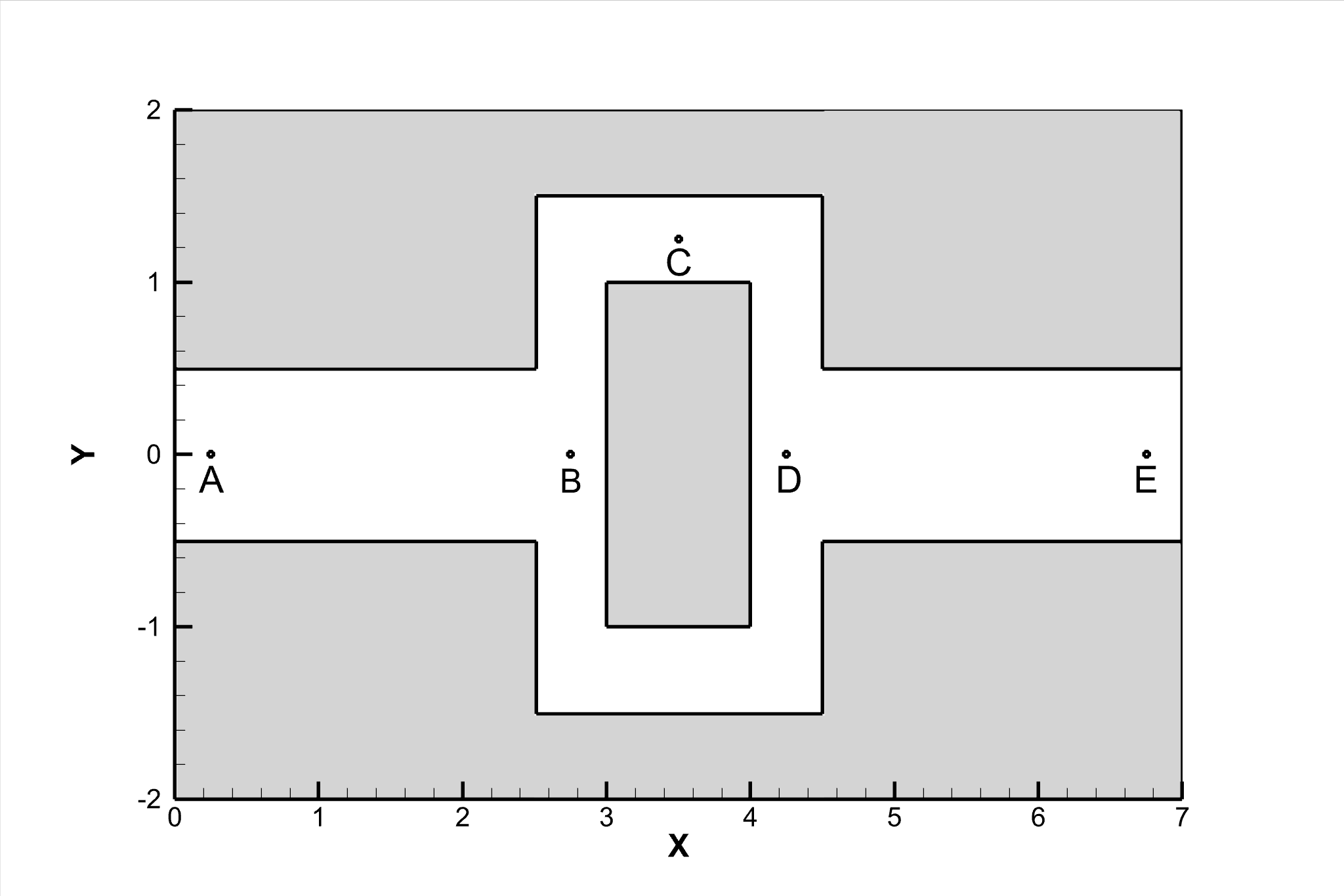}}
    \subfigure[Non-uniform mesh refined near material interface]{\includegraphics[width=0.58\textwidth,trim={40 40 40 140},clip]{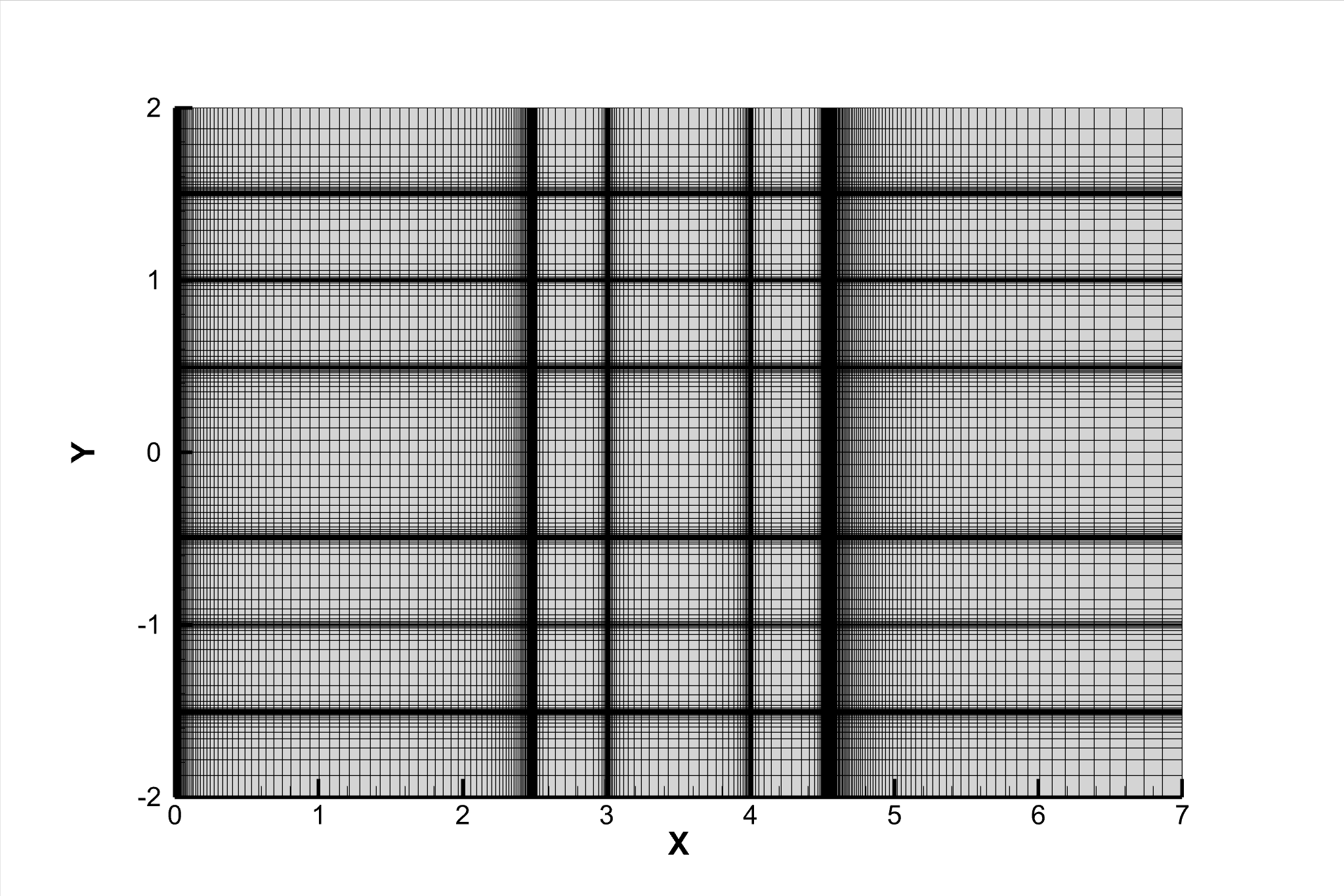}}
    \caption{(a) In the Tophat problem, the geometry consists of gray and white regions filled with optically thick and thin materials, respectively. The computational domain is discretized using two types of meshes: one with a uniform grid of $N_x \times N_y = 256 \times 128$ uniform meshes, and another with a non-uniform grid of $351\times201$ non-uniform meshes, where refined meshes are located near the interface between optically thick and thin materials, see Fig.~\ref{fig:Tophat_mesh}(b). To be specific, the whole computational domain is divided into $5\times 7$ rectangular blocks; for each side length,  there are 50 non-uniform grid points per centimeter. The commercial meshing software Pointwise is used, by setting the height of first layer at the interface to be  $10^{-4}$~cm, which is smaller than the mean free path of photons in the optically thick material ($5\times10^{-4}$~cm). Then, the built-in Spacing Constraints feature is adopted to generate the non-uniform Cartesian mesh.
    }
    \label{fig:Tophat_mesh}
\end{figure}

As shown in Fig.~\ref{fig:Tophat_mesh}(a), the computational domain is a rectangular region located in $x\times y=[0,7]~\text{cm} \times [-2,2]~\text{cm}$. A dense, optical thick material with $\sigma_a = 2000~\text{cm}^{-1}$ and $C_v = 1~\text{GJ}/(\text{keV} \cdot \text{cm}^3)$ occupies the gray regions: $[3,4]~\text{cm} \times [-1,1]~\text{cm}$, $[0,2.5]~\text{cm} \times [-2,-0.5]~\text{cm}$, $[0,2.5]~\text{cm} \times [0.5,2]~\text{cm}$, $[4.5,7]~\text{cm} \times [-2,-0.5]~\text{cm}$, $[4.5,7]~\text{cm} \times [0.5,2]~\text{cm}$, $[2.5,4.5]~\text{cm} \times [-2,-1.5]~\text{cm}$, and $[2.5,4.5]~\text{cm} \times [1.5,2]~\text{cm}$. The white zone is filled with optical thin material with $\sigma_a = 0.2~\text{cm}^{-1}$ and $C_v = 0.001~\text{GJ}/(\text{keV} \cdot \text{cm}^3)$. Initially, radiation and material are in equilibrium at a temperature of $0.05~\text{keV}$. An isotropy heating source with a temperature $0.5~\text{keV}$ is applied at the left boundary with $-0.5~\text{cm} < y < 0.5~\text{cm}$. The rest boundaries are vacuum. 

Two types of spatial grids are used. The first is a uniform mesh with $N_x \times N_y = 256 \times 128$, and the second is a $351\times201$ non-uniform mesh, as shown in Fig.~\ref{fig:Tophat_mesh}(b). 
In both meshes, a growing time step is set as $\varDelta t = 10^{-3}~\text{ns}+3\times10^{-5}t$, and the simulation is carried on until $t=1000$ ns, which means that about 100,000 time steps are simulated. The solid angle is discretized by Gauss-Legendre quadrature with $N_{\theta} \times N_{\varphi} = 32 \times 32$. 

\begin{figure}[p]
    \centering
    {\includegraphics[width=0.45\textwidth,trim={60 60 40 120},clip]{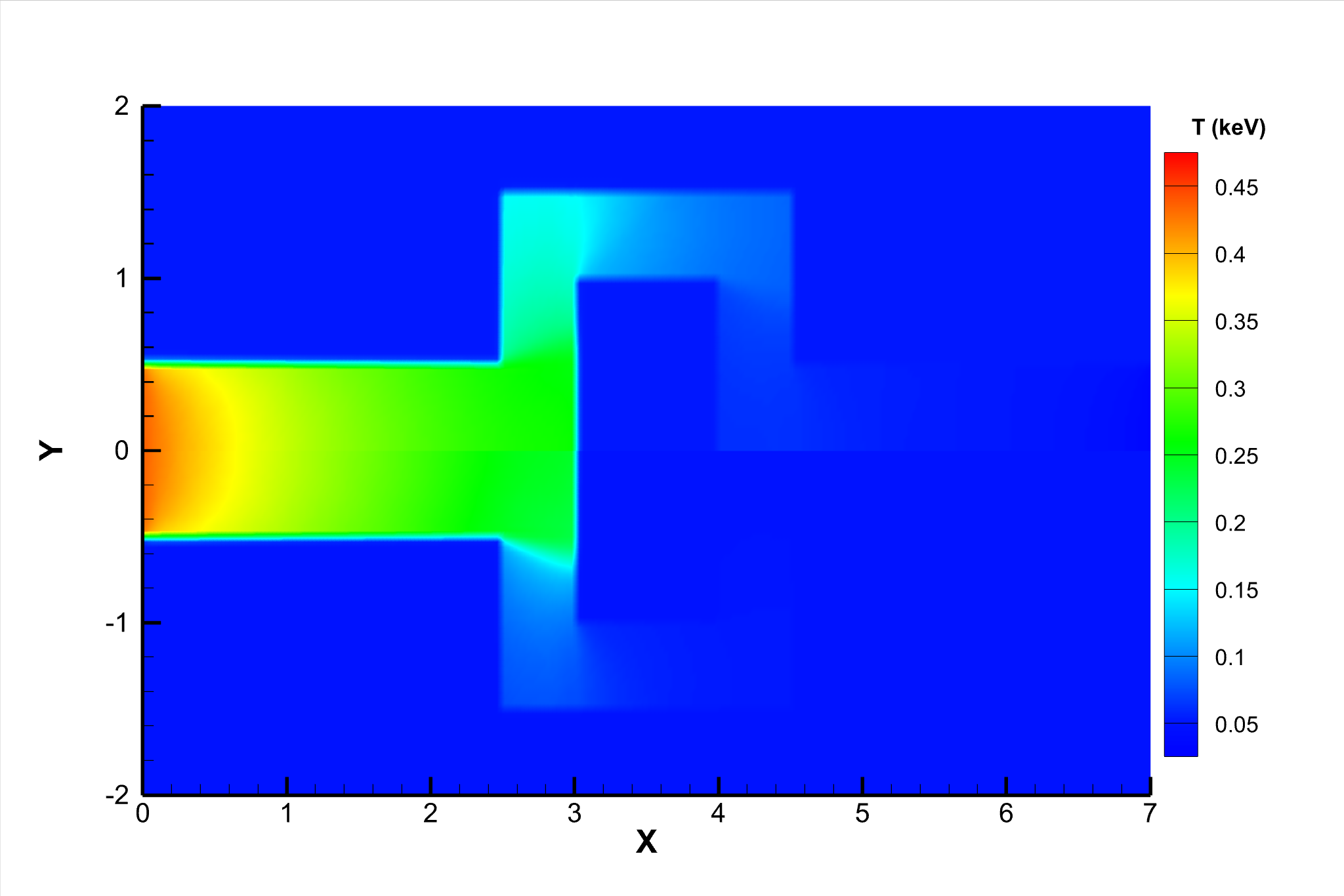}}
   {\includegraphics[width=0.45\textwidth,trim={60 60 40 120},clip]{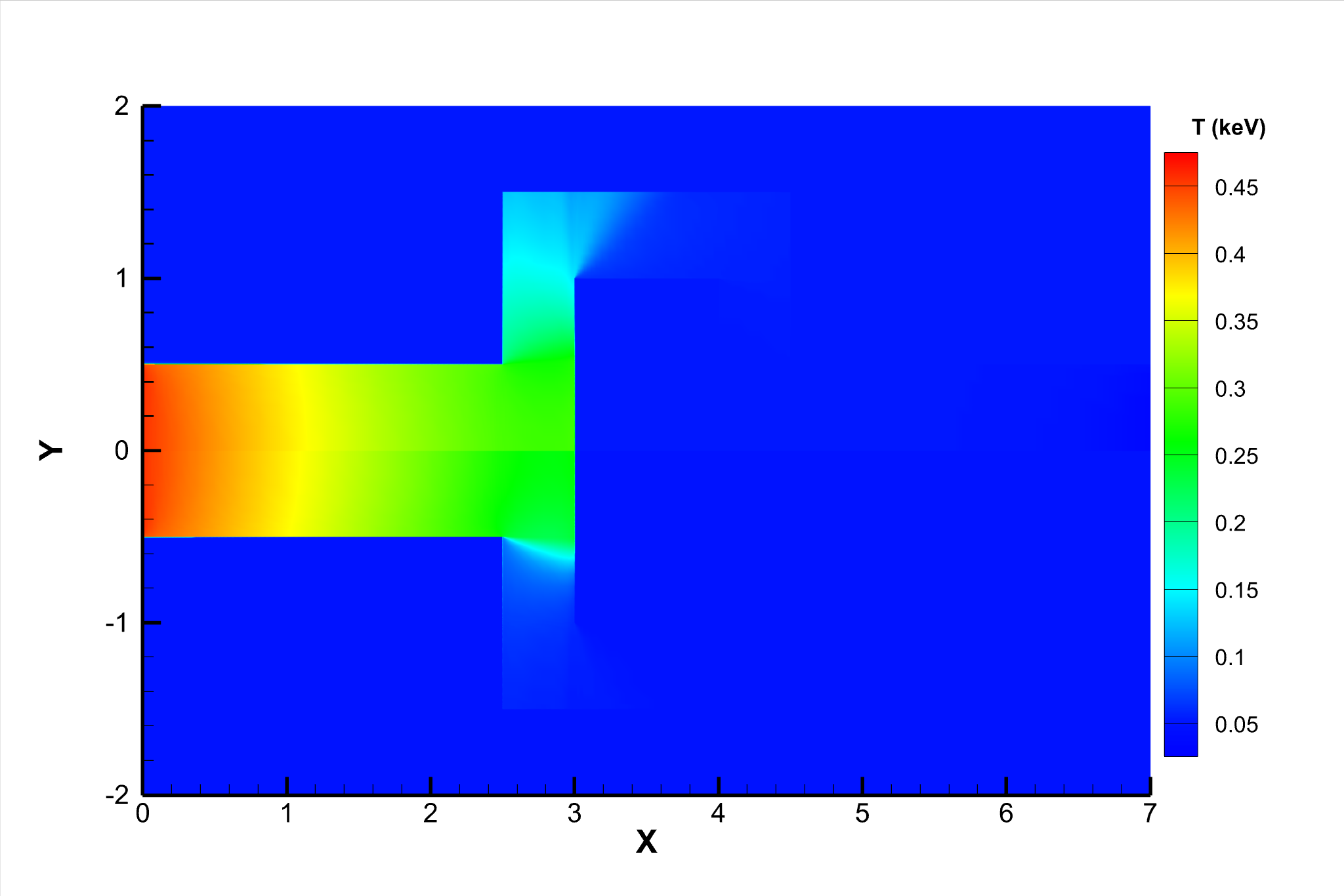}}\\
    {\includegraphics[width=0.45\textwidth,trim={60 60 40 120},clip]{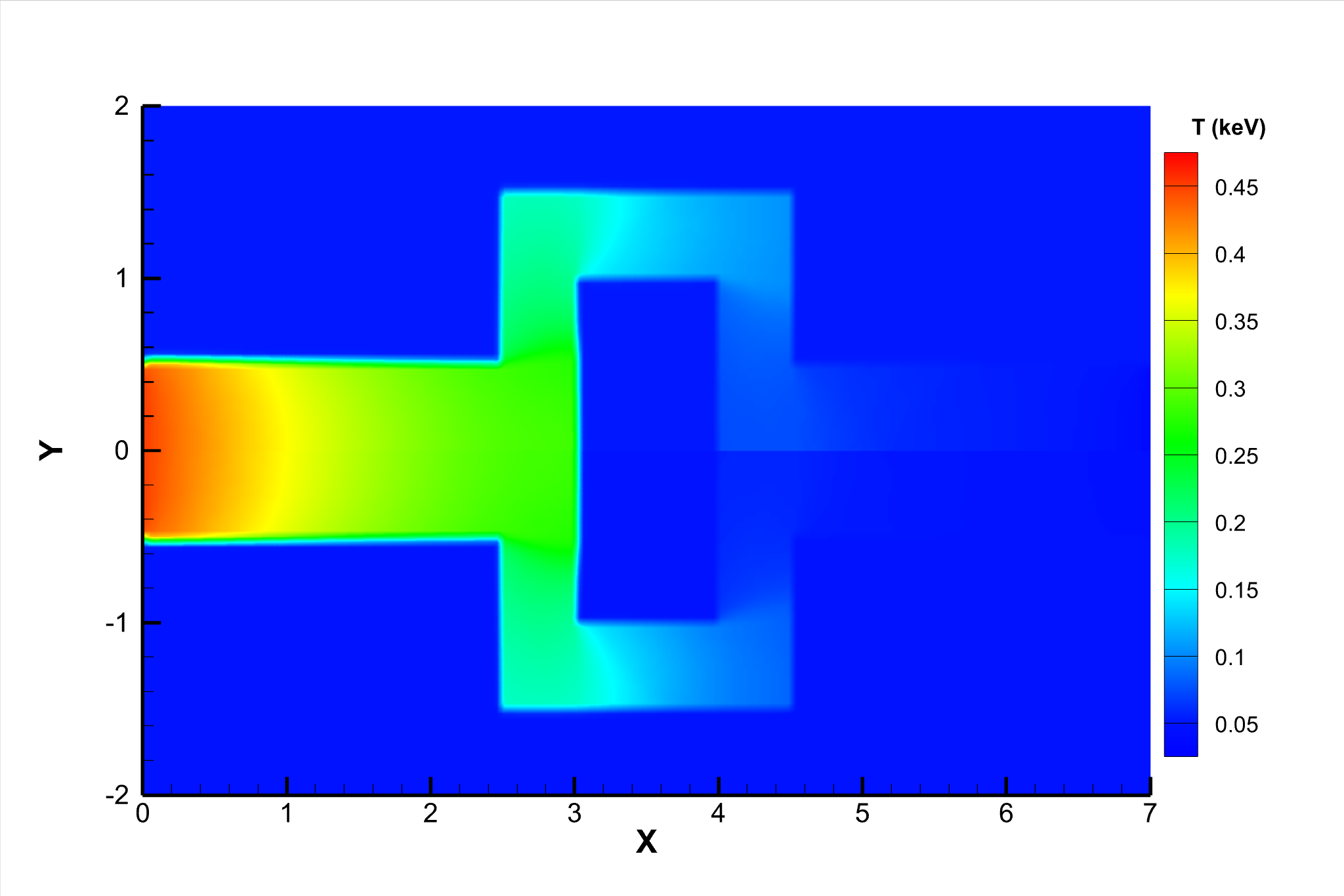}}
   {\includegraphics[width=0.45\textwidth,trim={60 60 40 120},clip]{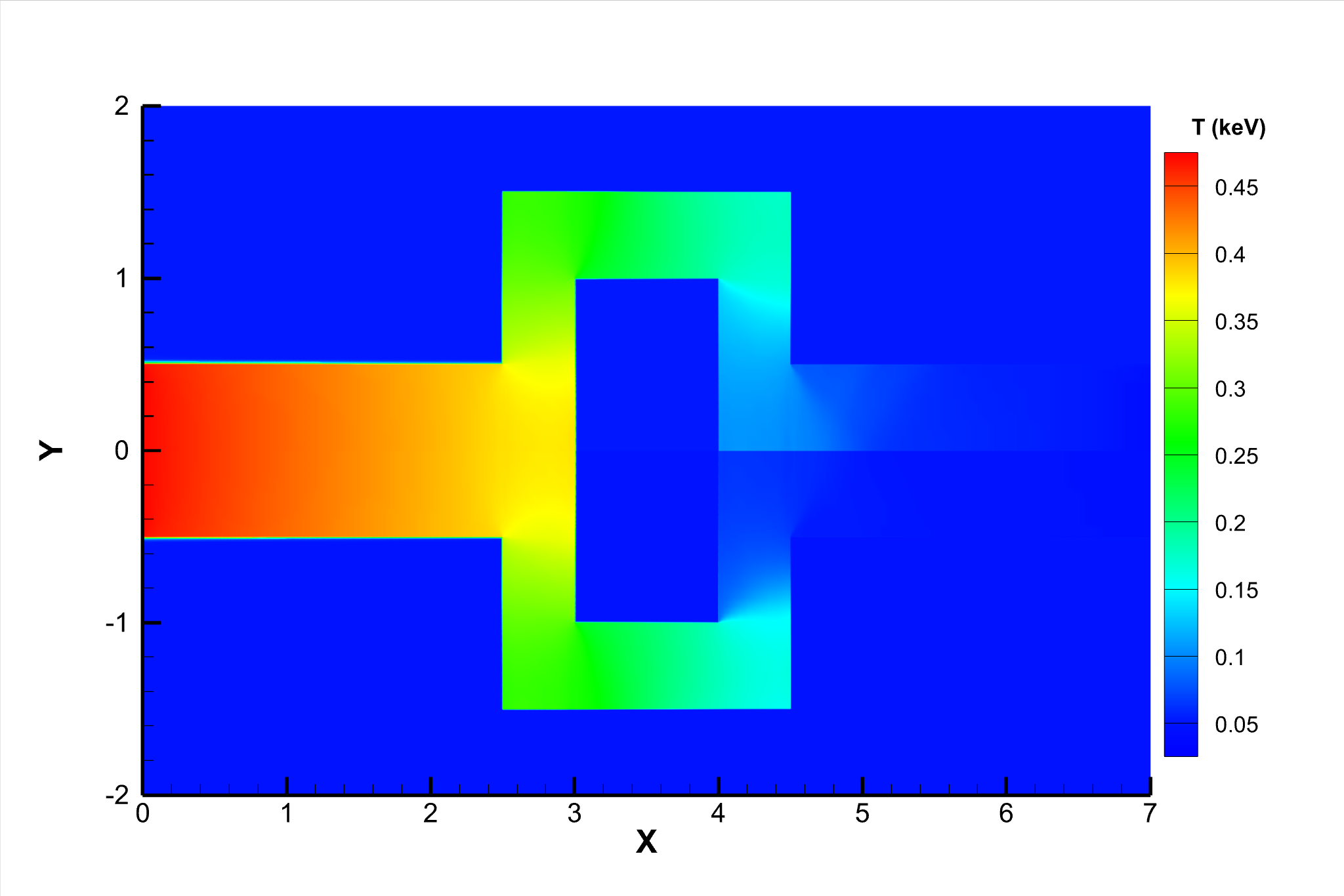}}\\
    {\includegraphics[width=0.45\textwidth,trim={60 60 40 120},clip]{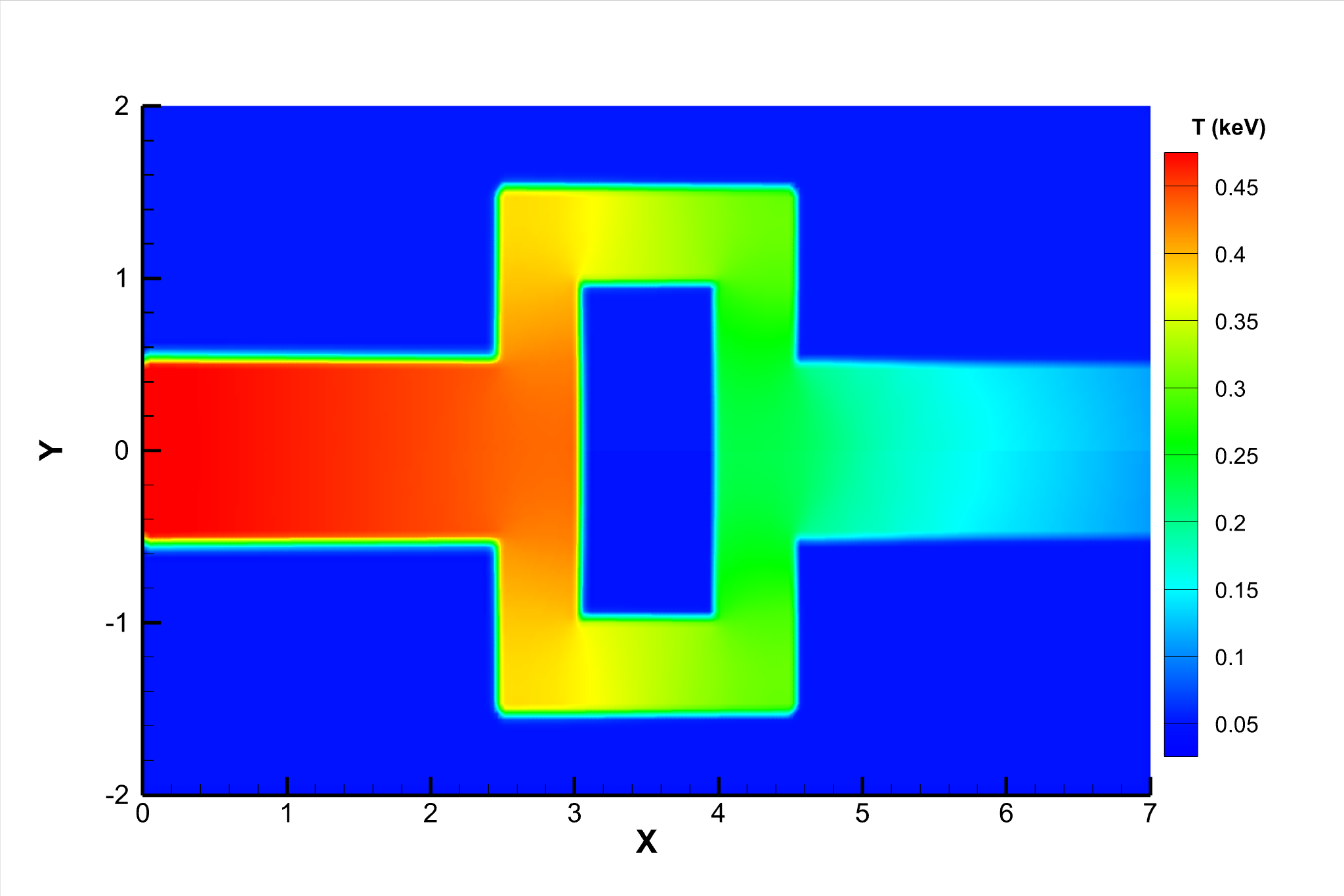}}
   {\includegraphics[width=0.45\textwidth,trim={60 60 40 120},clip]{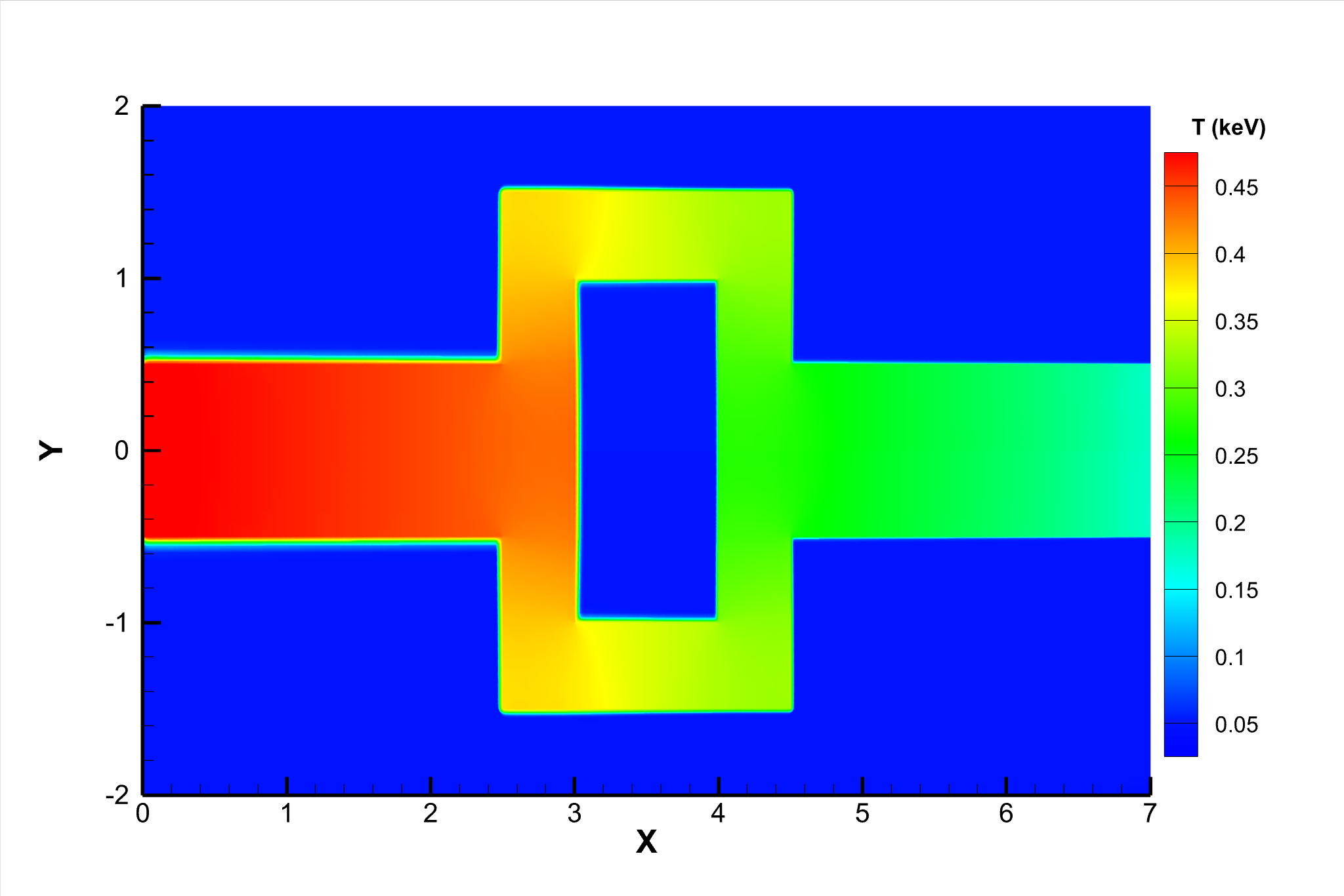}}\\
    \subfigure[Uniform mesh]{\includegraphics[width=0.45\textwidth,trim={60 60 40 120},clip]{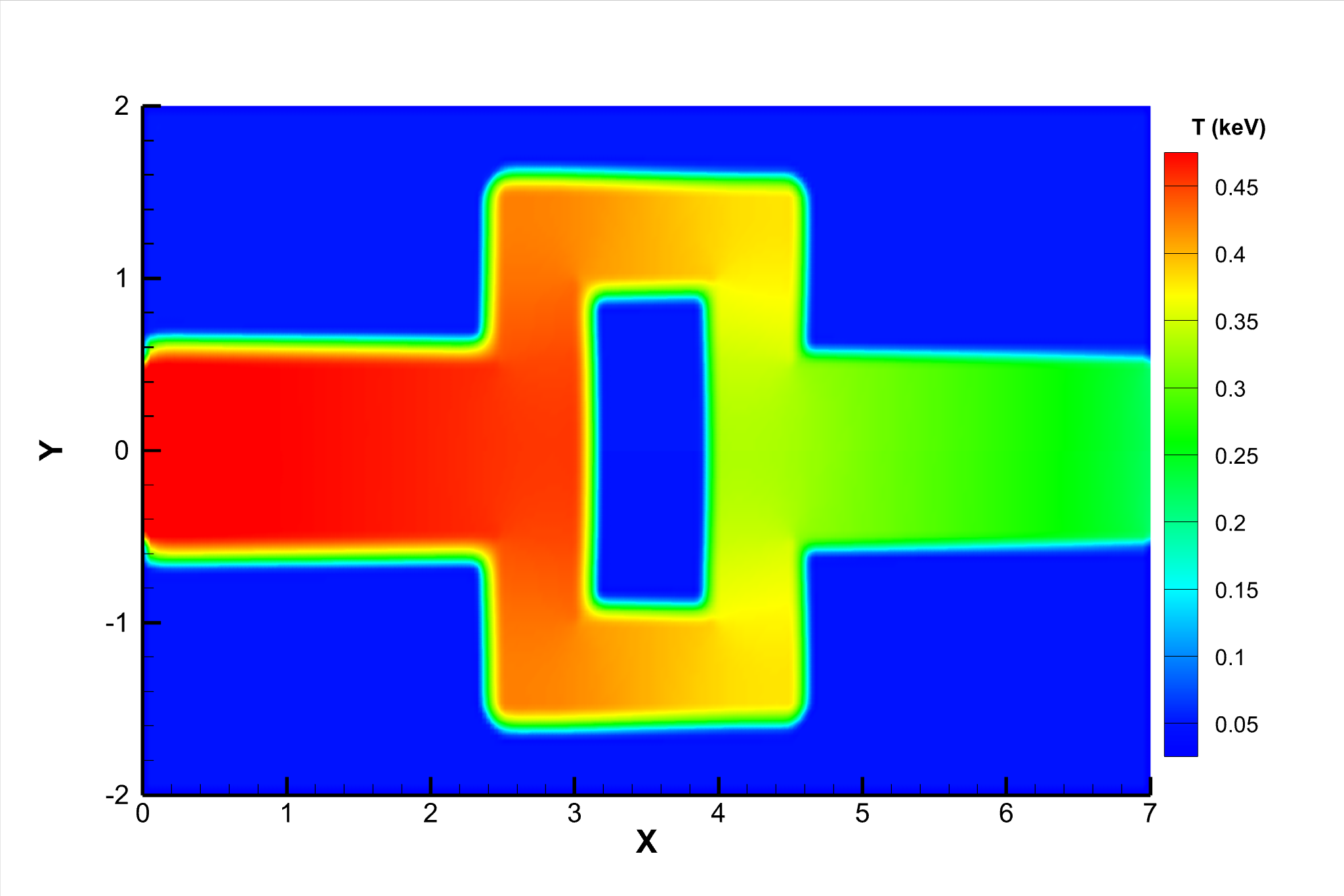}}
   \subfigure[Non-uniform mesh refined in the interface]{\includegraphics[width=0.45\textwidth,trim={60 60 40 120},clip]{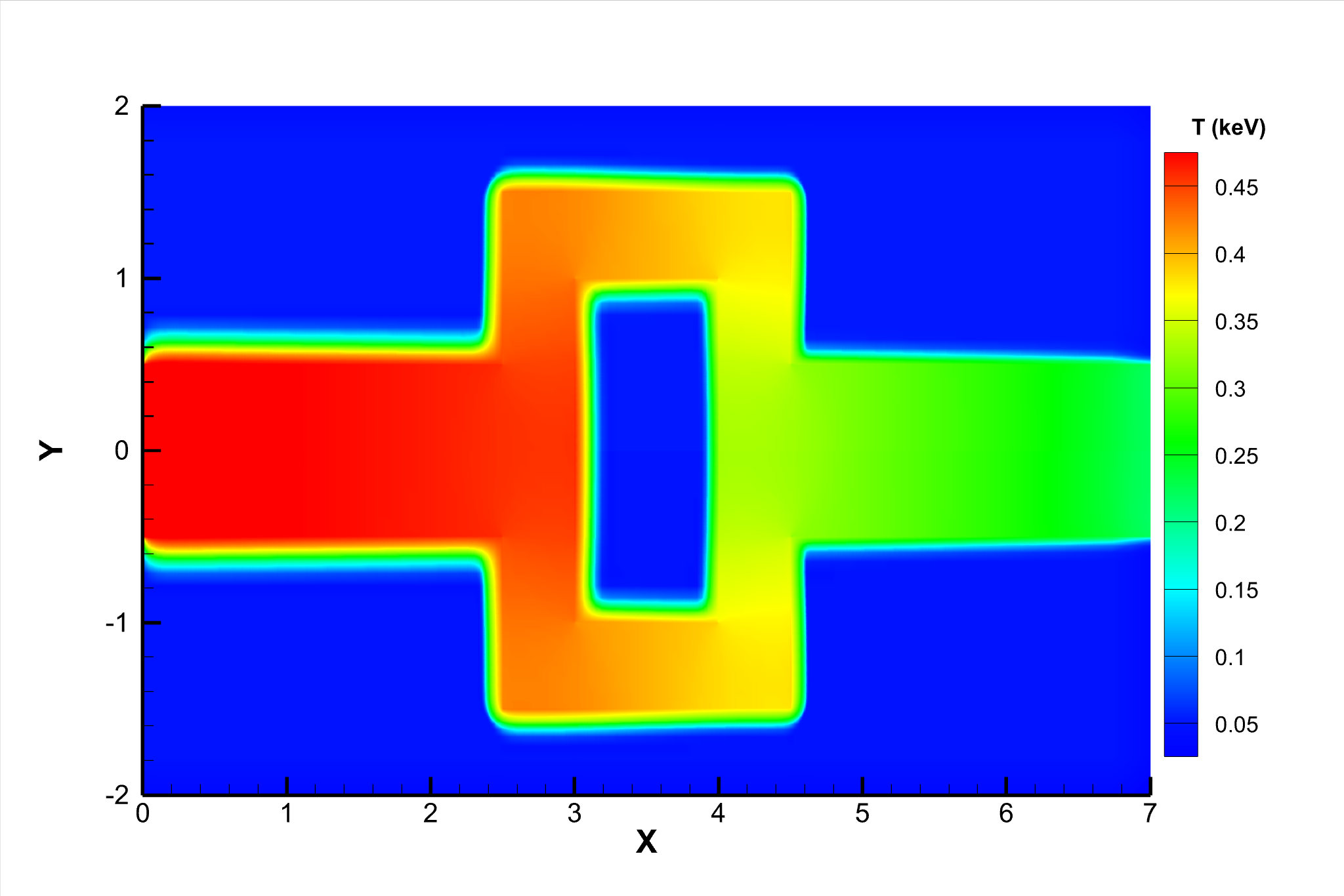}}
    \caption{
    Temperature contours in the Tophat problem. From the top to bottom, the time is $t=1, 8, 94$, and 1000~ns, respectively. In each subfigure, the top and bottom half regions show the radiation and material temperatures, respectively.
    }
    \label{fig:Tophat_color_maps}
\end{figure}



\begin{figure}[p]
    \centering
\subfigure[at point A: $ (x,y) = (0.25, 0)~\text{cm}$]{\includegraphics[width=0.4\textwidth,trim={40 60 40 160},clip]{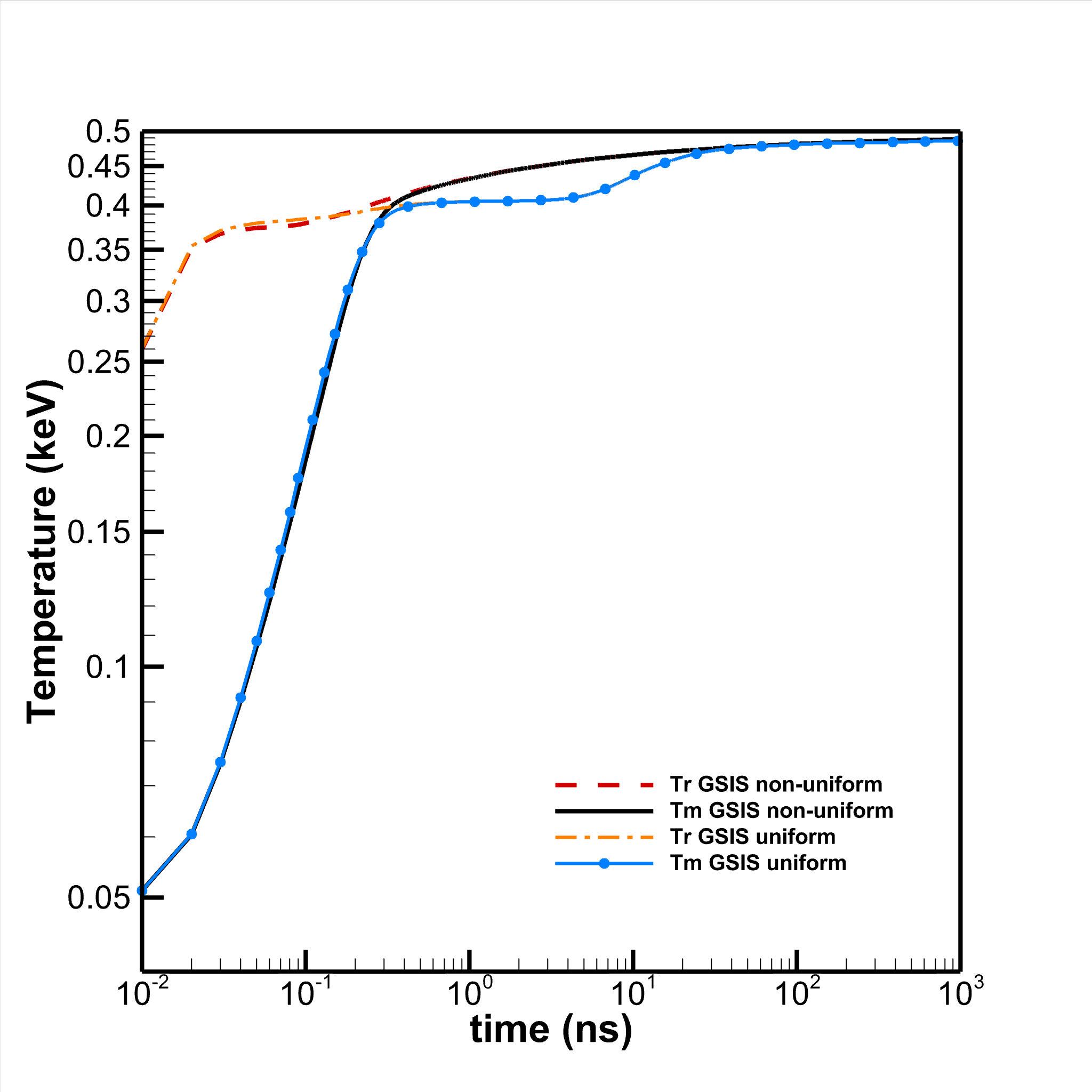}}
\subfigure[at point B: $ (x,y) = (2.75, 0)~\text{cm}$]{\includegraphics[width=0.4\textwidth,trim={40 60 40 160},clip]{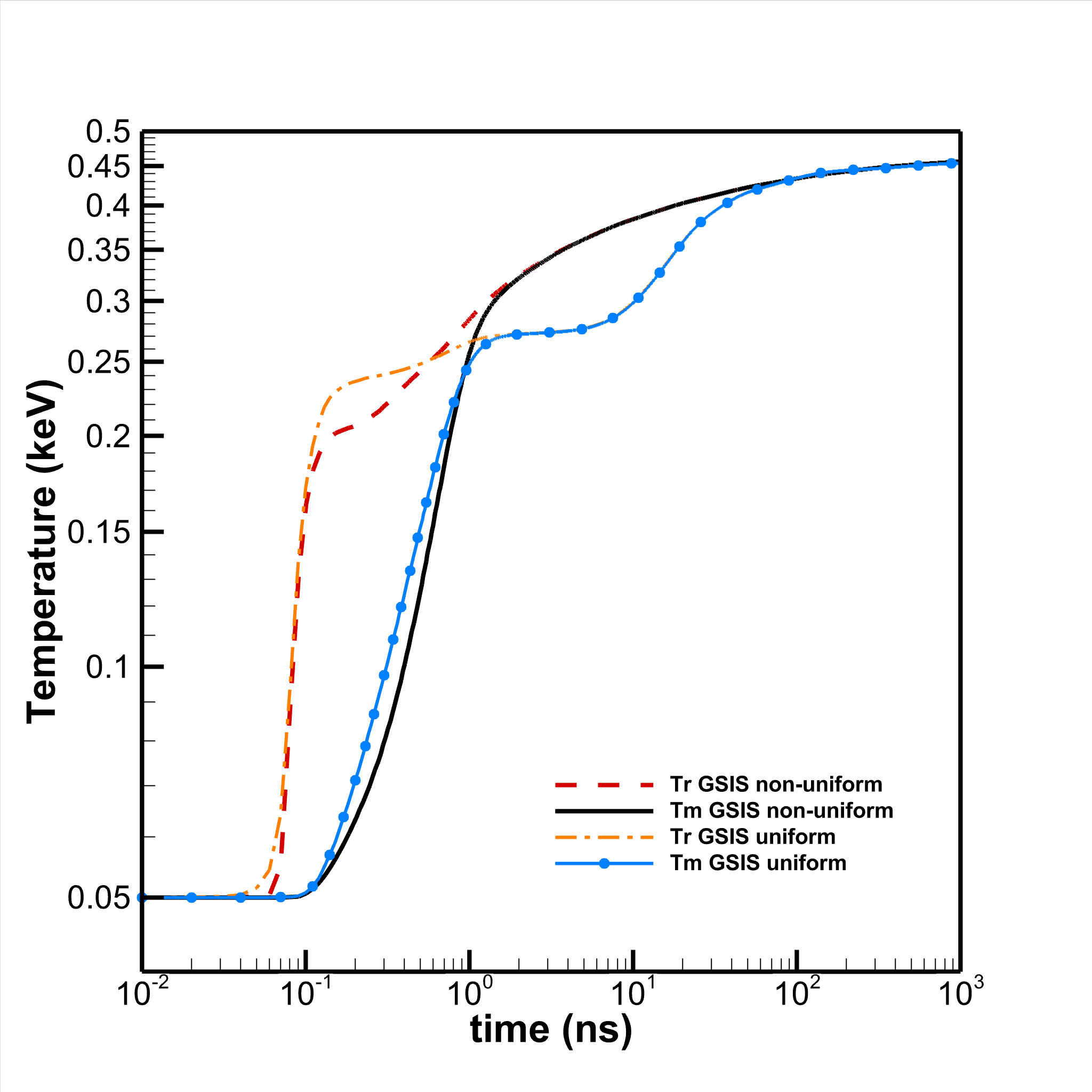}}\\
\subfigure[at point C: $ (x,y) = (3.5, 1.25)~\text{cm}$]{\includegraphics[width=0.4\textwidth,trim={40 60 40 160},clip]{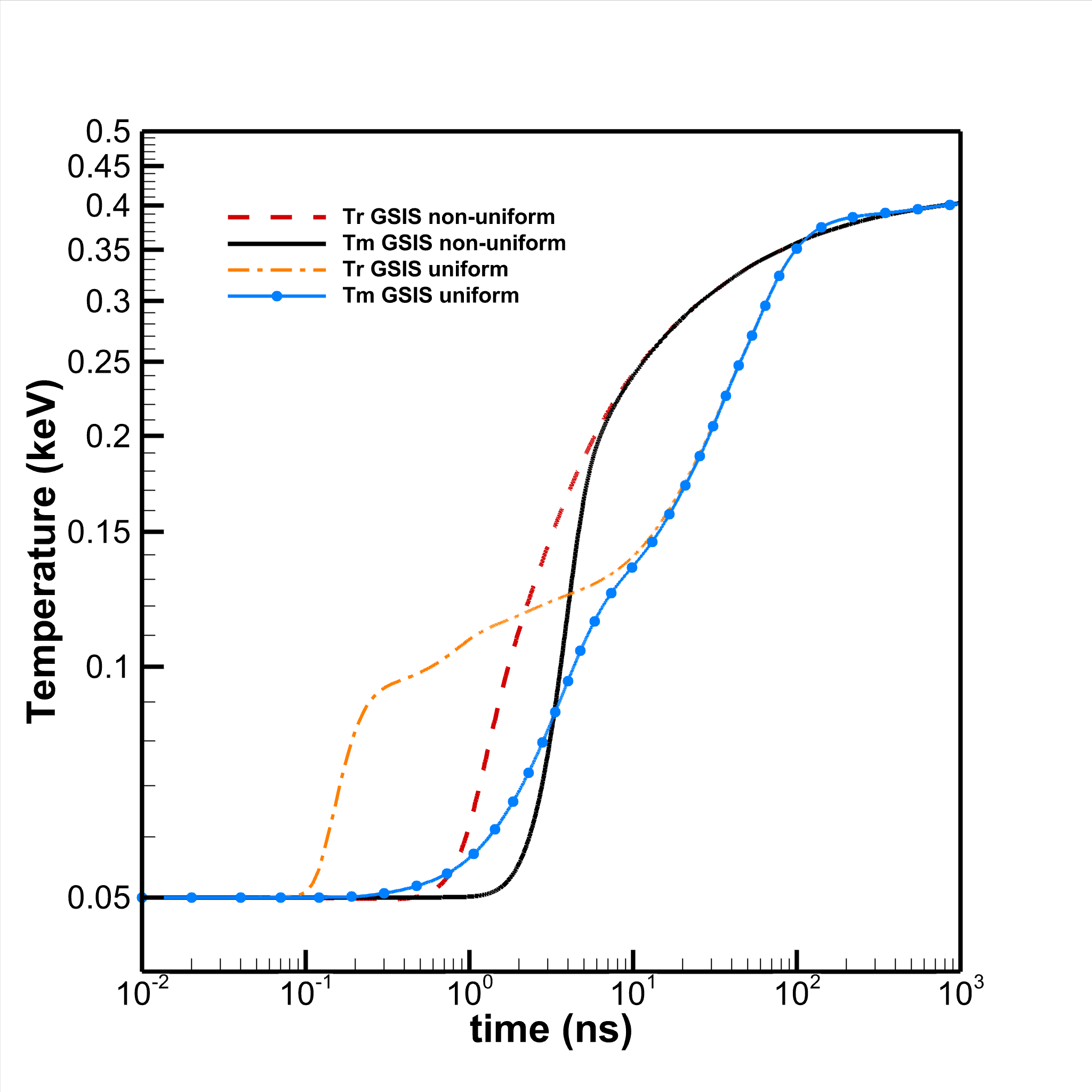}}
\subfigure[at point D: $ (x,y) = (4.25, 0)~\text{cm}$]{\includegraphics[width=0.4\textwidth,trim={40 60 40 160},clip]{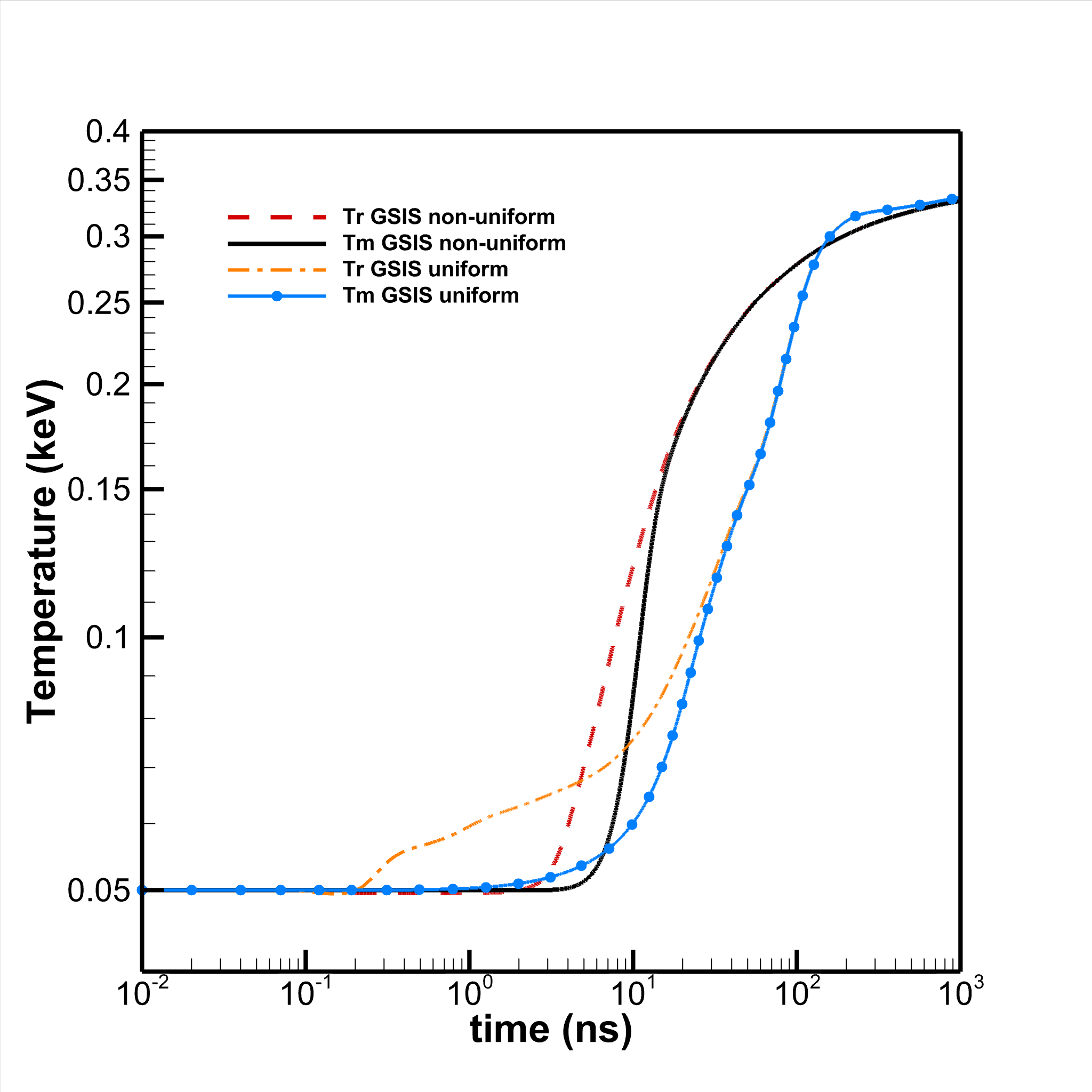}}\\
\subfigure[at point E: $ (x,y) = (6.75, 0)~\text{cm}$]{\includegraphics[width=0.4\textwidth,trim={40 60 40 160},clip]{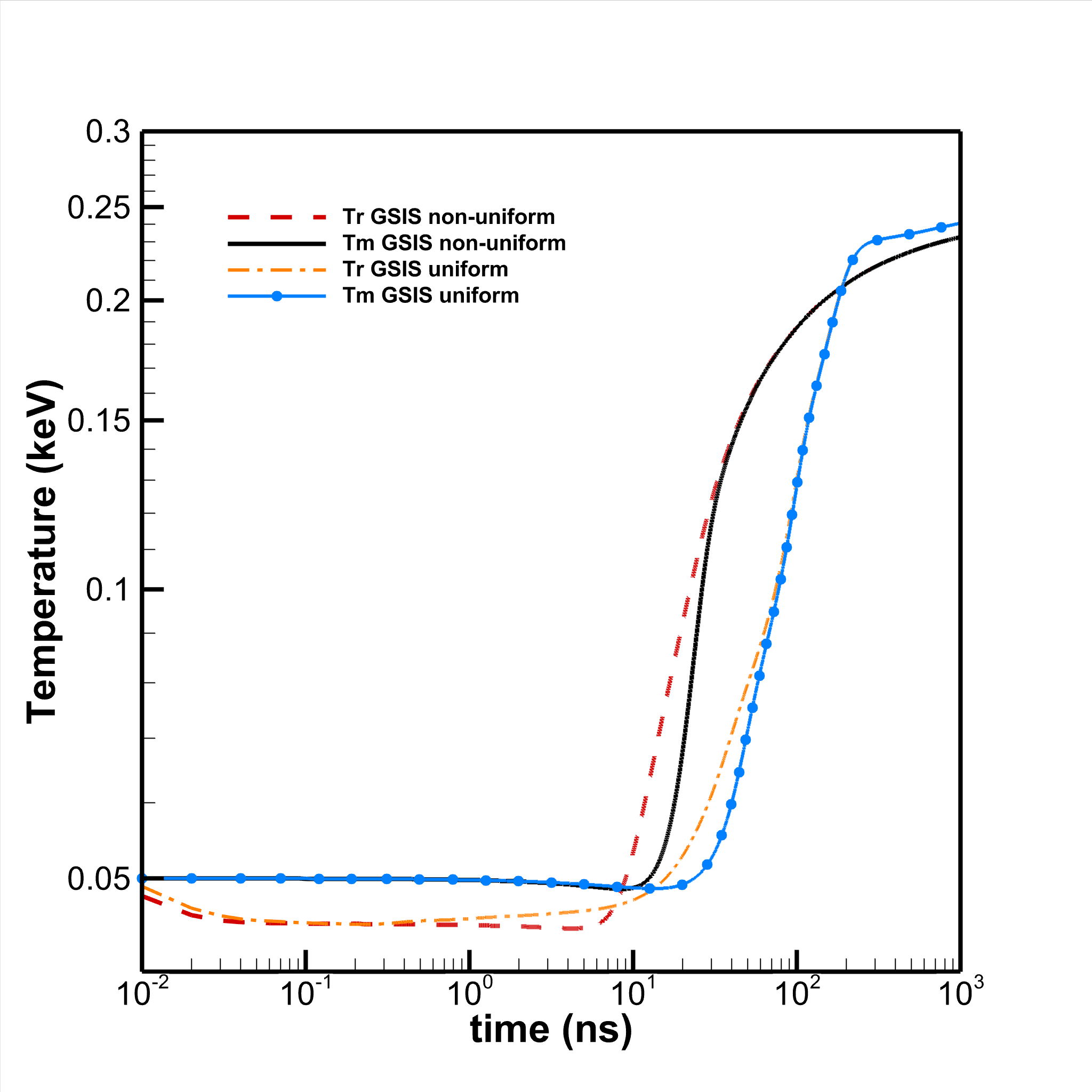}}
    \caption{The evolution of material and radiation temperatures at five probe points. 
    }
    \label{fig:Tophat}
\end{figure}

When a refined mesh is employed, the synthetic equation requires a substantial number of iterations to achieve convergence, although this iteration count can be significantly reduced by applying the multigrid method to the diffusion equation. Here, only the successive over-relaxation method is used, and the sub-inner iteration number in Fig.~\ref{fig:flowrate} is set to $s=1000$, or the process is terminated when Err in Eq.~\eqref{Err_W} falls below $10^{-7}$. 
Parallel computing utilizing 20 cores (AMD EPYC 7763 processor, 2.45GHz) is employed, see similar implementation for rarefied gas flows~\cite{zhang2024efficient}. The wall-clock time is 40.5 hours for the uniform mesh and 153 hours for the non-uniform mesh. The computational cost is high because our goal is to provide benchmark solutions by using dense mesh and tiny time step. 

Figure~\ref{fig:Tophat_color_maps} presents the material and radiation temperatures at 
$t =1, 8, 94$ and 1000~ns. At early times, radiation transport in the optically thin region (where probe points A and B are located) is dominated by free-streaming, resulting in a rapid rise in radiation temperature, followed by a gradual increase in material temperature. At the material interfaces, radiation is initially absorbed. Emission begins once the shallow surface of the optically thick material becomes sufficiently heated; subsequently, energy is transferred deeper into the optically thick region via diffusion. As a result, the optically thin region where probe points D and E are located is indirectly heated.
In the left column of Fig.~\ref{fig:Tophat_color_maps}, the calculation using the uniform mesh exhibits the basic behaviors, which can also be observed in the literature~\cite{GENTILE2001543,UGKS_SUN2015,AP-IMEX,FEM-UGKS_XU2020,IUGKWP_photon}. 
However, distinct features are observed in the right column of the figure, where a non-uniform mesh refined near the material interface is used. At $t=1$ ns, the energy is more concentrated near the entrance, whereas in the uniform mesh it spreads more widely to the location of point C. At $t=1$, 8, and 94 ns, the heated layer at the material interface is captured with greater sharpness and clarity using the refined mesh. By $t=94$ ns, more energy has been transferred to the exit region, at probe points D and E. The physics of this problem suggest that the differences in temporal evolution are closely tied to the heated layer within the optically thick material during the initial 100 ns. By $t=1000$ ns, both simulations converge to the same final state.

We further compare in Fig.~\ref{fig:Tophat} the time evolution of material and radiation temperatures at five probe points, between the two spatial discretizations. At the beginning, radiation directly propagates from the source at the entrance to the probe points A and B, leading a rapid increase in radiation temperature and a significant non-equilibrium with material temperature. In this free-transport of photons, both grid resolutions are adequate; therefore, there is no visible difference between the two results. However, a noticeable difference emerges in the range  $1\lesssim{}t\lesssim 10$ ns, where the coarse uniform mesh predicts lower temperatures compared to the refined non-uniform mesh. This is because, near the material interface, the coarse mesh contains more regions of optically thick material that are heated compared to the refined mesh, resulting in greater energy absorption. Consequently, the temperature rise in A and B is slower. As time progresses, the continuous heat from the left source increases and causes the temperature in regions C and D to saturate, so that after approximately 50 ns, both the coarse and refined meshes reach the same temperature. For probe points C and D, the coarse mesh initially absorbs more energy and subsequently emits more, leading to a faster temperature rise at these points. As times go on, however, the coarse mesh predicts lower radiation/material temperature in C (at $t \gtrapprox 2$ ns) and D (at $t \gtrapprox 5$ ns) than the refined mesh. 
This is because, in the range  $1\lesssim{}t\lesssim 10$ ns, the course mesh has lower temperature around the material interface located at $x=2.5$, thus less energy input.
At the probe point E, which is located near the exit, radiation loss leads to cooling for a considerable period. Subsequently, the radiation and material temperatures rise due to the arrival of radiation from the left. Since the temperatures at probe points C and D are higher in the refined mesh than in the coarse mesh, the temperature rise occurs earlier in the refined mesh.

It should be noted that refining the uniform mesh by halving the spatial cell length leads to minimal changes in the numerical results. Significant differences in the early evolution of radiative transfer are observed only when a non-uniform spatial discretization is used to generate cell sizes smaller than the photon mean free path near the material interfaces. This clearly demonstrates that accurately capturing the Knudsen layer is crucial for this problem. And the solution obtained from the non-uniform mesh can be considered a benchmark.

Finally, despite the differences during the early phase of evolution, both spatial meshes yield the same radiation and material temperature distributions in the entire domain.

%

%% file: data/sec06.tex
\section{Conclusions}
\label{Conclusions} 

In summary, we have developed a general synthetic iterative scheme for solving the radiative transfer equation within the finite volume framework. The transport equation is solved using source iteration, which is boosted by a straightforwardly designed synthetic equation. Unlike the popular DSA method, where the synthetic equation is designed for the increment of radiation energy density, our synthetic equation directly addresses the evolution of radiation energy density. Also, unlike high-order/low-order methods, our approach does not require additional consistent terms. We have analytically proven that the GSIS possesses fast convergence and asymptotic-preserving properties.
Furthermore, to address the issue of disparate opacities in neighboring cells, we have developed an adaptive least square method for gradient approximation, which is applicable to both the kinetic and synthetic equations. Additionally, to enhance numerical stability, we have proposed three strategies of pseudo-time stepping methods. 

We have performed numerical simulations on steady-state problems, as well as transient Marshak waves and Tophat problem. Our results have demonstrated that GSIS provides a substantial speed-up over conventional source iteration schemes in optically thick regimes. Specifically, despite difficulties such as discontinuous opacity coefficients (abrupt changes by five orders of magnitude), non-orthogonal meshes, and large spatial gradients, appropriate control of the iteration process through pseudo-time stepping can achieve significant acceleration while ensuring stability.

Finally, it should be emphasized that although the primary motivation for developing asymptotic-preserving schemes is to enable the use of large spatial cell sizes in regions of high opacity, our numerical simulations of the time evolution of the Tophat problem reveal the importance of resolving the Knudsen layer near interfaces where opacity changes rapidly\footnote{In this case, the use of implicit scheme becomes essential.}, in the initial state of evolution. In contrast, the steady-state solution is far less sensitive to the numerical grid. The developed GSIS for the gray radiative transfer equation, executed on a refined mesh, provides a benchmark solution for validation. 

Moreover, we believe that the present method can be readily extended to study multi-frequency radiative transfer phenomena. In such cases, the advantages of using GSIS become even more pronounced, as solving the synthetic equation is considerably less computationally expensive than solving the full kinetic equation, in which the distribution function is defined in a six-dimensional space.

\section*{Declaration of competing interest} 
The authors declare that they have no known competing financial interests or personal relationships that could have appeared to influence the work reported in this paper.

\section*{Acknowledgments} 
This work is supported by the National Natural Science Foundation of China (12172162, 12202177). 
Special thanks are given to the Center for Computational Science and Engineering at the Southern University of Science and Technology. LW appreciates the discussions with Dr. Wenjun Sun and Dr Chang Liu regarding the Tophat problem.

%% file: data/appendix.tex
\section{The impact of spatial discretizaitons in DSA and GSIS}
\label{Appdx:vrf}

The discretizations of the transport and synthetic equations affect the numerical stability and convergence rate for the traditional DSA. However, the GSIS maintain fast convergence and asymptotic preserving properties in different kinds of discretizations.
To demonstrate this, we consider a one-dimensional steady-state problem with the source term:
\begin{equation}
  \label{1d_constSource}
  \cos \theta \frac{\partial I}{\partial x} = \frac{1}{\epsilon} \left ( \frac{\rho}{4 \pi}  - I \right ) + \frac{\epsilon}{4\pi}.
\end{equation}
The computational domain is $x\in [0,1]$,  and vacuum boundary conditions are applied. When $\epsilon \to 0$, the radiation transfer is described by the following diffusion equation:
\begin{equation}
  \label{1d_constSource_diffu}
   -\frac{\partial}{\partial x} \left ( \frac{1}{3} \frac{\partial \rho}{\partial x} \right ) =1,
\end{equation}
and the radiation energy distribution has the analytical solution:
\begin{equation}
  \label{1d_constSource_diffu_analy}
  \rho (x) = -\frac{3}{2} \left ( x - \frac{1}{2}\right )^2 + \frac{3}{8}.
\end{equation}

In DSA, the increment $\Phi=\rho^{m+1}-\rho^{m+\frac{1}{2}}$ satisfies 
\begin{equation}
  -\frac{\epsilon}{3}\frac{\partial^2 \Phi ^{m+1} }{\partial x^2}  = \frac{1}{\epsilon}\left ( \rho^{m+\frac{1}{2}} -  \rho^{m}\right ),
\end{equation}
while in GSIS (note that GSIS in this case degenerates to Alcouffe's DSA for neutron transport~\cite{Alcouffe01101977}), $\rho$ satisfies the following synthetic equation: 
\begin{equation}
   -\frac{\epsilon}{3}\frac{\partial^2 \rho ^{m+1} }{\partial x^2}  = {\epsilon}  -\frac{\partial}{\partial x} \left (
  F_x^{m+\frac{1}{2}}+\frac{\epsilon}{3} \frac{\partial \rho^{m+\frac{1}{2}}}{\partial x} \right ),
\end{equation}
where the flux is
\begin{equation}
    F_x= 2\pi \int_{0}^{\pi} I \cos \theta\sin \theta d \theta.
\end{equation}

Discretizing Eq.~\eqref{1d_constSource}, at each control volume $x_j$ we have:
\begin{equation}
  \label{1d_constSource_d}
   \frac{\cos \theta}{\varDelta x_j} \left ( I_{j+\frac{1}{2}} - I_{j-\frac{1}{2}} \right ) = \frac{1}{\epsilon} \left ( \frac{\rho_j}{4 \pi}  - I_j \right ) + \frac{\epsilon}{4\pi}.
\end{equation}
The transport equation is solved using the following schemes (assuming $\cos \theta>0$):
\begin{equation}
    \begin{aligned}
        I_{j+\frac{1}{2}} = I_j, \quad 
        \text{First-order upwind (UW1),} \\
        I_{j+\frac{1}{2}} = I_j +\frac{I_{i+1} - I_{j-1}}{4}, \quad \text{Second-order upwind (UW2),}\\
        I_j = \frac{1}{2} \left ( I_{j+\frac{1}{2}} + I_{j-\frac{1}{2}} \right ),\quad \text{Diamond difference (DD).}
    \end{aligned}
\end{equation}

\begin{table}[t]
  \centering
  \small
  \setlength{\tabcolsep}{3pt}
  \caption{Comparisons in iteration numbers between the GSIS and DSA for different discretization schemes. ``--'' means the iteration takes too long or the iteration is unstable. 
  } 
  \label{table:1d_constSource}
  \begin{tabular}{lccccccc}
      \toprule
      \multirow{2}{*}{$\epsilon$} & {SI} & \multicolumn{2}{c}{GSIS } & \multicolumn{3}{c}{DSA} \\
      \cmidrule(lr){2-2} \cmidrule(lr){3-4} \cmidrule(lr){5-7} 
      & UW2 & UW1 & UW2 & UW1 & UW2 & DD \\
      \midrule
      $1$ & 29 & 22 & 22 & 22 & 22 & 22\\
      $10^{-1}$ & 394 & 24 & 64 & 23 & 23 & 23  \\
      $10^{-2}$ & 16206 & 65 & 53 & 29 & 26 & 36  \\
      $10^{-4}$ & -- & 118 & 48 & -- & -- & --  \\
      $10^{-6}$ & -- & 1415 & 50 & -- & -- & --  \\
      \bottomrule
  \end{tabular}
  \label{tab:1d_constSource}
\end{table}

\begin{figure}[t!]
  \centering
    \subfigure[first-order upwind, GSIS]{\includegraphics[width=0.43\textwidth,trim={20 80 20 200},clip]{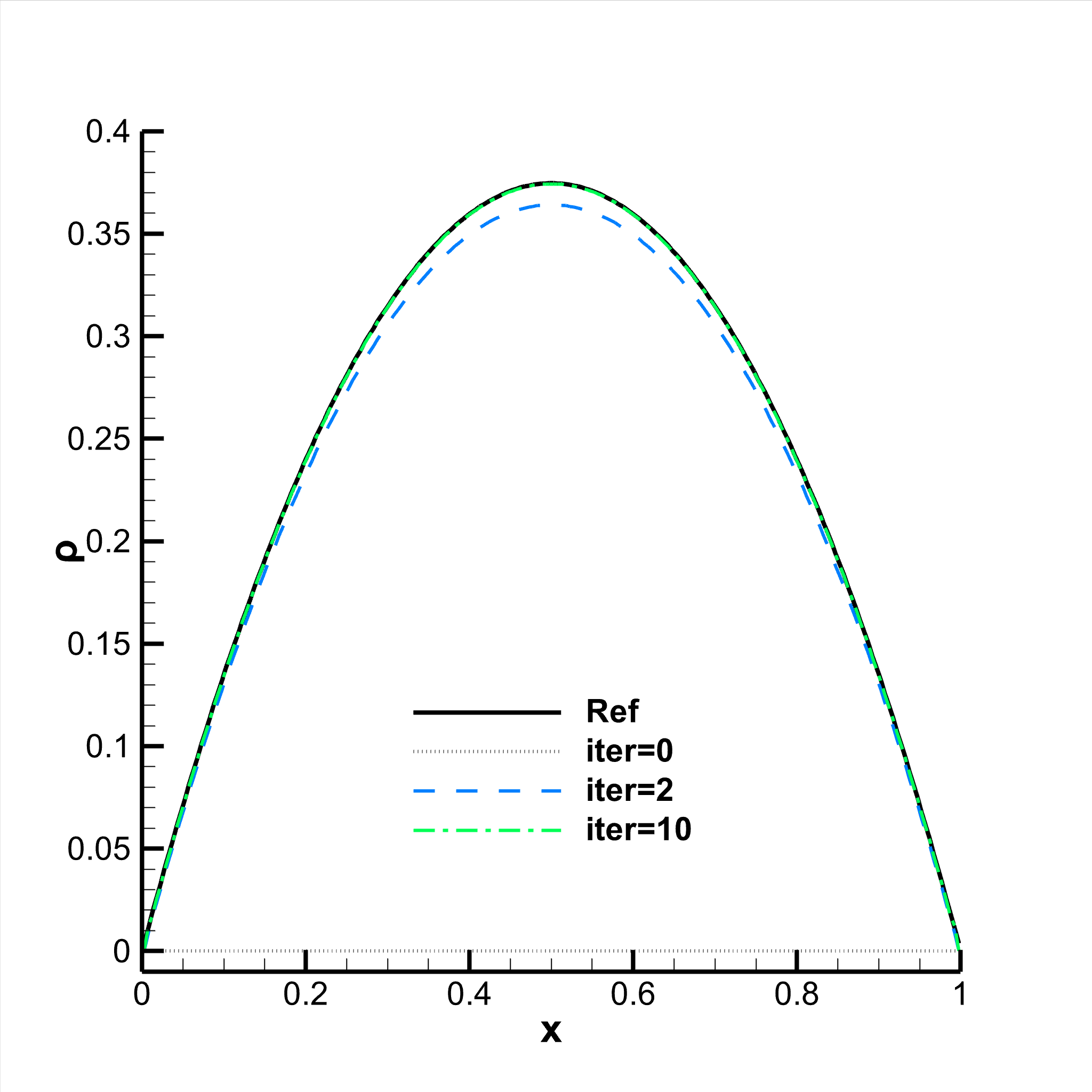}}
   \subfigure[second-order upwind, GSIS]{ \includegraphics[width=0.43\textwidth,trim={20 80 20 200},clip]{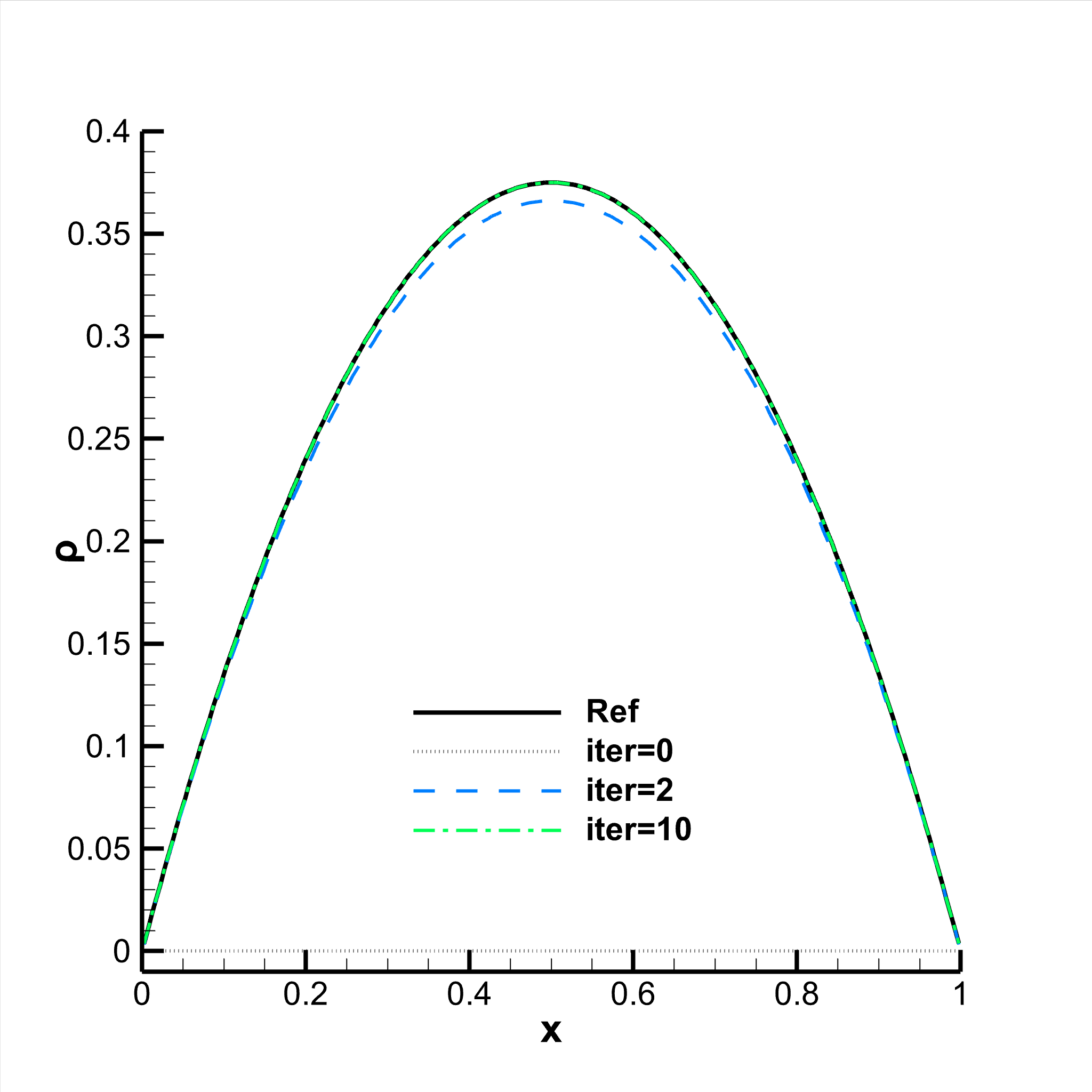}}\\
     \subfigure[second-order upwind, DSA]{\includegraphics[width=0.43\textwidth,trim={20 80 20 200},clip]{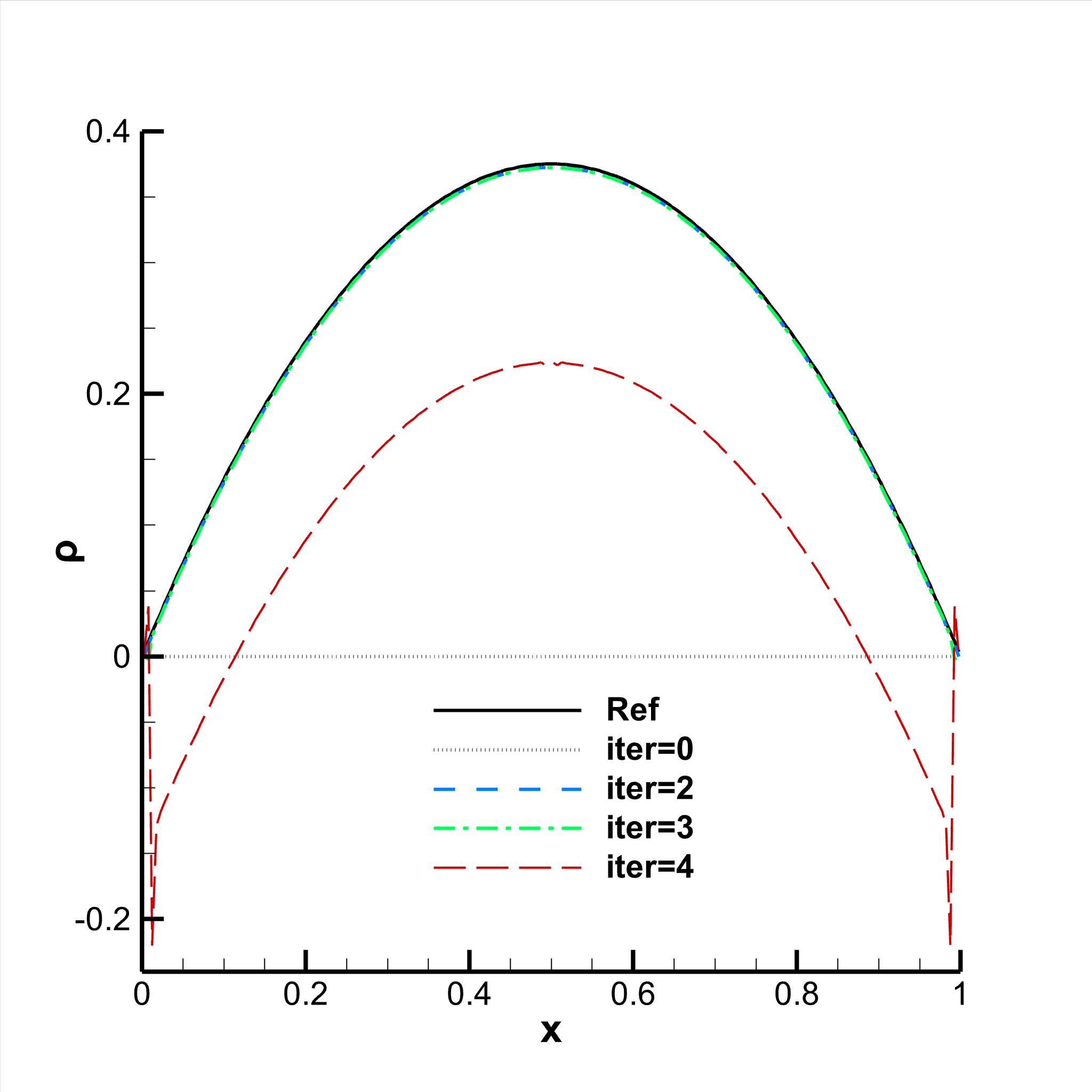}}
     \subfigure[diamond difference, DSA]{ \includegraphics[width=0.43\textwidth,trim={20 80 20 200},clip]{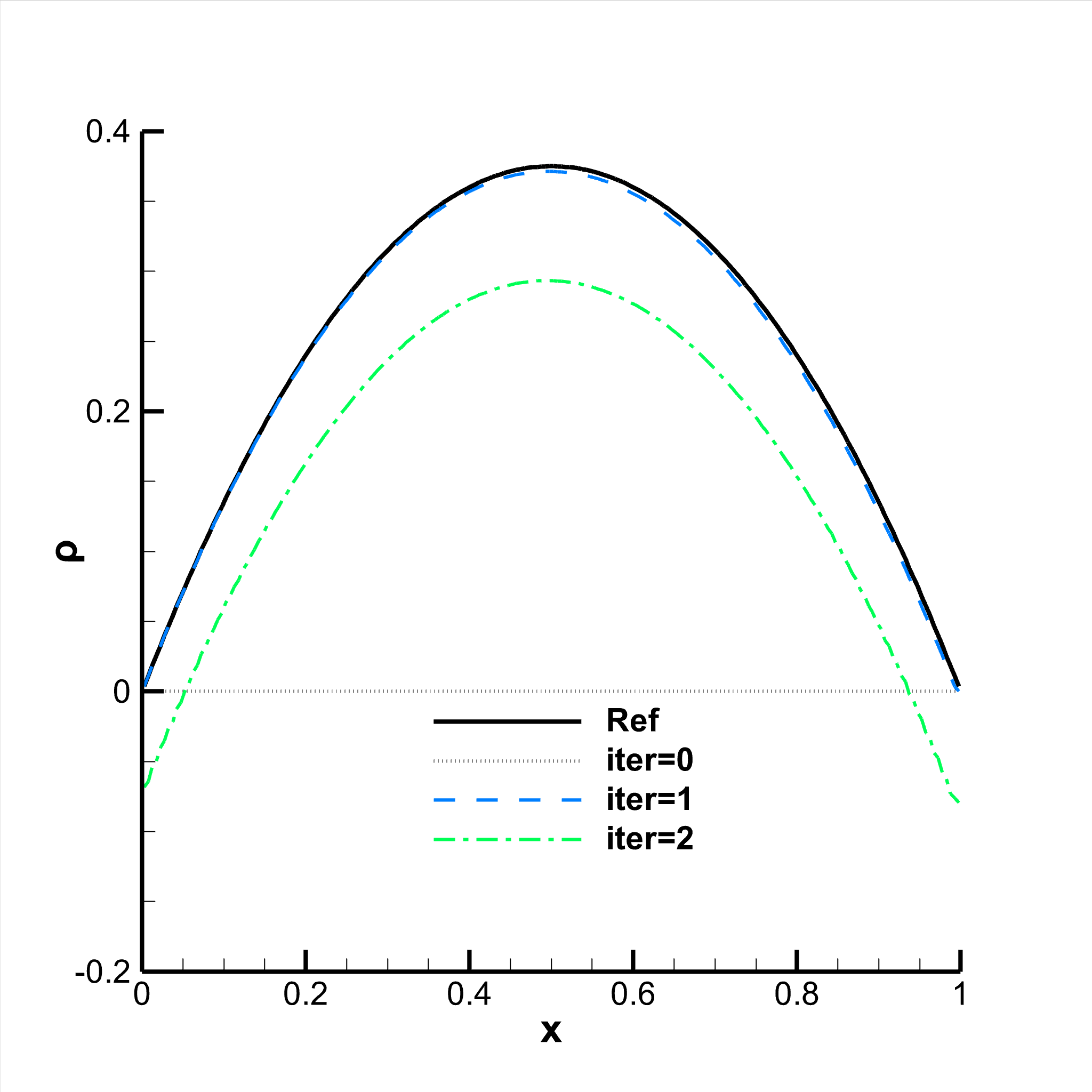}}
  \caption{Iteration history when the transport equation  \eqref{1d_constSource} with  $\epsilon=10^{-4}$ is discretized by different schemes, accelerated by GSIS and DSA.
 }
  \label{fig:1d_constSource_1e-4_hsitory1}
\end{figure}

Upwind schemes allow the discontinuities in $I$ at interfaces, while the diamond difference ensures continuity. In implementation, upwind schemes compute cell boundary values according to cell center values, whereas the diamond difference updates cell boundary values and then cell center values.


A uniform mesh of $N_x=200$ is used, with GSIS and DSA employed to accelerate the iterations. The diamond difference recovers the central difference scheme of the diffusion equation at the mesh nodes in the asymptotic limit \cite{larsen1987asymptotic}, and its acceleration equation also uses the central difference scheme:
\begin{equation}
  \label{DSA_cd}
   \frac{\partial^2 \Phi}{\partial x^2} \bigg|_{j+\frac{1}{2}} = \frac{\Phi_{j+\frac{3}{2}} - 2\Phi_{j+\frac{1}{2}} + \Phi_{j-\frac{1}{2}}}{\varDelta x^2},
\end{equation}
while others use the finite-volume diffusion flux scheme, equivalent to the central difference scheme for cell centers under these conditions:
\begin{equation}
  \label{1d_diffu_flux}
  \begin{aligned}
  \frac{\partial^2 \rho}{\partial x^2} \bigg|_{j} &= \frac{\rho_{j+1} - 2\rho_{j} + \rho_{j-1}}{\varDelta x^2},\\
  \frac{\partial F_x}{\partial x} \bigg|_{j} &=  \frac{F_{x,j+1} - F_{x,j-1}}{2\varDelta x}. 
  \end{aligned}
\end{equation}

The number of iterations required for convergence ($\text{Err}<10^{-6}$) is shown in Table \ref{tab:1d_constSource}. When the grid size is smaller than $\epsilon$, i.e., $\epsilon = 1, 0.1, 0.01$, DSA is effective for all difference schemes. However, when the grid size is larger than $\epsilon$, the DSA is unable to converge. 
GSIS with the second-order upwind scheme converges within tens of iterations, while GSIS with the first-order upwind scheme requires more iterations for small $\epsilon$. 

\begin{figure}[t!]
  \centering
     \subfigure[$\epsilon=1$]{\includegraphics[width=0.48\textwidth,trim={20 80 20 200},clip]{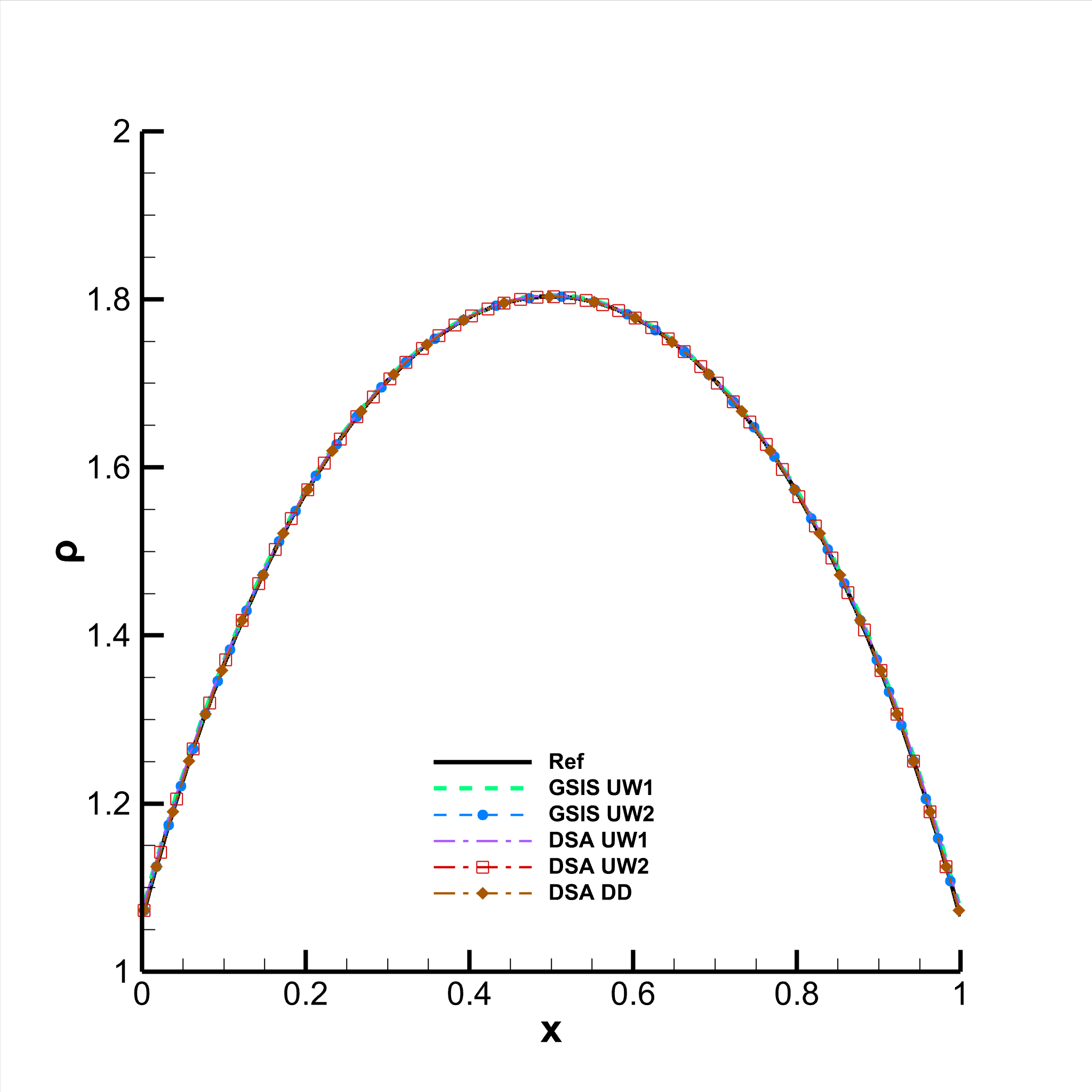}}
      \subfigure[$\epsilon=10^{-2}$]{\includegraphics[width=0.48\textwidth,trim={20 80 20 200},clip]{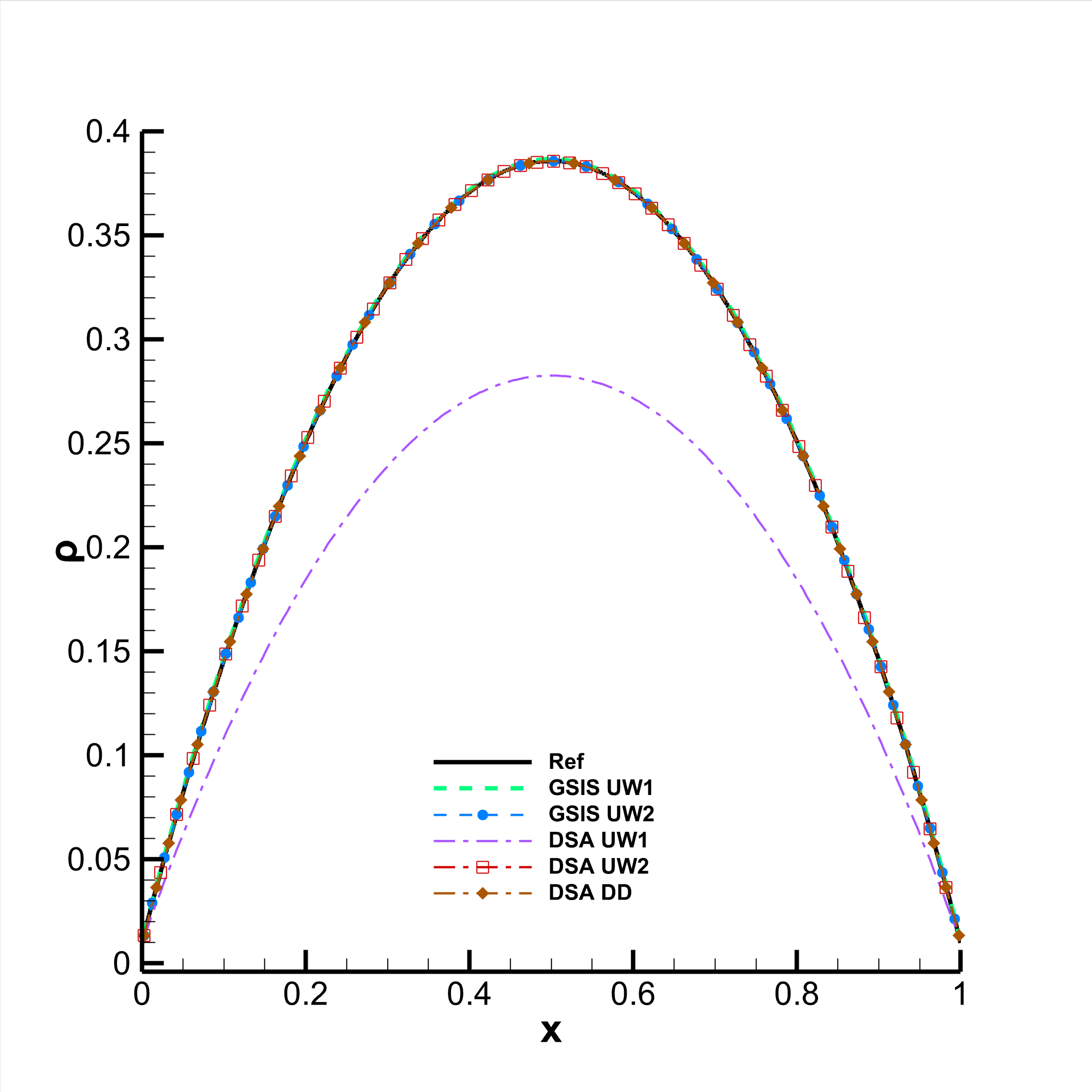}}\\
      \subfigure[$\epsilon=10^{-4}$]{\includegraphics[width=0.48\textwidth,trim={20 80 20 200},clip]{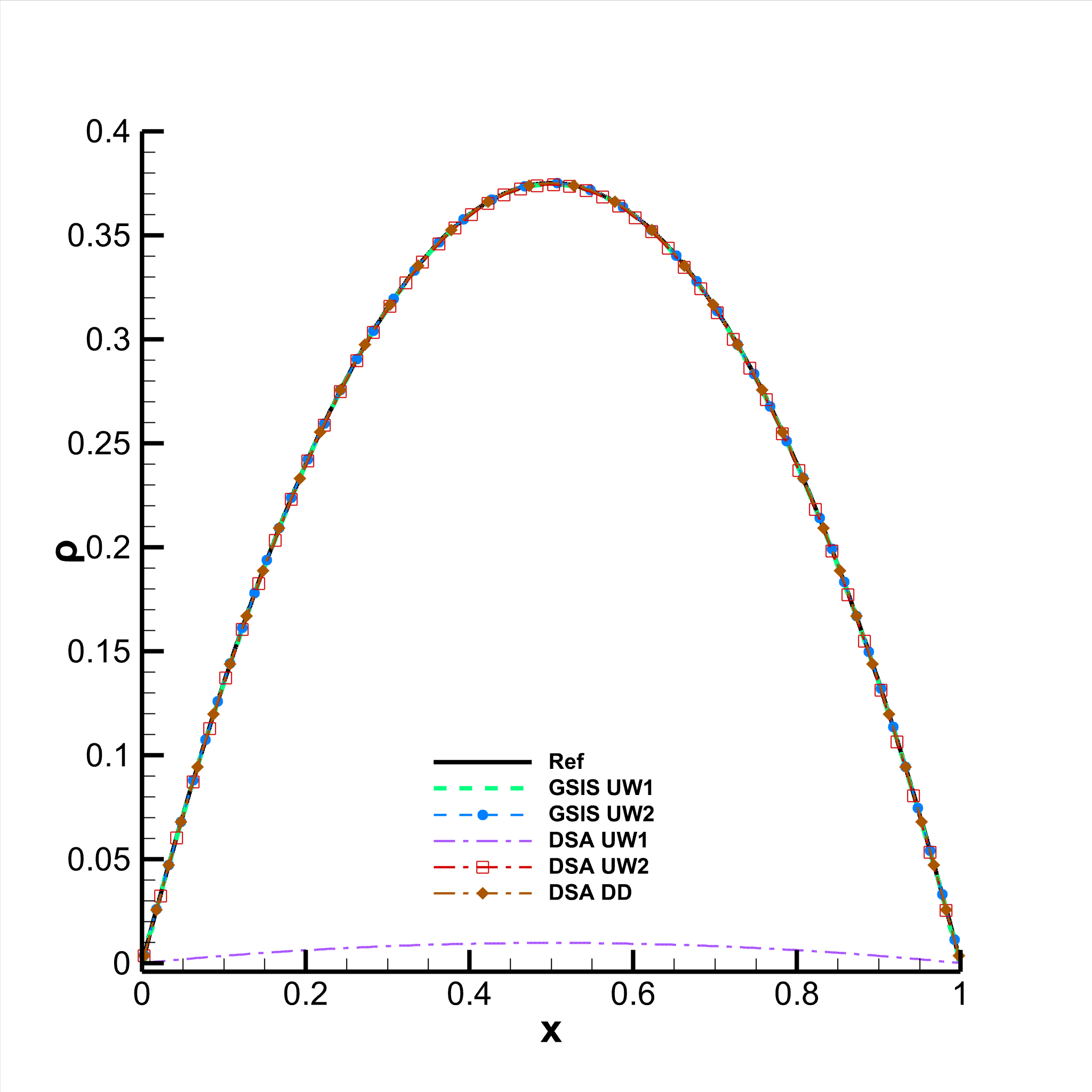}}
      \subfigure[$\epsilon=10^{-6}$]{\includegraphics[width=0.48\textwidth,trim={20 80 20 200},clip]{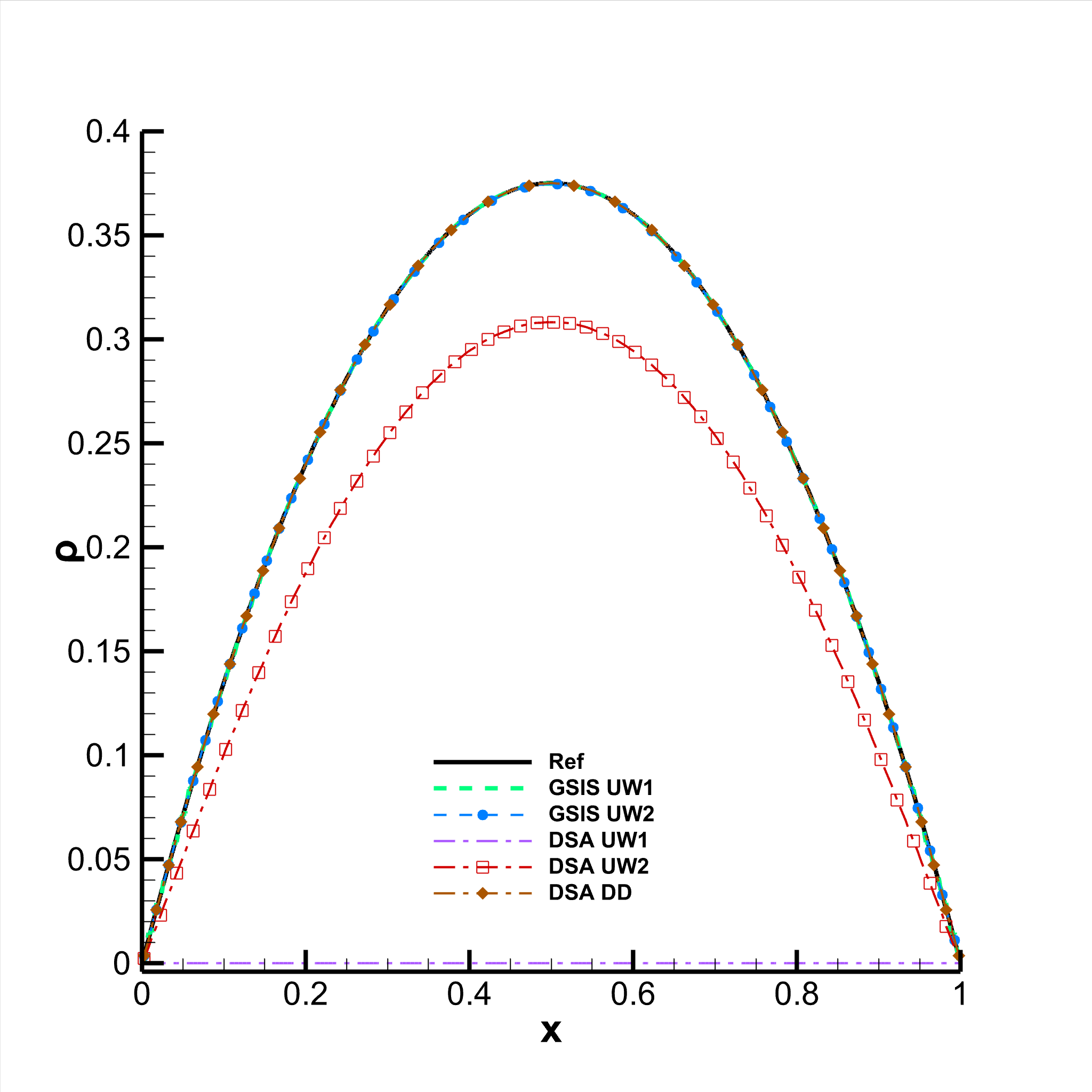}}
  \caption{Radiation energy computed using different difference schemes. The reference solutions in (a, b)  are obtained via mesh refinement, while these in (c, d)  are analytical approximations  \eqref{1d_constSource_diffu_analy}.}
  \label{fig:1d_constSource}
\end{figure}

The iteration histories for GSIS and DSA at $\epsilon=10^{-4}$ are shown in Fig.~\ref{fig:1d_constSource_1e-4_hsitory1}. After initial acceleration, both evolve significantly towards the reference solution. GSIS remains stable and nearly overlaps with the reference solution after 10 iterations, while DSA exhibits oscillations and negative values, leading to divergence.  The correction originates from the error equation, making it more sensitive to small perturbations as $\epsilon\to 0$.

To verify the asymptotic preserving property, i.e., the impact of the transport equation discretization on the final results, the DSA correction $\Phi$ is multiplied  by a factor at the order of $\epsilon$ for small $\epsilon$ to stabilize the calculation, and a sufficiently small error criterion is set to ensure full convergence, although this introduces significant computational overhead, so iteration counts are not recorded. As shown in Fig.~\ref{fig:1d_constSource}, DSA converges to the discrete solution of the transport equation. As $\epsilon\to 0$, numerical dissipation of upwind schemes causes nonphysical solutions, and it is more pronounced in the first-order upwind scheme. The diamond difference, which has the diffusion limit, aligns with the reference solution. Remarkably, GSIS corrects the asymptotic behavior even for the first-order upwind scheme.


\section{The direction independency of numerical gradient operator}
\label{Appdx:proof}

In cell-centered finite-volume method, the values of a flow variable $U$ at all cell centers are denoted by $\bm{U}=[U_1,U_2,...,U_N]$, corresponding one-to-one with the cell indices $\bm{N}=[1,2,...,N]$. The values at the common face between any two adjacent cells $i$ and $j$, namely $U_{i,j}$ and $U_{j,i}$, can be expressed as linear combinations of the cell center values:
\begin{equation}
  \label{commonface_linear_rec}
  U_{i,j}=\mathcal{F}_{i,j} \bm{U},\quad U_{j,i}=\mathcal{F}_{j,i} \bm{U},
\end{equation}
The linear mapping $\mathcal{F}_{i,j}$ depends solely on the mesh and discretization scheme. Considering the upwind property, the part of $U_{ij}$ that contributes to the flux along the direction $\bm{\Omega}$ is:
\begin{equation}
  \label{commonface_upwind}
  U_{ij} \left (\bm{\Omega}  \right ) = \frac{1}{2}\left ( U_{i,j} + U_{j,i}\right ) + \gamma \left ( U_{i,j} - U_{j,i}\right ) \text{sign} \left (\bm{\Omega} \cdot \bm{n}_{ij} \right ),
\end{equation}
where $0\le \gamma \le \frac{1}{2}$ is a fixed constant, and $\bm{n}_{ij}$ is the interface normal vector pointing from cell $i$ to cell $j$. By iterating over all adjacent cells $j$, the effective average gradient of control cell $i$ is obtained by summing the fluxes:
\begin{equation}
  \label{grad_eff}
  \nabla_{\bm{\delta}}\left (\bm{\Omega}\right )U_i = \frac{1}{V_i}\sum_{j \in N(i)} U_{ij}\left (\bm{\Omega}\right ) A_{ij},
\end{equation}
where $V_i$ is the volume of control cell $i$, and $A_{ij}$ is the area of interface $ij$. The average numerical gradient in the opposite direction of cell $U_i$ has a unique representation:
\begin{equation}
  \label{grad_mean}
  \begin{aligned}
  \bar{\nabla}_{\bm{\delta}} U_i &=  \frac{1}{V_i}\sum_{j \in N(i)} \frac{1}{2}\left [U_{ij}\left (\bm{\Omega}\right ) + U_{ij}\left (\bm{-\Omega}\right )\right ] A_{ij}\\
   &= \frac{1}{V_i}\sum_{j \in N(i)} \frac{1}{2}\left ( \mathcal{F}_{i,j} + \mathcal{F}_{j,i} \right )\bm{U} A_{ij},
  \end{aligned}
\end{equation}
which is direction-independent. Thus Eq.~\eqref{grad_operator_mean} is proved.

%% file: main.bbl
\begin{thebibliography}{10}
\expandafter\ifx\csname url\endcsname\relax
  \def\url#1{\texttt{#1}}\fi
\expandafter\ifx\csname urlprefix\endcsname\relax\def\urlprefix{URL }\fi
\expandafter\ifx\csname href\endcsname\relax
  \def\href#1#2{#2} \def\path#1{#1}\fi

\bibitem{Zheleznyakov1996}
V.~V. Zheleznyakov, Transfer of radiation in astrophysical plasmas, Springer
  Netherlands, Dordrecht, 1996, pp. 155--197.

\bibitem{Kallman2023}
T.~R. Kallman, An Overview of Astrophysical Plasmas, Springer Nature Singapore,
  Singapore, 2023, pp. 151--172.

\bibitem{grinstein2025transition}
F.~F. Grinstein, V.~P. Chiravalle, B.~M. Haines, R.~K. Greene, F.~S. Pereira,
  Transition in {ICF} capsule implosions, Flow, Turbulence and Combustion 114
  (2025) 801--825.

\bibitem{James1979}
W.~R.~M. James J.~Duderstadt, Transport theory, Wiley, New York, 1979.

\bibitem{osti_4376236}
B.~G. Carlson, Solution of the transport equation by {$S_n$} approximations,
  Tech. rep., Los Alamos National Lab., Los Alamos, NM (United States) (02
  1955).

\bibitem{OU1982271}
S.-C.~S. Ou, K.-N. Liou, Generalization of the spherical harmonic method to
  radiative transfer in multi-dimensional space, Journal of Quantitative
  Spectroscopy and Radiative Transfer 28 (1982) 271--288.

\bibitem{Raithby1990}
G.~D. Raithby, E.~H. Chui, A finite-volume method for predicting a radiant heat
  transfer in enclosures with participating media, Journal of Heat Transfer 112
  (1990) 415--423.

\bibitem{Razzaque1983}
M.~M. Razzaque, D.~E. Klein, J.~R. Howell, Finite element solution of radiative
  heat transfer in a two-dimensional rectangular enclosure with gray
  participating media, Journal of Heat Transfer 105 (1983) 933--936.

\bibitem{FLECK1971313}
J.~Fleck, J.~Cummings, An implicit monte carlo scheme for calculating time and
  frequency dependent nonlinear radiation transport, Journal of Computational
  Physics 8 (1971) 313--342.

\bibitem{howell1998monte}
J.~R. Howell, The {Monte Carlo} method in radiative heat transfer, Journal of
  Heat Transfer 120 (1998) 547--560.

\bibitem{larsen1987asymptotic}
E.~W. Larsen, J.~E. Morel, W.~F. Miller~Jr, Asymptotic solutions of numerical
  transport problems in optically thick, diffusive regimes, Journal of
  Computational Physics 69 (1987) 283--324.

\bibitem{larsen1989asymptotic}
E.~W. Larsen, J.~E. Morel, Asymptotic solutions of numerical transport problems
  in optically thick, diffusive regimes {II}, Journal of Computational Physics
  83 (1989) 212–236.

\bibitem{Larsen01121992}
E.~W. Larsen, The asymptotic diffusion limit of discretized transport problems,
  Nuclear Science and Engineering 112 (1992) 336--346.

\bibitem{farmer1998comparison}
J.~T. Farmer, J.~R. Howell, Comparison of {Monte Carlo} strategies for
  radiative transfer in participating media, in: Advances in heat transfer,
  Vol.~31, Elsevier, 1998, pp. 333--429.

\bibitem{Erturk2018}
H.~Erturk, J.~Howell, Monte Carlo Methods for Radiative Transfer, Springer
  Cham, 2018.

\bibitem{ADAMS20023}
M.~L. Adams, E.~W. Larsen, Fast iterative methods for discrete-ordinates
  particle transport calculations, Progress in Nuclear Energy 40 (2002) 3--159.

\bibitem{Kopp01091963}
H.~J. Kopp, Synthetic method solution of the transport equation, Nuclear
  Science and Engineering 17 (1963) 65--74.

\bibitem{GOLDIN1964136}
V.~Gol'din, A quasi-diffusion method of solving the kinetic equation, USSR
  Computational Mathematics and Mathematical Physics 4 (1964) 136--149.

\bibitem{Wieselquist2014}
W.~A. Wieselquist, D.~Y. Anistratov, J.~E. Morel, A cell-local finite
  difference discretization of the low-order quasidiffusion equations for
  neutral particle transport on unstructured quadrilateral meshes, Journal of
  Computational Physics 273 (2014) 343–357.

\bibitem{Ramone01031997}
G.~L. Ramone, M.~L. Adams, P.~F. Nowak, A transport synthetic acceleration
  method for transport iterations, Nuclear Science and Engineering 125 (1997)
  257--283.

\bibitem{Alcouffe01101977}
R.~E. Alcouffe, Diffusion synthetic acceleration methods for the
  diamond-differenced discrete-ordinates equations, Nuclear Science and
  Engineering 64 (1977) 344--355.

\bibitem{Larsen01011984}
E.~W. Larsen, Diffusion-synthetic acceleration methods for discrete-ordinates
  problems, Transport Theory and Statistical Physics 13 (1984) 107--126.

\bibitem{Lorence01041989}
L.~J. Lorence, J.~E. Morel, E.~W. Larsen, An {$S_2$} synthetic acceleration
  scheme for the one-dimensional {$S_n$} equations with linear discontinuous
  spatial differencing, Nuclear Science and Engineering 101 (1989) 341--351.

\bibitem{Morel01091982}
J.~E. Morel, A synthetic acceleration method for discrete ordinates
  calculations with highly anisotropic scattering, Nuclear Science and
  Engineering 82 (1982) 34--46.

\bibitem{ALCOUFFE1985}
R.~E. Alcouffe, B.~A. Clark, E.~W. Larsen, 4 - the diffusion-synthetic
  acceleration of transport iterations, with application to a radiation
  hydrodynamics problem, in: J.~U. Brackbill, B.~I. Cohen (Eds.), Multiple Time
  Scales, Academic Press, 1985, pp. 73--111.

\bibitem{Larsen01091982}
E.~W. Larsen, Unconditionally stable diffusion-synthetic acceleration methods
  for the slab geometry discrete ordinates equations. {Part I}: Theory, Nuclear
  Science and Engineering 82 (1982) 47--63.

\bibitem{McCoy01091982}
D.~R. McCoy, E.~W. Larsen, Unconditionally stable diffusion-synthetic
  acceleration methods for the slab geometry discrete ordinates equations.
  {Part II}: Numerical results, Nuclear Science and Engineering 82 (1982)
  64--70.

\bibitem{AZMY2002213}
Y.~Azmy, Unconditionally stable and robust adjacent-cell diffusive
  preconditioning of weighted-difference particle transport methods is
  impossible, Journal of Computational Physics 182 (2002) 213--233.

\bibitem{Warsa2004}
J.~S. Warsa, T.~A. Wareing, J.~E. Morel, Krylov iterative methods and the
  degraded effectiveness of diffusion synthetic acceleration for
  multidimensional sn calculations in problems with material discontinuities,
  Nuclear Science and Engineering 147 (2004) 218--248.

\bibitem{Park01052012}
H.~Park, D.~A. Knoll, R.~M. Rauenzahn, A.~B. Wollaber, J.~D. Densmore, A
  consistent, moment-based, multiscale solution approach for thermal radiative
  transfer problems, Transport Theory and Statistical Physics 41 (2012)
  284--303.

\bibitem{CHACON201721}
L.~Chacón, G.~Chen, D.~Knoll, C.~Newman, H.~Park, W.~Taitano, J.~Willert,
  G.~Womeldorff, Multiscale high-order/low-order ({HOLO}) algorithms and
  applications, Journal of Computational Physics 330 (2017) 21--45.

\bibitem{Park01112020}
H.~Park, Toward asymptotic diffusion limit preserving high-order, low-order
  method, Nuclear Science and Engineering 194 (2020) 952--970.

\bibitem{Klar1998}
A.~Klar, An asymptotic-induced scheme for nonstationary transport equations in
  the diffusive limit, SIAM Journal on Numerical Analysis 35 (1998) 1073--1094.

\bibitem{Jin2000}
S.~Jin, L.~Pareschi, G.~Toscani, Uniformly accurate diffusive relaxation
  schemes for multiscale transport equations, SIAM Journal on Numerical
  Analysis 38 (2000) 913--936.

\bibitem{FILBET20107625}
F.~Filbet, S.~Jin, A class of asymptotic-preserving schemes for kinetic
  equations and related problems with stiff sources, Journal of Computational
  Physics 229 (2010) 7625--7648.

\bibitem{XU20107747}
K.~Xu, J.-C. Huang, A unified gas-kinetic scheme for continuum and rarefied
  flows, Journal of Computational Physics 229 (2010) 7747--7764.

\bibitem{MIEUSSENS2013138}
L.~Mieussens, On the asymptotic preserving property of the unified gas kinetic
  scheme for the diffusion limit of linear kinetic models, Journal of
  Computational Physics 253 (2013) 138--156.

\bibitem{UGKS_SUN2015}
W.~Sun, S.~Jiang, K.~Xu, An asymptotic preserving unified gas kinetic scheme
  for gray radiative transfer equations, Journal of Computational Physics 285
  (2015) 265--279.

\bibitem{SUN2015222}
W.~Sun, S.~Jiang, K.~Xu, S.~Li, An asymptotic preserving unified gas kinetic
  scheme for frequency-dependent radiative transfer equations, Journal of
  Computational Physics 302 (2015) 222--238.

\bibitem{Sun_Jiang_Xu_2017}
W.~Sun, S.~Jiang, K.~Xu, An implicit unified gas kinetic scheme for radiative
  transfer with equilibrium and non-equilibrium diffusive limits,
  Communications in Computational Physics 22 (2017) 889–912.

\bibitem{Guo2013}
Z.~Guo, K.~Xu, R.~Wang, Discrete unified gas kinetic scheme for all {Knudsen}
  number flows: Low-speed isothermal case, Physical Review E 88 (2013) 033305.

\bibitem{SONG2023123799}
X.~Song, Y.~Zhang, X.~Zhou, C.~Zhang, Z.~Guo, Modified steady discrete unified
  gas kinetic scheme for multiscale radiative heat transfer, International
  Journal of Heat and Mass Transfer 203 (2023) 123799.

\bibitem{SONG2023120349}
X.~Song, Y.~Zhang, X.~Zhou, Z.~Guo, A diffusion synthetic acceleration method
  for steady discrete unified gas kinetic scheme in radiative heat transfer,
  Applied Thermal Engineering 227 (2023) 120349.

\bibitem{Olivier2017}
S.~S. Olivier, J.~E. Morel, Variable {Eddington} factor method for the {$S_n$}
  equations with lumped discontinuous {Galerkin} spatial discretization coupled
  to a drift-diffusion acceleration equation with mixed finite-element
  discretization, Journal of Computational and Theoretical Transport 46 (2017)
  6--7.

\bibitem{Lou2019JCP}
J.~J. Lou, J.~E. Morel, N.~A. Gentile, A variable {Eddington} factor method for
  the {1-D grey radiative transfer equations with discontinuous Galerkin} and
  mixed finite-element spatial differencing, Journal of Computational Physics
  393 (2019) 258--277.

\bibitem{SU2020jcp}
W.~Su, L.~Zhu, P.~Wang, Y.~Zhang, L.~Wu, Can we find steady-state solutions to
  multiscale rarefied gas flows within dozens of iterations?, Journal of
  Computational Physics 407 (2020) 109245.

\bibitem{ZHU2021110091}
L.~Zhu, X.~Pi, W.~Su, Z.-H. Li, Y.~Zhang, L.~Wu, General synthetic iterative
  scheme for nonlinear gas kinetic simulation of multi-scale rarefied gas
  flows, Journal of Computational Physics 430 (2021) 110091.

\bibitem{Su2020siam}
W.~Su, L.~Zhu, L.~Wu, Fast convergence and asymptotic preserving of the general
  synthetic iterative scheme, SIAM Journal on Scientific Computing 42 (2020)
  B1517--B1540.

\bibitem{liu2024further}
W.~Liu, Y.~Zhang, J.~Zeng, L.~Wu, Further acceleration of multiscale simulation
  of rarefied gas flow via a generalized boundary treatment, Journal of
  Computational Physics 503 (2024) 112830.

\bibitem{GSISPolyGas_ZENG2023}
J.~Zeng, R.~Yuan, Y.~Zhang, Q.~Li, L.~Wu, General synthetic iterative scheme
  for polyatomic rarefied gas flows, Computers and Fluids 265 (2023) 105998.

\bibitem{zhang2024efficient}
Y.~Zhang, J.~Zeng, R.~Yuan, W.~Liu, Q.~Li, L.~Wu, Efficient parallel solver for
  rarefied gas flow using {GSIS}, Computers and Fluids 281 (2024) 106374.

\bibitem{Valougeorgis2003}
D.~Valougeorgis, S.~Naris, Acceleration schemes of the discrete velocity
  method: Gaseous flows in rectangular microchannels, SIAM Journal on
  Scientific Computing 25 (2003) 534--552.

\bibitem{SHI2025113501}
Z.~Shi, Y.~Zhang, L.~Wu, General synthetic iterative scheme for non-equilibrium
  dense gas flows, Journal of Computational Physics 520 (2025) 113501.

\bibitem{Zhangchuang_2021}
C.~Zhang, S.~Chen, Z.~Guo, L.~Wu, {A fast synthetic iterative scheme for the
  stationary phonon Boltzmann transport equation}, International Journal of
  Heat and Mass Transfer 174 (2021) 121308.

\bibitem{LARSEN1988459}
E.~W. Larsen, A grey transport acceleration method for time-dependent radiative
  transfer problems, Journal of Computational Physics 78 (1988) 459--480.

\bibitem{MOREL1996445}
J.~Morel, T.~A. Wareing, K.~Smith, A linear-discontinuous spatial differencing
  scheme for {$S_n$} radiative transfer calculations, Journal of Computational
  Physics 128 (1996) 445--462.

\bibitem{LARSEN1983285}
E.~Larsen, G.~Pomraning, V.~Badham, Asymptotic analysis of radiative transfer
  problems, Journal of Quantitative Spectroscopy and Radiative Transfer 29
  (1983) 285--310.

\bibitem{Emerson2010}
B.~John, X.-J. Gu, D.~R. Emerson, Investigation of heat and mass transfer in a
  lid-driven cavity under nonequilibrium flow conditions, Numerical Heat
  Transfer, Part B 52 (2010) 287--303.

\bibitem{ShiYi2023JCP}
Y.~Shi, A maximum principle preserving implicit {Monte-Carlo} method for
  frequency-dependent radiative transfer equations, Journal of Computational
  Physics 495 (2023) 112552.

\bibitem{Brath1989}
T.~Barth, D.~Jespersen, The design and application of upwind schemes on
  unstructured meshes, 1989, p. 366.

\bibitem{Chang01041992}
K.~C. Chang, U.~J. Payne, Numerical treatment of diffusion coefficients at
  interfaces, Numerical Heat Transfer, Part A: Applications 21~(3) (1992)
  363--376.

\bibitem{RayEffect_ZHU2020}
Y.~Zhu, C.~Zhong, K.~Xu, Ray effect in rarefied flow simulation, Journal of
  Computational Physics 422 (2020) 109751.

\bibitem{FEM-UGKS_XU2020}
X.~Xu, W.~Sun, S.~Jiang, An asymptotic preserving angular finite element based
  unified gas kinetic scheme for gray radiative transfer equations, Journal of
  Quantitative Spectroscopy and Radiative Transfer 243 (2020) 106808.

\bibitem{LARSEN201382}
E.~W. Larsen, A.~Kumar, J.~E. Morel, Properties of the implicitly
  time-differenced equations of thermal radiation transport, Journal of
  Computational Physics 238 (2013) 82--96.

\bibitem{IUGKWP_photon}
C.~Liu, W.~Li, Y.~Wang, P.~Song, K.~Xu, An implicit unified gas-kinetic
  wave–particle method for radiative transport process, Physics of Fluids 35
  (2023) 112013.

\bibitem{AP-IMEX}
J.~Fu, W.~Li, P.~Song, Y.~Wang, An asymptotic-preserving {IMEX} method for
  nonlinear radiative transfer equation, Journal of Scientific Computing 92
  (2022) 112013.

\bibitem{GENTILE2001543}
N.~Gentile, Implicit {Monte Carlo} diffusion—an acceleration method for
  {Monte Carlo} time-dependent radiative transfer simulations, Journal of
  Computational Physics 172 (2001) 543--571.

\end{thebibliography}
